\documentclass{report}
\usepackage{amsmath}
\usepackage{graphicx}
\usepackage{rotating}
\usepackage{epsfig}
\begin{document}
\begin{titlepage}
\begin{center}
\vspace*{1.0cm}
\vskip 2cm
\vskip 2cm
{\LARGE \bf Letter of Intent\\
\vspace*{6mm}
 to build an Off-axis Detector\\
\vspace*{6mm}
 to study $\nu_{\mu}\rightarrow\nu_e$ oscillations \\
\vspace*{6mm}
 with the NuMI Neutrino Beam} 
\vskip 1cm
{\it \large Version 6.0}\\     
\vspace*{5mm}
{\it \large \today}\\
\vspace*{1.0cm}
Argonne~-~Athens~-~Berkeley~-~Boston~-~Caltech~-~Chicago~-
~\mbox{College de France}~-~Fermilab~-~
Harvard~-~ITEP~-~Lebedev~-
~\mbox{UC-London}~-~LSU~-~MIT~-~MSU~-~\mbox{Minnesota-Duluth}~-~
\mbox{Minnesota-Minneapolis}~-~
\mbox{TUM-M\"unchen}~-~NIU~-~\mbox{Ohio-Athens}~-~Oxford~-~Pittsburgh~-~
Princeton~-~Rochester~-~Rutherford~-~S\~ao~Paulo~-~
Stanford~-~\mbox{Stony~Brook}~-~
Sussex~-~\mbox{Texas-Austin}~-~\mbox{TMU-Tokyo}~-~Tufts~-~UCLA~-~
\mbox{Virginia Tech}~-~\mbox{York-Toronto}

\end{center}
\end{titlepage}

\def\sinsqthot{$\sin^2(2\theta_{13})$}
\def\dmsqotw{$\Delta m_{12}^2$}
\def\dmsqoth{$\Delta m_{13}^2$}
\def\dmsqtt{$\Delta m_{23}^2$}
\def\nubar{\bar \nu}
\def\dmsq23{\Delta m^2_{32}}
\def\delmsq{\Delta m^2}
\def\numu{\nu_\mu}
\def\nutau{\nu_\tau}
\def\nue{\nu_e}
\def\sinsq2t23{\sin^2 2\theta_{23}}
\def\ssq2t13{\sin^2 2\theta_{13}}
\def\th13{\theta_{13}}
\def\numutonue{\nu _{\mu }\rightarrow \nu _{e}}
\def\numutotau{\nu _{\mu }\rightarrow \nu _{\tau}}
\def\pizero{\pi^0}
\def\numubar{\bar{\nu}_\mu}
\def\nuebar{\bar{\nu}_e}
\def\etal{{\it et al.}}
\newpage
\begin{center}
D. Ayres, G. Drake, M. Goodman, V. Guarino, T.~Joffe-Minor, D~Reyna,
   R.~Talaga, J.~Thron\\
{\it Argonne National Laboratory, IL}\\
\vspace{5mm}
N. Giokaris, G. Tzanakos\\
{\it University of Athens, Athens, Greece}\\
\vspace{5mm}
E. Kearns, C. Walter\\
{\it Boston University, Boston, MA}
% ...
\\
\end{center}
{\bf{$^{\dag}$ Contact person}} 
\newpage
\tableofcontents
\newpage
\newpage
\noindent
\vspace{1cm}
\begin{quote}
I apologize for this letter being so long but I had no time to make it 
shorter. {\em Blaise Pascal}
\end{quote}
\vspace{2cm}
\begin{center}
{\LARGE \bf Executive summary}
\end{center}
\vspace{1cm}

\addcontentsline{toc}{chapter}{Executive summary}

The question of neutrino masses is of fundamental importance. Neutrino 
oscillations seem to be the only tool available to us to unravel the pattern
of  neutrino masses and, perhaps, shed some light on the origin of
masses in general.

The NuMI neutrino beam line and the MINOS experiment represent a major 
investment of US High Energy Physics 
in the area of neutrino physics. The forthcoming
results could decisively establish  neutrino oscillations as  the  underlying
physics mechanism for the atmospheric $\numu$ deficit and  provide a 
precise measurement of the corresponding oscillation parameters, $\dmsq23$
and $\sinsq2t23$.

This, however, is just a beginning of a long journey into  uncharted
territories. The key to these new territories is the detection of  
$\numutonue$ oscillations associated with the atmospheric $\numu$ deficit, 
controlled
by the  little known mixing angle $\ssq2t13$. A precise measurement of
the amplitude of these oscillations will enable a determination of the pattern 
of the mass hierarchy of neutrinos. If the solar neutrino
experiments  determine that the value  $ \Delta m^2_{12}$ is in the 
range of $10^{-5}-10^{-4}~eV^2$ then the measurement of the CP violation in the
neutrino sector  may well be  within our reach.

The full potential of the NuMI neutrino beam  can be exploited by 
complementing
the MINOS detector, under construction, with a new detector(s) placed at some
off-axis position and collecting data in parallel with MINOS. The first
phase of the proposed program includes a  new detector,
optimized for  $\nue$ detection, with a fiducial mass of the order of
20 kton and exposed to  neutrino and antineutrino beams. In a five year run its
sensitivity to the $\numutonue$ oscillations will be at least a factor of ten
 beyond the current limit.

 The future direction of the program will depend on the results of this
first  phase,
but it is very likely that it will be a combination of a significant 
increase of the neutrino beam intensity via an upgraded proton source
and an increase of the detector mass by a factor of five or so. Depending
on the circumstances, the goals of  Phase II may be a further increase
of the sensitivity of a search for $\numutonue$ oscillations, or, perhaps,
a measurement of the CP violating phase $\delta$ in the lepton sector.

This document is organized as follows:
\begin{itemize}
\item Chapter 1 summarizes the physics motivation for the proposed 
 experiment

\item Chapter 2 briefly reviews the role of the MINOS experiment in the
investigations of  neutrino oscillations

\item Chapter 3 outlines the  possible evolution of the neutrino program
in the US and world-wide

\item Chapter 4 discusses the properties of the NuMI neutrino beam, especially 
 at the off-axis positions

\item Chapter 5 discusses  experimental considerations for a $\numutonue$
oscillation experiment

\item Chapter 6 discusses   possible detectors. Details of specific 
detector designs and various technical aspects are described in the 
Appendix A.

\item Chapter 7 addresses the issues of  possible sites for the new
experiment. Details of a specific site: the LTV mine are described in the 
Appendix B.

\item Chapter 8 discusses possible scenarios of the evolution of the neutrino 
program at Fermilab

\item Chapter 9 outlines a possible schedule and milestones

\end{itemize}
\renewcommand{\thepage}{\arabic{chapter}-\arabic{page}}

\newpage

\chapter{Motivation}

 The SuperKamiokande\cite{SKatm} experiment has recently provided 
very strong evidence that the muon neutrino undergoes  flavor 
changing transitions.
These transitions are seen for neutrinos whose path length
divided by energy (L/E) is greater than $\sim 10^2$ km/GeV.
SuperK also has some supporting evidence that these muon neutrinos 
are transformed primarily into tau neutrinos.  
Although the SuperKamiokande detector
has some sensitivity to flavor transitions of electron neutrinos
their data provides no evidence that electron neutrinos are
involved in these transitions.  
In fact, the Chooz \cite{Chooz} reactor experiment provides 
a tighter constraint on the 
upper limit on the probability of electron neutrino flavor 
transitions of 5\%, at the 
values  of L/E for which SuperKamiokande sees muon neutrino
flavor transitions.
This leaves open the interesting and important question: 
What is the role of the
electron neutrino in  flavor transitions at these values of L/E? 
A measurement or stringent limit on the probability of
$\nu_\mu \rightarrow \nu_e$ for such values of L/E 
($10^2$ to $10^3$ km/GeV) is an important
step in understand these neutrino flavor transitions in atmospheric 
neutrinos.
As the NuMI beam is primarily a $\nu_\mu$ beam, the observation of
$\nu_e$ appearance would address this question directly.
This is the primary goal of the experiment described by this letter
of intent.

The SNO\cite{SNO} experiment  has recently reported large  transitions
of solar electron 
neutrinos  to muon and/or tau neutrinos. 
The onset of these transitions occurs for an L/E which is a least
one order of magnitude larger than the L/E for which SuperKamiokande
sees flavor transitions in atmospheric muon neutrinos.
If these transitions occur near the lower allowed values of
L/E then the transition probability $\nu_\mu \rightarrow \nu_e$
measured by an experiment in the NuMI beam can have some sensitivity
to the flavor transitions associated with solar neutrinos.

The LSND\cite{LSND} experiment  has reported small muon antineutrino
to electron antineutrino transitions for values of L/E which are less than
two orders of magnitude smaller than the transitions seen 
in atmospheric neutrinos.  
However this transition probability is very small, compared to
that observed in atmospheric and solar neutrinos, on the order of 0.3\%.
If this result is confirmed by the up coming mini-Boone \cite{miniB} 
experiment, this could be an important background for a measurement of
$\nu_\mu \rightarrow \nu_e$ transitions at the larger values of L/E associated
with atmospheric neutrinos.

To explain the above phenomena, extensions to the Standard Model are
required. The simplest and most widely accepted extension is to 
allow the neutrinos to have masses and mixings such that the above
phenomena are explained by neutrino oscillations.  
The masses and mixing of the neutrinos in these extensions
would be the low energy remnant of some yet to be determined high energy 
physics.
Thus, neutrino masses and mixing provide a unique window on
physics that is inaccessible to current or near future collider experiments.
One popular theory is the so called ``seesaw'' scenario where the active
left handed neutrinos seesaw off their heavier right handed (sterile)
partners leaving three very light Majorana neutrinos.
It is already clear that the masses and mixings in the neutrino
sector are very different than the masses and mixing in the quark
sector and that a detailed understanding of the neutrino masses and mixings
will be important in differentiating fermion mass theories.
Also, they may provide the key to advancing our theoretical
understanding of this fundamental question.

If the neutrinos have masses and mixings then the neutrino mass eigenstates,
$\nu_i = (\nu_1, ~\nu_2, ~\nu_3, ~\dots)$ with masses 
$m_i = (m_1, ~m_2, ~m_3, \dots)$ are related to the 
flavor eigenstates,
$\nu_\alpha = (\nu_e, ~\nu_\mu, ~\nu_\tau, ~\dots)$ 
by a matrix $U_\nu$,
\begin{equation}
  |\nu_\alpha\rangle = \sum_i \left(U_\nu\right)_{\alpha i}|\nu_i\rangle
\label{mix}
\end{equation}
The charged weak current is given by

\begin{equation}
 J_\lambda = \bar\nu_L U  \gamma_\lambda \ell_L
\end{equation}
where  $\ell = (e, \mu, \tau)$ is the  vector of charged lepton mass
eigenstates and $U = U_\nu U_\ell^\dagger$ with  $U_\ell$ being the
mixing matrix for chargad leptons in analogy with Eq.~\ref{mix} .  
In the absence of light sterile
neutrinos, the $3 \times 3$ lepton mixing matrix $U$ is unitary.  Lepton flavor
mixing was first discussed (for the $2 \times 2$ case) by Maki, Nakagawa, and
Sakata. 

If we restrict the light neutrino sector to the three known active flavors
and set aside the LSND results
\footnote{In the 3+1 neutrino mass hierarchy the LSND result can be
accommodated as a perturbation on the pure active 3 neutrino hierarchy.
The 2+2 mass hierarchy would require major modifications.}
then the unitary matrix MNS matrix, $U$,
can be written as 

\begin{equation}
\hspace*{-0.5cm}
U_{\alpha i} =
\left( \begin{array}{ccc}
  c_{13} c_{12}       & c_{13} s_{12}  & s_{13} e^{-i\delta} \\
- c_{23} s_{12} - s_{13} s_{23} c_{12} e^{i\delta}
& c_{23} c_{12} - s_{13} s_{23} s_{12} e^{i\delta}
& c_{13} s_{23} \\
    s_{23} s_{12} - s_{13} c_{23} c_{12} e^{i\delta}
& - s_{23} c_{12} - s_{13} c_{23} s_{12} e^{i\delta}
& c_{13} c_{23} 
\end{array} \right) \, 
\end{equation}
where $c_{jk} \equiv \cos\theta_{jk}$ and $s_{jk} \equiv \sin\theta_{jk}$.

With this labeling the atmospheric neutrinos oscillations are primarily
determined by $\theta_{23}$ and $\Delta m^2_{32}$ whereas the solar
neutrino oscillations depend on  $\theta_{12}$ and $\Delta m^2_{21}$,  
where $\Delta m^2_{ij}=m^2_i-m^2_j$.  
From SuperKamiokande we already have some knowledge of
$|\Delta m^2_{32}| = (1.4-4.0) \times 10^{-3} {\rm eV}^2$ 
and $\theta_{23} \approx \pi/4$.
SNO and the other solar neutrino experiments indicate
$\Delta m^2_{21} = +(2-10)\times 10^{-5} {\rm eV}^2$ and 
$\theta_{12} \approx \pi/6$. 
 Furthermore, Chooz (and SuperK) provide us with a limit
on $\sin^2(2 \theta_{13})$.  The Chooz limit is dependent on the input value
used for $|\Delta m^2_{atm}|$; for the current central value $2.5 \times
10^{-3}$ eV$^2$, this is $\sin^2(2 \theta_{13}) < 0.11$, while for $|\Delta
m^2_{atm}|=2.0 \times 10^{-3}$ eV$^2$, it is $\sin^2(2 \theta_{13}) < 0.18$
\cite{Chooz}.  Thus the proposed long-baseline neutrino oscillation experiment
to search for $\nu_\mu \to \nu_e$ will be able to search over a substantial
range below this upper bound. 
  
The appearance probability of $\nu_e$ in a $\nu_\mu$ beam  in vacuum
 is given,  to  leading order, by
\begin{equation}
P_{vac}(\nu_\mu \rightarrow \nu_e) = 
\sin^2 \theta_{23} ~\sin^2 2\theta_{13} 
~\sin^2 \Delta_{atm}
\end{equation}
where $ \Delta_{atm} \approx 1.27\left({\Delta m^2_{32} L \over E}\right)$. 
If the experiment is performed at the  peak of this probability,
that is, when $\Delta_{atm} = \frac{\pi}{2}+n\pi$, then 
\begin{equation}
P_{vac}(\nu_\mu \rightarrow \nu_e) \approx
\frac{1}{2} \sin^2 2\theta_{13}. 
\end{equation}
The first peak occurs at neutrino energy,
\begin{equation}
E = 1.8 ~{\rm GeV} 
\left({\Delta m^2_{32} \over 3 \times 10^{-3} {\rm eV}^2}\right)
\left({L \over 732 {\rm km}}\right).
\end{equation}
Using the constraint from Chooz of $\sin^2 2\theta_{13} < 0.1$
gives a maximum appearance probability of 5\%.
To be effective any $\nu_e$ appearance experiment has to aim to
exclude or convincingly see a signal at least an order of magnitude
below this 5\% limit.

\begin{figure}[t!]
\vspace*{6cm}
\includegraphics{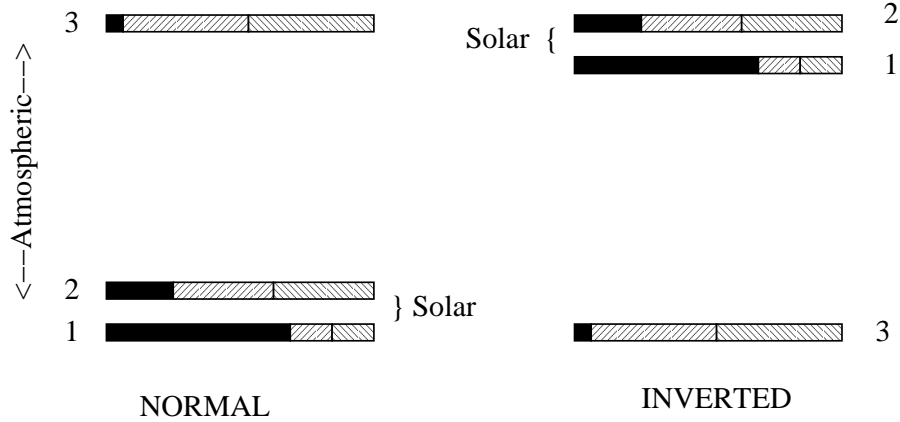}
\vspace{0.5cm}
\caption[]{
The two allowed three-neutrino mass squared spectrums that account for
the oscillations of solar and atmospheric neutrinos. The normal spectrum
has $\Delta m^2_{32} > 0$ and the inverted has $\Delta m^2_{32} < 0$. 
The $\nu_e$ fraction of each mass eigenstate is indicated by the solid
region whereas the $\nu_\mu$ ($\nu_\tau$) fraction is indicated by
the right-leaning (left-leaning) hatching.  The $\nu_e$ fraction in the 
mass eigenstate labeled, 3, has been enhanced for clarity.}
\label{hierfig}
\end{figure}

The  neutrinos in the NuMI beam propagate through the Earth
and matter induced contribution to the propagation amplitude
 are non-negligible . 
These matter effects have opposite sign for neutrinos and anti-neutrinos
and for the normal versus inverted neutrino mass hierarchies. 
The matter effects can be thus  used to distinguish the
two possible three neutrino mass hierarchies, see Fig. \ref{hierfig}.
If the experiment is performed at the
first peak in the oscillation, as above, the matter effects are primarily a
function of the energy of the neutrino beam and the transition probability
in matter can be approximated by
\begin{eqnarray}
P_{mat}(\nu_\mu \rightarrow \nu_e) 
\approx \left(1 \pm 2{E \over E_R} \right) 
~P_{vac}(\nu_\mu \rightarrow \nu_e)
\end{eqnarray}
where $E_R$ is the matter resonance energy associated with the atmospheric
$\Delta m^2$, that is 
\begin{eqnarray}
E_R = {\Delta m^2_{32} \over 2 \sqrt{2} G_F N_e } = 13 ~\rm{GeV} 
\left({\Delta m^2_{32} \over 3 \times 10^{-3} {\rm eV}^2}\right)
\left({ 1.5 ~{\rm g cm}^{-3} \over Y_e \rho}\right).
\end{eqnarray}
where $N_{e}$ is the electron number density in the earth, $\rho$ is the matter density and $Y_{e}$ is the electron fraction.

For the normal hierarchy, matter effects enhance 
(suppress) the transition probability for 
neutrinos (anti-neutrinos) and vice versa 
for the inverted hierarchy.
For a  2 GeV neutrino energy, matter effects give a $30~\%$ enhancement
or suppression in the transition probability.

If the solution to the solar neutrino puzzle is determined to be
neutrino oscillations with the parameters in the ``Large Mixing Angle''
(LMA) region then the $\nu_\mu \rightarrow \nu_e$ transition probability
is sensitive to sub-leading effects and in particular to the 
CP violating phase $\delta$.

In vacuum the shift in the transition probability associated with
the CP violating phase is given by
\begin{eqnarray}
\Delta P_{\delta}(\nu_\mu \rightarrow \nu_e)  \approx 
\frac{1}{2} \cos \delta ~J_r ~\Delta_\odot ~\sin 2 \Delta_{atm}
 \pm  \sin \delta ~J_r ~\Delta_\odot ~\sin^2 \Delta_{atm} 
\end{eqnarray}
where the plus (minus) sign is for neutrinos (anti-neutrinos),\\
$J_r  =  \sin 2\theta_{12}\sin 2\theta_{23}
\sin 2\theta_{13}\cos \theta_{13}$ and
\begin{eqnarray}
\Delta_\odot & = & 1.27 {{\delta m^2_{\odot} L} \over {E}}
= {{\delta m^2_{\odot}} \over {\delta m^2_{atm}}} ~\Delta_{atm} 
\approx \frac{1}{60} ~\Delta_{atm}.
\end{eqnarray}
In the maximum of oscillations at the atmospheric $\delmsq$ scale 
the shift in the transition probability dependent on $\delta$ is 
of order 
\begin{eqnarray}
|\Delta P_{\delta}(\nu_\mu \rightarrow \nu_e)|  \sim 
0.5\% ~\sqrt{{\sin^2 2\theta_{13} \over 0.05 }} 
~\left( {{ \delta m^2_\odot \sin 2 \theta_{12}} 
\over {5 \times 10^{-5} eV^2 }}\right).
\end{eqnarray}
This shift is smaller but of similar size to the shift 
associated with matter effects.

 The full transition probability, in vacuum,  is then given  by~\cite{full}: 

\begin{eqnarray}
P(\nu_\mu\to \nu_e)&=&2 \sin(2\theta_{13}) s_{23}c_{13}s_{12}
(s_{12}s_{23}s_{13}-c_{12}c_{23} c_\delta)\sin^2\phi_{32}
+\nonumber \\
&+&2 \sin(2\theta_{13}) s_{23}c_{13}c_{12}
(c_{12}s_{23}s_{13} +s_{12}c_{23} c_\delta)\sin^2\phi_{31}
-\nonumber \\
&-& 2 \sin(2\theta_{12}) c_{13}^2 \biggl [
s_{12} c_{12}(s_{13}^2 s_{23}^2-c_{23}^2)+s_{13} s_{23} c_{23}
(s_{12}^2-c_{12}^2)c_\delta \biggr ] \sin^2\phi_{21}
\nonumber \\
&+& \frac{1}{2} \sin(2\theta_{12}) \sin(2\theta_{13})\sin(2\theta_{23})c_{13}
s_\delta \bigg[ \sin \phi_{32}\cos \phi_{32} \nonumber \\
&-&\sin\phi_{31}\cos\phi_{31} +\sin\phi_{21}\cos\phi_{21} \bigg]
\end{eqnarray} 
where $\phi_{ij} = \Delta m^2_{ij} L/(4E)$.

\begin{figure}[h!] 
\includegraphics{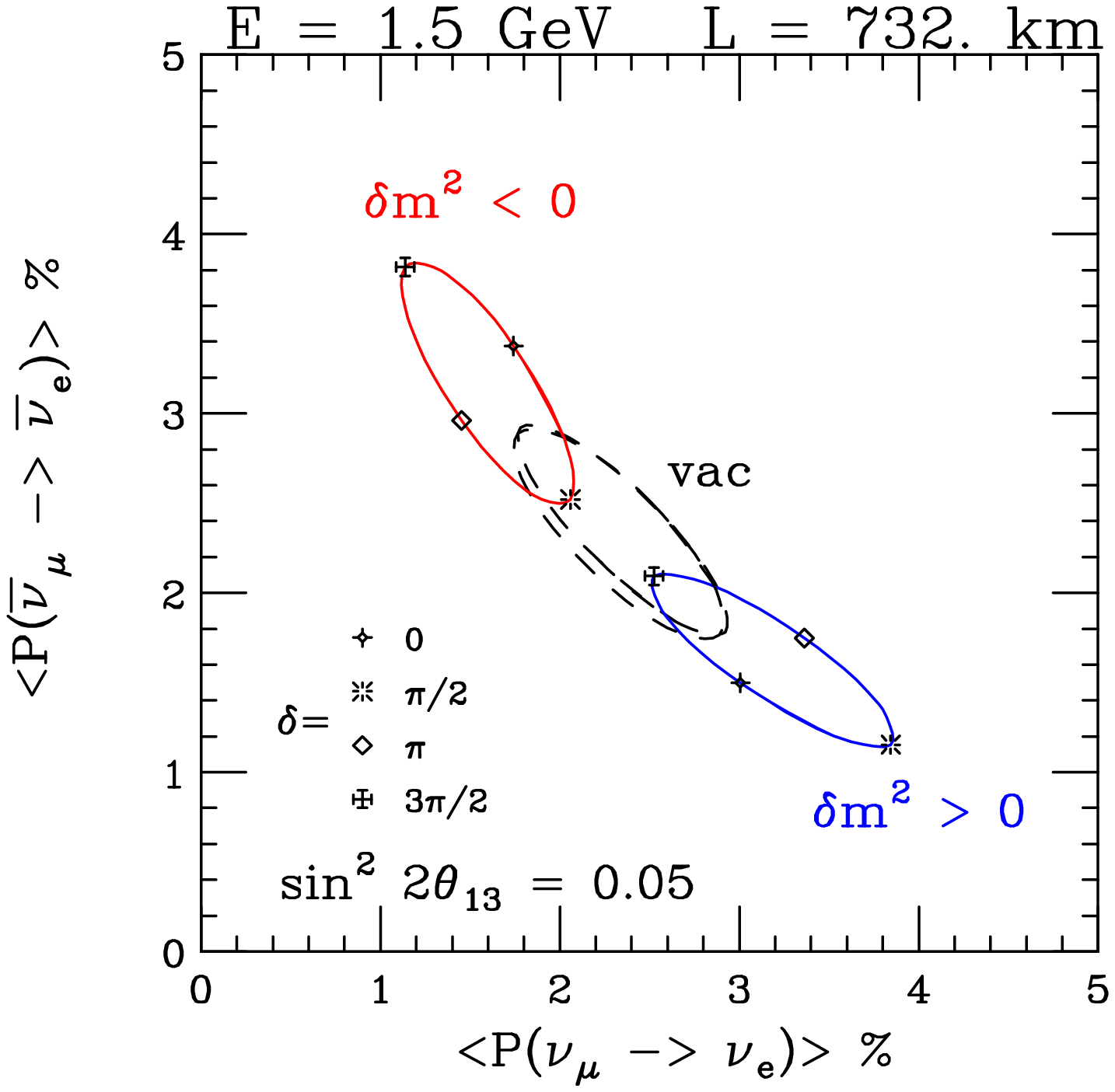}
\includegraphics{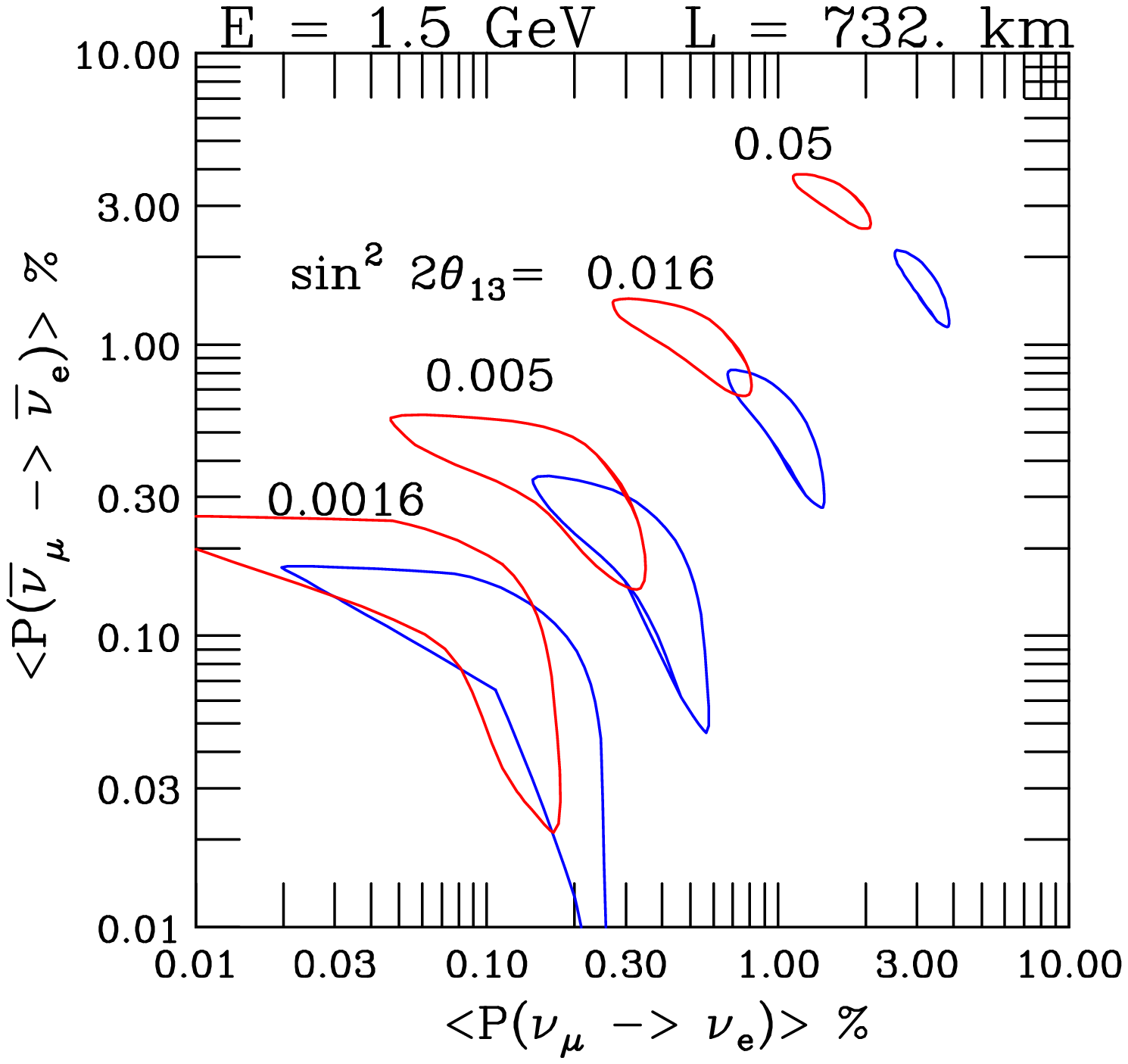}
\vspace{3.0in}
\caption{
The bi-probability plots $P(\nu_\mu \rightarrow \nu_e)$ versus 
$P(\bar{\nu}_\mu \rightarrow \bar{\nu}_e)$ assuming a constant matter
density of $\rho = 3.0 g cm^{-3}$ for an L/E of 500 km/GeV. 
The mixing parameters are fixed to be 
$|\Delta m^2_{31}| = 3 \times 10^{-3} eV^2$,
$\sin^2 2\theta_{23}=1.0$,
$\Delta m^2_{21} = +5 \times 10^{-5} eV^2$,
$\sin^2 2\theta_{12}=0.8$ with the labeled values of 
$\sin^2 2\theta_{13}$ and $\delta$.
}
\label{probsfig}
\end{figure}

A useful and instructive way to present the combined effects of matter
and sub-leading terms is in the bi-probability plots of
$P(\nu_\mu \rightarrow \nu_e)$ versus 
$P(\bar{\nu}_\mu \rightarrow \bar{\nu}_e)$,
invented by Minakata and Nunokawa \cite{MinNun}.
Fig.~\ref{probsfig} shows an example of such a plot for a NuMI case.
At the larger values of $\sin^2 2\theta_{13}$
the ellipses associated with the two possible mass hierarchies
separate in matter, whereas they are approximately degenerate in vacuum.
There is also a significant sensitivity to the CP violating phase, $\delta$.
It is the sensitivity to the sign of $\Delta m^2_{32}$  
and the CP violating phase in these
plots which allows for the determination of these parameters in 
a sufficiently accurate experiment.
For a single experiment there can be a degeneracy in the determined
parameters but this degeneracy can be broken by further experimentation.

A high precision measurement of $\nu_\mu \rightarrow \nu_\mu$ at 
the atmospheric $\Delta m^2$ can also be used to determine the 
difference from maximal mixing in $\theta_{23}$. 
This difference is a measure of the breaking of a 
$\nu_\mu \leftrightarrow \nu_\tau$ symmetry at some high energy scale.
Since matter effects are suppressed in the channel  
$\nu_\mu \rightarrow \nu_\mu$ compared to $\nu_\mu \rightarrow \nu_e$,
a comparison of $\nu_\mu \rightarrow \nu_\mu$ to 
$\bar{\nu}_\mu \rightarrow \bar{\nu}_\mu$ is a sensitive test of CPT
in the neutrino sector.

In summary the important measurements that could be made 
by the proposed experiment are 
\begin{itemize}
\item Observation of $\nu_\mu \rightarrow \nu_e$ at an L/E of
in the range of $10^2$ to $10^3$ km/GeV which would determine the $\nu_e$ role
in atmospheric neutrino flavor transitions.  In the neutrino
oscillation scenario this is a measure of $\sin^2 2\theta_{13}$.
\item Matter effects can be used to distinguish the two mass
hierarchies and therefore determine the sign of $\Delta m^2_{32}$.
\item For the Large Mixing Angle solution to the solar neutrino
puzzle there is sensitivity to the CP violating phase in
the channel $\nu_\mu \rightarrow \nu_e$.
\item Precision measurements in $\nu_\mu \rightarrow \nu_\mu$ channel
can measure how close $\theta_{23}$ is to $\pi/4$, that is maximal mixing. 
A comparison of $\nu_\mu \rightarrow \nu_\mu$ to 
$\bar{\nu}_\mu \rightarrow \bar{\nu}_\mu$ is a sensitive test of CPT 
violation since matter effects are suppressed in this channel.
\end{itemize}
Thus, there is a very rich neutrino physics program to be explored
in a $\nu_e$ appearance experiment using the NuMI beam. 
Details of experimental and beam possibilities will be explored in
subsequent sections.

\chapter{The Role of the  MINOS Experiment}

The zenith angle distribution of  atmospheric $\nu _{\mu }$ interactions
observed in the SuperKamiokande experiment gives a very strong indication of
 neutrino oscillations.  A 50\% deficit of upcoming neutrino flux
suggests that the mixing angle is very large. A fit to the angular
distribution yields the information on the corresponding mass difference $%
\Delta m_{23}^{2}$ in the range $1.4-4.0\times 10^{-3}~ eV^{2}$. 
This result is in a
good agreement with the initial results of the subsequent accelerator-based
long baseline oscillation experiment K2K\cite{K2K}. 
Further results from  SuperK
indicate that the dominant oscillation mode is $\nu _{\mu }\rightarrow \nu
_{\tau }$, although a significant admixture of the oscillations into a
sterile neutrino is not excluded\cite{SKtau}.

 The Fermilab Main Injector neutrino beam (NuMI)\cite{NUMI}
 and the long baseline neutrino
oscillations experiment MINOS\cite{MINOS} were proposed in early 90's to study the
'atmospheric neutrino anomaly' indicated by the results of the IMB\cite{IMB}
 and
Kamiokande\cite{Kamiokande} water Cherenkov experiments. An intense neutrino 
beam is produced
from decays of pions and kaons produced by 120 GeV protons interacting with
a carbon target. Two parabolic horns focus the produced mesons into a nearly
parallel secondary beam of particles which decay inside a 675 meter long
evacuated decay pipe.

 The MINOS experiment consists of two functionally identical detectors;
magnetized iron and solid scintillator calorimeters. The near detector is
located at Fermilab $250~m$ downstream of the end of the decay pipe. The far
detector is located in a Soudan mine at a distance of 735 km from Fermilab.

Charged current $\nu _{\mu }$ interactions will be recognized in the
MINOS detectors via the presence of a long penetrating muon track. 
Muon momenta
will be measured from range with an accuracy of $\sim 6\%$ for stopping
tracks and from bending in the toroidal magnetic field with an accuracy of $%
\symbol{126}12\%$. The iron-scintillator sandwich provides a calorimetric
measurement of the energy of the hadronic final state with the resolution $%
\Delta E/E\approx 60\%/\sqrt{E}$. The parent neutrino energy is calculated as a
sum of the muon and hadronic final state energies. For a neutrino energy
around $5~GeV$ the typical resolution is of the order of $\sim 20\%$.

Charged current $\nu _{\mu }$ interactions observed in the near MINOS 
detector will provide a basis for the prediction of the neutrino spectrum
expected in the far detector. Owing to the finite size of the decay volume
the neutrino spectra in these two detectors are not identical, but the
difference is small and can be corrected for. The systematic error on the
predicted neutrino spectrum in the far detector is expected to be smaller
than $4-5\%$. An additional systematic cross check of the understanding of
the neutrino beam will be provided by precise measurement of the pion and
kaon production spectra in the experiment E907, which should enable
reliable prediction of the neutrino spectrum in the near detector.

\begin{figure}[t!]
\begin{center}
\includegraphics[scale=0.5]{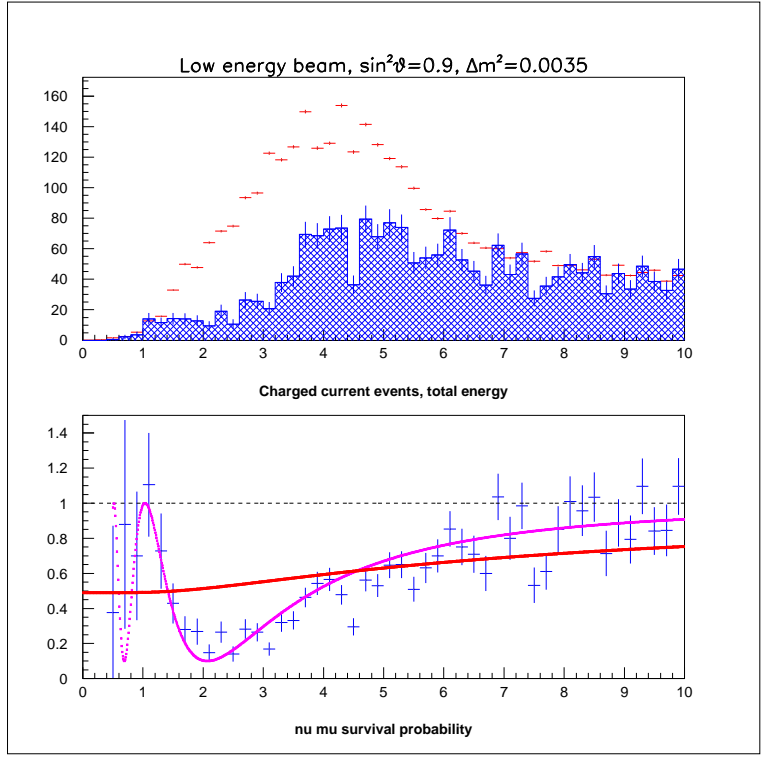}
\vspace{0.5cm}
\caption[]{Top: expected spectrum of $\numu$ charged current events in the far
MINOS detector for $8\times10^{20}$ protons and the observed spectrum in case 
of oscillations  with $\dmsq23=0.0035~eV^2$ and $\sinsq2t23=0.9$ , shaded 
histogram. Bottom: ratio of the observed and the expected energy 
distributions, i.e. $\numu$ survival probability, as a function of the 
neutrino energy. Magenta curve is the shape expected for the neutrino 
oscillations, red curve is the expectation of the neutrino decay model
consistent with the SuperK data.}

\label{oscill_proof}
\end{center}
\end{figure}

The ratio of the number of charged current $\nu _{\mu }$ interactions observed in
the far MINOS detector to the number expected from the spectrum observed at
the near detector as a function of the neutrino energy will determine the $%
\nu _{\mu }$ survival probability at a distance of $L=735~km$ as shown in
Fig. \ref{oscill_proof}.The functional shape of the observed depletion, if any, will be a
direct test of the oscillation hypothesis as an underlying physics
mechanism. The energy range available to MINOS is well suited to distinguish
the oscillation hypothesis from other ad-hoc models put forward to describe
the SuperK atmospheric results.

\begin{figure}[t!]
\begin{center}
\includegraphics[scale=0.5]{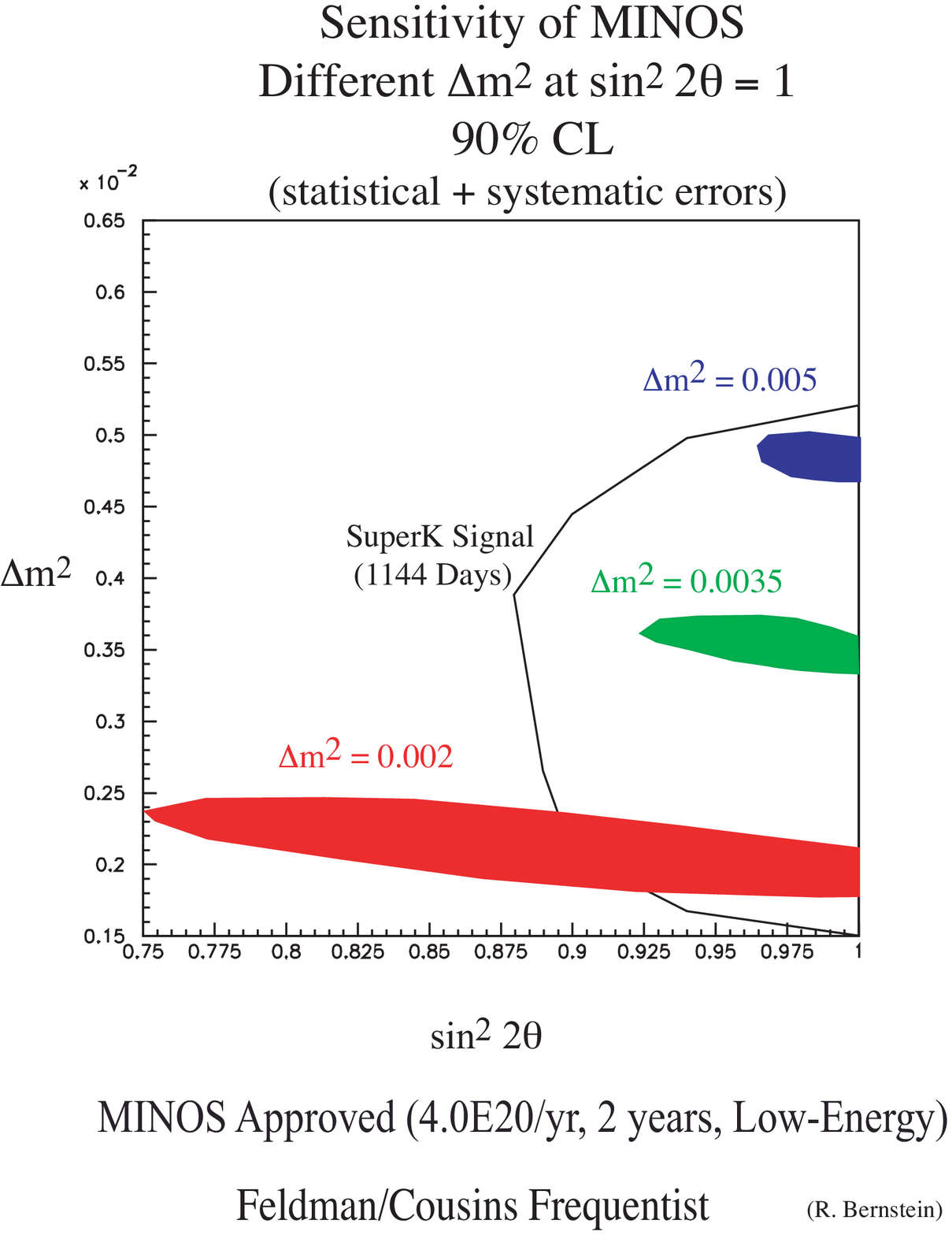}
\vspace{0.5cm}
\caption[]{Expected precision of the determination of the oscillations 
parameters, $\dmsq23$ and $\sinsq2t23$ with  low 
energy beam for different values of  $\dmsq23$. Horizontal axis is   
$\sinsq2t23$, vertical axis is $\dmsq23$ in $eV^2$.}
\label{parameters}
\end{center}
\end{figure}

If the observed survival probability will follow  the expected 
$\sin^{2}\Delta m^{2}L/E$
behavior then the position of the minimum will yield the information on the 
$\Delta m^{2}$, whereas a deviation of the minimum from one will measure the
mixing angle $\sin ^{2}2\theta _{23}$. Precise values of the parameters will
be determined  from a fit to the distribution, after unfolding experimental 
resolution. Fig. \ref{parameters} 
shows the expected precision of such a determination.

The MINOS detector is designed for a $\nu _{\mu }$ disappearance experiment 
and
the detector design is optimized for the detection, identification and
measurement of  $\nu _{\mu }$ charged current interactions. At the same
time, it offers a considerable potential for a  search for  $\nu _{\mu
}\rightarrow \nu _{e}$ oscillations. The longitudinal sampling of the MINOS
calorimeter,$\ 2.5~cm$ of iron, represents $1.5~X_{0}$ whereas the transverse
granularity of the readout $4~cm$, is of the order of the Moliere radius.
Electrons produced in    $\nu _{e}$ charged current interactions produce
a characteristic  energy deposition with small spatial extent
and high energy density. This pattern is quite different from the pattern
produced by charged hadrons, but indistinguishable from the energy
deposition of a neutral pion. 

Several studies have demonstrated that MINOS
detector can identify $\nue$ charged current interactions\cite{MINOS_nue} with
efficiency of the order of $25\%$ while providing a rejection  against
neutral current background by a factor of 65. This will allow the MINOS
experiment to detect $\numutonue$ oscillations, if they
occur at the level of few per cent,  or set a limit on 
$\left| U_{e3}\right|^{2}$ 
of the order of 0.01. This would represent a factor of 3
improvement over the current limit from the CHOOZ experiment for the central
value of $\Delta m^2_{atmos}$ and even bigger improvement for lower  values of 
$\Delta m^2_{atmos}$ still allowed by SuperK and other
experiments. Table \ref{MINOS_nue} 
shows a
composition of the expected background sample in
the MINOS detector
for an  exposure corresponding to $8\times 10^{20}$ POT  in the low 
energy beam.
\begin{table}
\begin{tabular}{|c|c|c|c|c|c|}
\hline
signal &beam $\nu _{e}$ & $\nu _{\mu }$ CC & $\nu _{\mu }\rightarrow \nu_{\tau }$ & NC$<$ 
10 GeV & NC $>$ 10 GeV \\ \hline 
8.5 & 5.6 & 3.9 & 3.0 & 15.7 & 11.5\\ \hline
\end{tabular}
\caption{Number of the expected signal $\nue$ events, with  $\left|U_{e3}\right|^2~=~0.01$, and contributions to the background samples for 
$8\times 10^{20}$ POT
exposure of the MINOS detector in the low energy beam.} 
\label{MINOS_nue}
\end{table}

Relatively high background levels, exceeding the unavoidable intrinsic $\nu
_{e}$ component of the beam are partly related to the relatively poor
granularity of the MINOS detector. A significant factor is due to the
mismatch between the neutrino spectrum of the NuMI beam and the oscillation
probability: a significant fraction of neutrinos present in the beam do
not contribute to the signal, whereas it produces background events. 
Improvement of the $\nu _{\mu }\rightarrow \nu _{e}$ oscillations limit 
with the exposure time will be very slow and it will be limited by the
systematic uncertainties in the background estimate.

To attain higher sensitivity to the $\numutonue$ oscillations the following 
steps, with respect to the MINOS experiment, are desirable:
\begin{itemize}
\item reduce the NC background contribution by a better match of the neutrino
beam spectrum to the oscillation probability curve
\item reduce/eliminate the background due to $\numutotau$ oscillations by
 reducing the neutrino flux above the $\tau$ production threshold 
\item improve efficiency for the signal detection and reduce the background
from NC interaction by an optimization of the detector for the $\nue$
detection and identification
\item improve statistics by an increase of the numbers of protons on the 
neutrino target and/or an increase of the detector size (preferably both)
\end{itemize}

\parindent=0mm
\parskip=5mm
\renewcommand{\textfraction}{.10}
\renewcommand{\topfraction}{.85}
\renewcommand{\bottomfraction}{.85}
\renewcommand{\floatpagefraction}{.85}
\newcommand{\eV}{\mbox{eV}}
\newcommand{\keV}{\mbox{keV}}
\newcommand{\MeV}{\mbox{MeV}}
\newcommand{\GeV}{\mbox{GeV}}
\newcommand{\GeVc}{\mbox{GeV/{\it c}}}
\newcommand{\MeVc}{\mbox{MeV/{\it c}}}

\newcommand{\nf}{Neutrino Factory}

\newcommand{\nub}{\overline{\nu}}
\newcommand{\muminus}{\mbox{$\mu^{-}$}}
\newcommand{\muplus}{\mbox{$\mu^{+}$}}
\newcommand{\dmonetwo}{\mbox{$\Delta{m^2_{12}}$}}
\newcommand{\dmonethree}{\mbox{$\Delta{m^2_{13}}$}}
\newcommand{\dmtwothree}{\mbox{$\Delta{m^2_{23}}$}}
\newcommand{\thonetwo}{\mbox{$\theta_{12}$}}
\newcommand{\thonethree}{\mbox{$\theta_{13}$}}
\newcommand{\thtwothree}{\mbox{$\theta_{23}$}}
\newcommand{\nuenumu}{\mbox{$\nu_{\rm e} \leftrightarrow \nu_{\mu}$}}
\newcommand{\numunue}{\mbox{$\nu_{\mu} \leftrightarrow \nu_{\rm e}$}}
\newcommand{\nuenutau}{\mbox{$\nu_{\rm e} \leftrightarrow \nu_{\tau}$}}

%
\chapter{Possible Evolution of  Neutrino Physics} 

While the preceding chapters gave the motivation for the proposed
neutrino oscillation experiment solely from 
the physics point of view, it may be useful to put this endeavor
into the broader context of the research programme of the USA,
and even of the world-wide programme.

In the USA, a strong tradition of successful experimentation with
accelerator neutrinos exists at Fermilab, which has culminated with the
high-precision results from CCFRR and NuTeV on the nucleon structure
functions and on the electroweak mixing angle. With the
enthusiasm for the prospects offered by the newly discovered neutrino
oscillations, MINOS and MiniBooNE will take over and guarantee 
interesting research and important results in the next five to ten
years. Yet, the development in neutrino physics has proven fast, 
many new initiatives are being discussed also in Europe and Japan.
It is important for the USA, and in particular for Fermilab, to
plan the future in accelerator neutrino physics beyond MINOS and
MiniBooNE now.   

With a view to complying with scarce resources 
 the concurrent use of the
NuMI beam for a second 'off-axis' detector appears as optimal
choice, combining an effective use of existing facilities with
both a significant and far-reaching extension of the physics agenda
and a relatively modest expenditure. While this program is expected
to face some healthy competition from overseas activities, it is fair to
say  that by and large it fits very well into a largely
complementary world-wide
effort to advance our understanding of the physics of neutrino
oscillations.

The first outstanding issue is pinning down the oscillation parameters
of solar neutrinos. The recent spectacular results from SNO\cite{SNO_1} 
(located
in the Sudbury mine, Canada) have 
finally confirmed that the solar neutrino deficit is not of astrophysical
origin, but due to the oscillation of electron-neutrinos into
other active neutrino flavors. Remaining ambiguities between different
domains of oscillation parameters are expected to be resolved within
a few months by the KamLAND\cite{Kamland} 
reactor neutrino experiment (located in the 
Kamioka mine, Japan), and subsequently 
confirmed by the BOREXINO\cite{Borexino} experiment (located in
the Gran Sasso Laboratory, Italy). From then on, the challenge will
be to measure the relevant mixing parameters $\Delta m^2_{12}$ and
$\sin^2 2\theta_{12}$ as precisely as possible, a
program to which
accelerator neutrinos are not expected to contribute in the foreseeable
future.

The main drive to measure the oscillation parameters of atmospheric
neutrinos, $\Delta m^2_{23}$ and $\sin^2 2\theta_{23}$, is expected
to come from MINOS. MiniBooNE is expected to rule out the evidence
from LSND, or else a new and equally complicated and interesting situation
will have to be faced, which is not anticipated here. The second
outstanding issue will be to confirm a non-zero value of the third mixing
parameter in a scheme with three and only three active neutrino flavors,
$\sin^2 2\theta_{13}$, or else reduce its current upper bound of $\sim
0.1$ to as
low a value as possible. In this area, parallel to the ideas put
forward in this LoI, several initiatives are currently
under discussion also in Europe and Japan, albeit with grossly
varying chances of realization.

Table~\ref{synopsis} attempts to summarize these initiatives and
compare their physics objectives. The table also gives
order-of-magnitude prices and
an estimate of the earliest time of realization. It should be understood
that this assessment constitutes at best an educated guess and can in no
way be taken as an commitment of the proponents themselves. The entry 'F2S' 
of the table refers to the experiment proposed in this letter.
The other
entries in the table are now discussed in turn:
\begin{itemize}
\item
An Expression of Interest was
recently put forth to look for $\numutonue$ oscillations using the CERN
neutrino beam aimed at Gran Sasso Laboratory\cite{cngs}, 
currently under
construction, with an off-axis  Cherenkov detector located deep
underwater in the
Gulf of Taranto, 1200 km from CERN\cite{taranto}. 
The off-axis angle of about 2 deg is
such that the corresponding neutrino energy of 0.8 GeV would place the second 
 oscillation maximum at this location. The detector would
be movable and would take data at several baselines. This initiative,
is listed in the tables as `C2GT'. The CNGS beam would have to be tuned
to a much lower energy than
currently being planned for the Gran Sasso program. This would imply
either a change in the current plans or a delay till 2011 or later when
the OPERA experiment is scheduled to be finished.

\item In Japan, a 50 GeV proton
accelerator
with very high intensity, the  Japanese Hadron Facility (JHF)\cite{JHF}, has
been approved and is under construction. This
facility is due to be completed around 2007 and has a first-stage design beam power
of 0.77 MW. There are now rather detailed plans to construct a neutrino
beam\cite{JHF}, 
most likely an off-axis beam at 2 deg with an energy centered around
0.7 GeV, aimed at the Super-K detector 295 km away.
Even though the neutrino program at the JHF 
is not officially approved as yet it is
anticipated that the formal approval will be obtained within a year. 
This proposed experiment is listed as `JHF2K ' in the table. A
more detailed comparison of this program with the proposed NuMI program is
given elsewhere in this document.
    The JHF program envisions also a second phase some time later. The
accelerator itself would have its intensity upgraded to 4.0 MW. In
addition, plans are being formulated to construct a new water Cherenkov
detector in the megaton range, Hyper-K, near the Super-K detector so that
the same
neutrino beam could be exploited by 
both detectors. The program based on such a detector with a higher
intensity accelerator
should be able to address the issues of CP violation if $\th13$ is not 
too
low. This second phase of the Japanese program is termed `JHF2K
II' in the table.

\item There has been a discussion at CERN about
building a superconducting proton linac re-using the LEP RF cavities.
It would deliver a very high flux of protons of 2.2 GeV kinetic 
energy\cite{SPL}.
Such an accelerator could produce an intense wide-band 
neutrino beam with an average neutrino energy of 300 MeV.
In parallel, the idea was put forth to construct a large water
Cherenkov detector of Super-K type in the Fr\'{e}jus tunnel, utilizing the
opportunity of doing
that relatively cheaply while other construction was proceeding in that
tunnel.
   The energy of the proposed beam and the CERN Fr\'{e}jus distance would be a
good match to search for the $\numutonue$ oscillations at the first maximum
of oscillations. Both of these
efforts are still in the design and discussion stage, they require
a considerable financial expenditure and face therefore
major hurdles arising from the very tight financial situation
at CERN which is imposed by the priority which is given to the LHC
program. They are termed `C2F' in the table.

\item The novel idea of a `Beta-Beam' 
was recently put forth to create pure $\nue$ and
$\nuebar$ beams from decays of radioactive nuclei, $^{18}$Ne and 
$^6$He\cite{betabeam}. 
The nuclei
would have to be accelerated to a $\gamma$ around 75 and would give
well-collimated wide-band neutrinos in the 300 MeV range. The plan would
be to aim this beam at the water
Cherenkov detector in the Fr\'{e}jus tunnel, discussed above.
This initiative is indicated as `C2F+BB' in the table.

Both a new detector and a new storage ring-accelerator with long
straight sections would have to be constructed to make such a program a
reality. While interesting in its own right, the realization of such
a project appears unrealistic because of the very high expenditure
associated with it.

\item The ultimate challenge in the physics of neutrino mixing is to determine
the existence and size of leptonic CP-violation. It is universally agreed
that the `neutrino factory´\cite{nufact}
 based on novel muon storage rings with
long straight
sections gives the best access to this potentially very important
phenomenon. It goes almost without saying that any other aspect of
the oscillation $\numutonue$ which would be left open from earlier
experiments, would be easily and decisively addressed by the neutrino
factory.

Over the last several years there have been
intensive studies of the physics potential of a high energy muon storage ring
with energy up to 50 GeV with long straight sections. 

The beam of the neutrino factory, for stored
$\mu^+$ would consist of $\nue$'s and $\numubar$'s with a well-defined energy spectrum.
The
$\numutonue$ oscillations would give a clean and experimentally well
accessible 
signature  of "wrong sign" muons.

   The technical studies of such a machine identified a number of
difficult challenges associated with obtaining the desired high intensity
of the stored muon beams. The projected cost appears to be in the range
of two billion dollars. Therefore, it is unlikely that there will be a
significant movement towards construction of a neutrino factory unless and
until $\th13$ is shown to be large enough to allow detailed studies of CP
violation. Neutrino factories are entered as `$\nu$F' in the table.

\end{itemize}
\begin{table}[h]
\begin{minipage}{\textwidth}
\begin{center}
\mbox{\begin{turn}{90}
\begin{tabular}{|c|c|c|c|c|c|c|c|c|}
\hline
\hline
\multicolumn{2}{|c|}{ }  & F2S  
&  C2GT  & JHF2K  & JHF2K-II & C2F & C2F+BB & $\nu$F \\
\hline
\hline
\multicolumn{2}{|c|}{$<{\rm E}_\nu>$ [GeV]} 
   & 2 & 0.8 & 1 & 1 & 0.3 & 0.3 & 10 \\
\hline
\hline
Fiducial mass & Water Cherenkov & & 1 Mt & 22.5 kt & 1 Mt & 40 kt & 1Mt & \\
\cline{2-9} 
              & Iron/scintillator & 20 kt & & & & & & 40 kt \\
\cline{2-9} 
              & Plastic/RPCs & 20 kt & & & & & &  \\
\hline
\hline
Physics reach & $\sigma (\Delta m^2_{23})$ [eV$^2$] 
              & $1 \times 10^{-4}$  &  $3 \times 10^{-5}$ 
              & $1 \times 10^{-4}$ & & $1 \times 10^{-4}$ & &  \\
\cline{2-9} 
              & $\sigma (\sin^2 2 \theta_{23})$ & 0.01 & 0.01 
              & 0.01 & & 0.01 & & \\
\cline{2-9} 
              & $\sin^2 \theta_{13}$ [90\% CL] & $1.5 \times 10^{-3}$ & 
              & $1.5 \times 10^{-3}$ &  $2.5 \times 10^{-4}$ 
              & $1.5 \times 10^{-3}$ & &  $2.5 \times 10^{-5}$ \\ 
\cline{2-9} 
              & $\theta_{13}$ [deg; 90\% CL] & 2.2 & 
              & 2.2 & 0.9 & 2.2 & & 0.3 \\
\cline{2-9} 
              & sgn $\Delta m^2_{23} $ & ? & No & No & ? & No & No & Yes \\
\cline{2-9} 
              & CP-violation & No & No & No & ? & No & ? & Yes \\
\hline
\hline
\multicolumn{2}{|c|}{Incremental material cost} & 0.1 
& 0.1 & 0.2 & 1.0 & 0.7 & 2.0 & 2.0 \\
\multicolumn{2}{|c|}{(facility + detector [$10^9$ US \$])} & &   
&     &     &     &     &     \\
\hline
\hline
\multicolumn{2}{|c|}{Year of earliest operation} & 2006 
& \multicolumn{2}{|c|}{2008} & \multicolumn{2}{|c|}{2015} 
& \multicolumn{2}{|c|}{2020} \\
\hline
\hline 
\end{tabular}
\end{turn}}
\end{center} 
\end{minipage}
\label{synopsis}
\end{table}

 The current best limits on $\numutonue$ oscillations
in the atmospheric $\delmsq$ range come from reactor experiments, 
CHOOZ\cite{Chooz} and
Palo Verde\cite{paloverde}.
They study a related process, i.e. disappearance of $\nue$'s
by measuring flux and energy spectrum of $\nue$'s some distance away from the
reactor and comparing it with the predicted one. Statistical and
systematic errors contribute roughly equally in both of these experiments;
thus improving the sensitivity will require advances on both of those
fronts. Searching for a small effect in a disappearance experiment
requires very good control of systematics.

   Recently there has been a proposal for a new experiment to
search for $\nue$ disappearance with a sensitivity roughly a factor of 8
better than the CHOOZ limit. This experiment\cite{kr2d}, 
based on the Krasnoyarsk
reactor, would utilize two identical detectors, 50 tons each, at two
different locations. The detection method would be similar to the one used
by the KamLAND and BOREXINO experiments.
   
It is not clear whether this proposal will be realized. The proponents
plan on performing first a pilot experiment, with smaller detectors at
shorter distances, to investigate the LSND effect. There is also a
possibility that the Krasnoyarsk reactor will be decommissioned sometimes
in the near future.

\chapter{Off-Axis NuMI Neutrino Beam}

Construction of the NuMI neutrino beam is currently underway at
Fermilab. This beam directs 120~GeV protons from the Main Injector
towards a 96~cm long graphite target. The secondary pions and kaons
produced on this target are captured and focused by two parabolic
magnetic horns pulsed at 200~kA.  They  decay in an evacuated  region 675~m 
in length and 2~m in diameter  to yield a very pure beam of muon
neutrinos. 

\begin{figure}
\begin{center}
\epsfig{figure=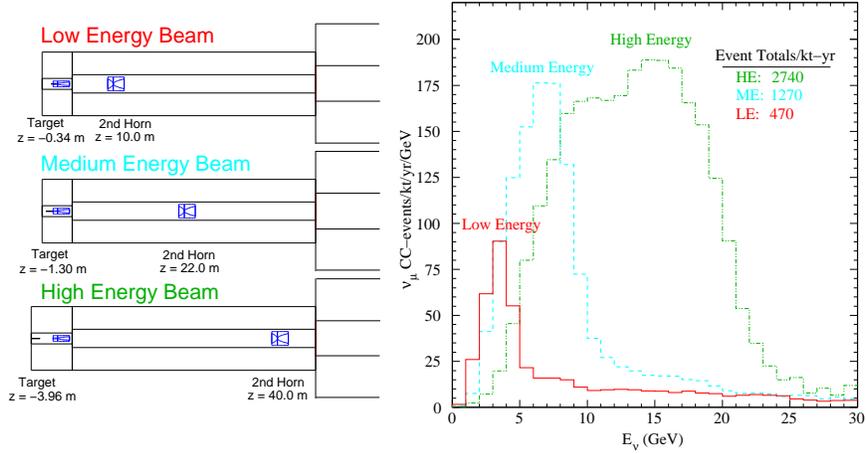,width=2.5in,angle=-90}
\caption{Left: The locations of the target and second horn for the
NuMI low, medium, and high energy wide-band configurations. Right:
Expected neutrino spectra at the MINOS far detector site in each of
the NuMI beam tunes.}
\label{fig:numi-spectra}
\end{center}
\end{figure}

The unique feature of the NuMI neutrino beam is its flexibility. The
neutrino energy spectrum can be selected as `low', `medium' or `high energy'
by adjusting the relative positions of the focusing horns and the 
target\cite{NUMI}.
The neutrino beam energy can be also changed in a continuous fashion by
moving the target with respect to the first horn\cite{variable_energy}.
 The expected neutrino spectra at the MINOS far
detector site are shown in Figure~\ref{fig:numi-spectra} for the NuMI
low, medium, and high energy beam tunes.

The NuMI beam line is designed for a proton intensity of
4$\times10^{13}$ protons per pulse every 1.9 sec.  - roughly 0.4~MW. 
At the design intensity,  NuMI should collect 3.7$\times10^{20}$
protons per year, which will produce 470~neutrino interactions,
in the absence of oscillations, per
year per kt of detector mass at the far MINOS detector location
in the low-energy beam configuration.

Excavation of the NuMI beam line is  complete. Commissioning of
the beam is expected to begin in the first months of 2005.

\begin{figure}
\begin{center}
\epsfig{figure=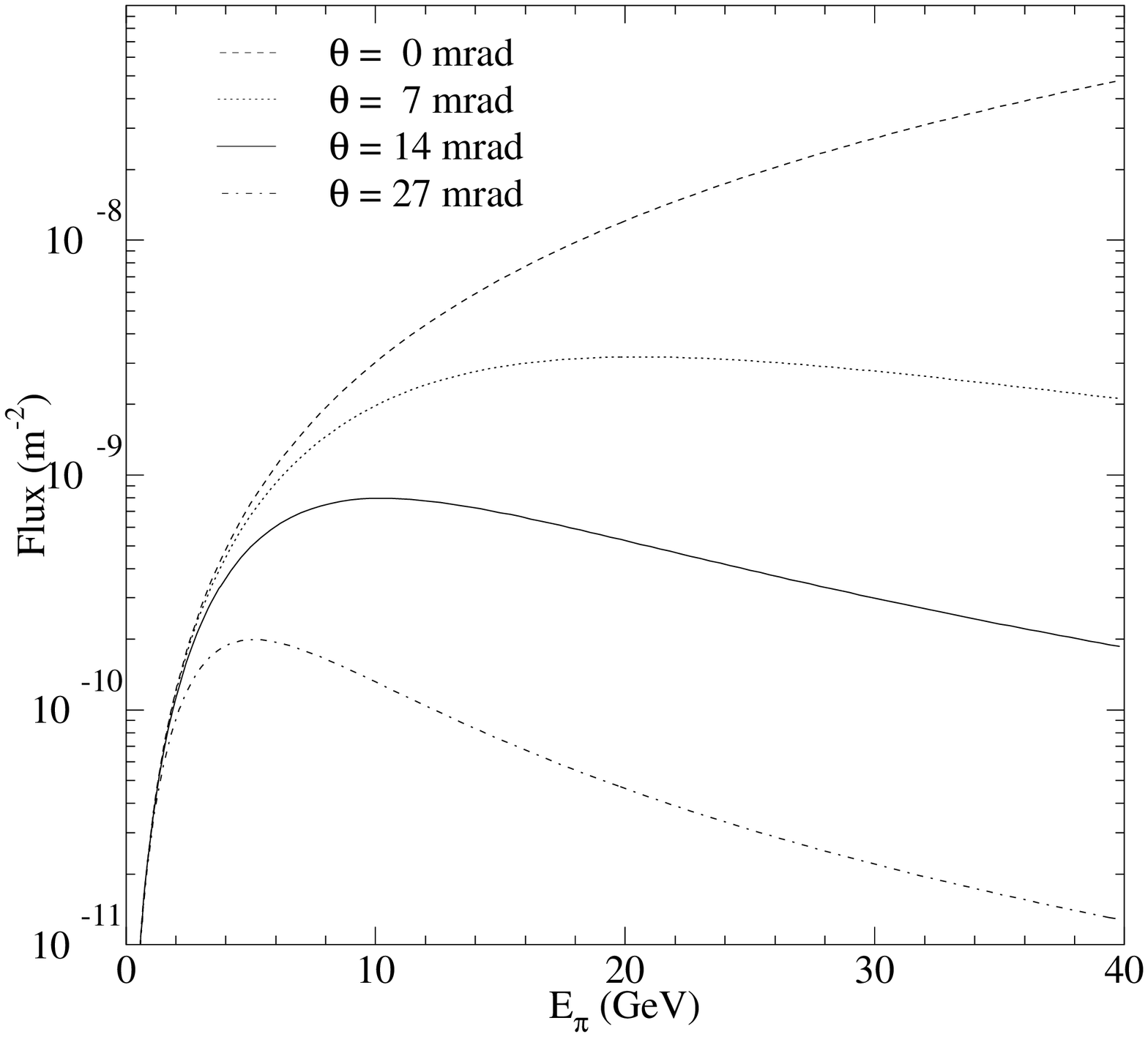,width=2.25in}~
\epsfig{figure=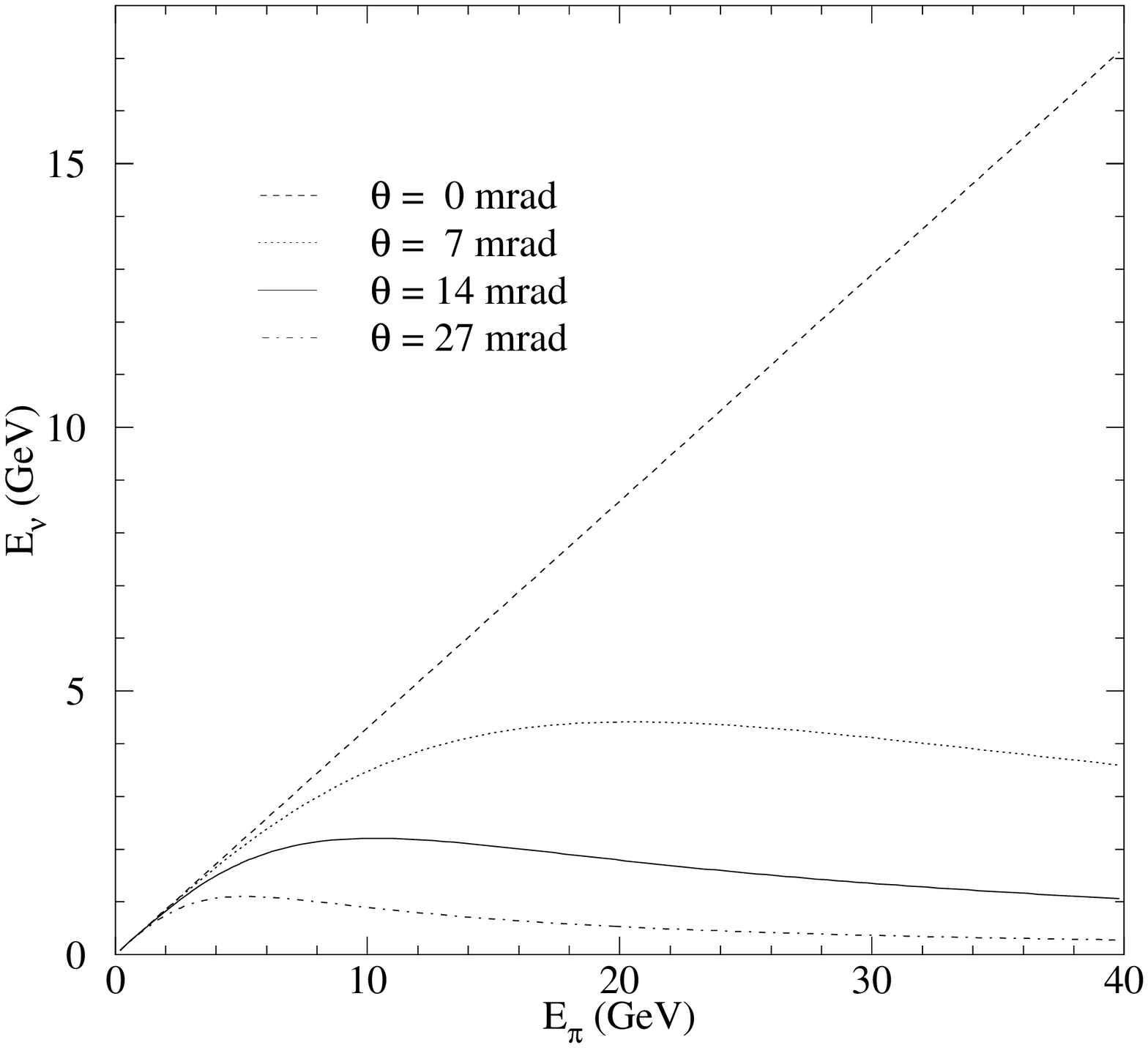,width=2.25in}
\caption{Left: The neutrino flux from a pion of energy $E_\pi$ as
viewed from a site located at an angle $\theta$ from the beam
axis. The flux has been normalized to a distance of
$z$=735~km. Right: The energy of the neutrino produced at angle
$\theta$ relative to the pion beam direction as a function of the pion
energy.}
\label{fig:oaeandflux}
\end{center}
\end{figure}

\begin{figure}
\begin{center}
\epsfig{figure=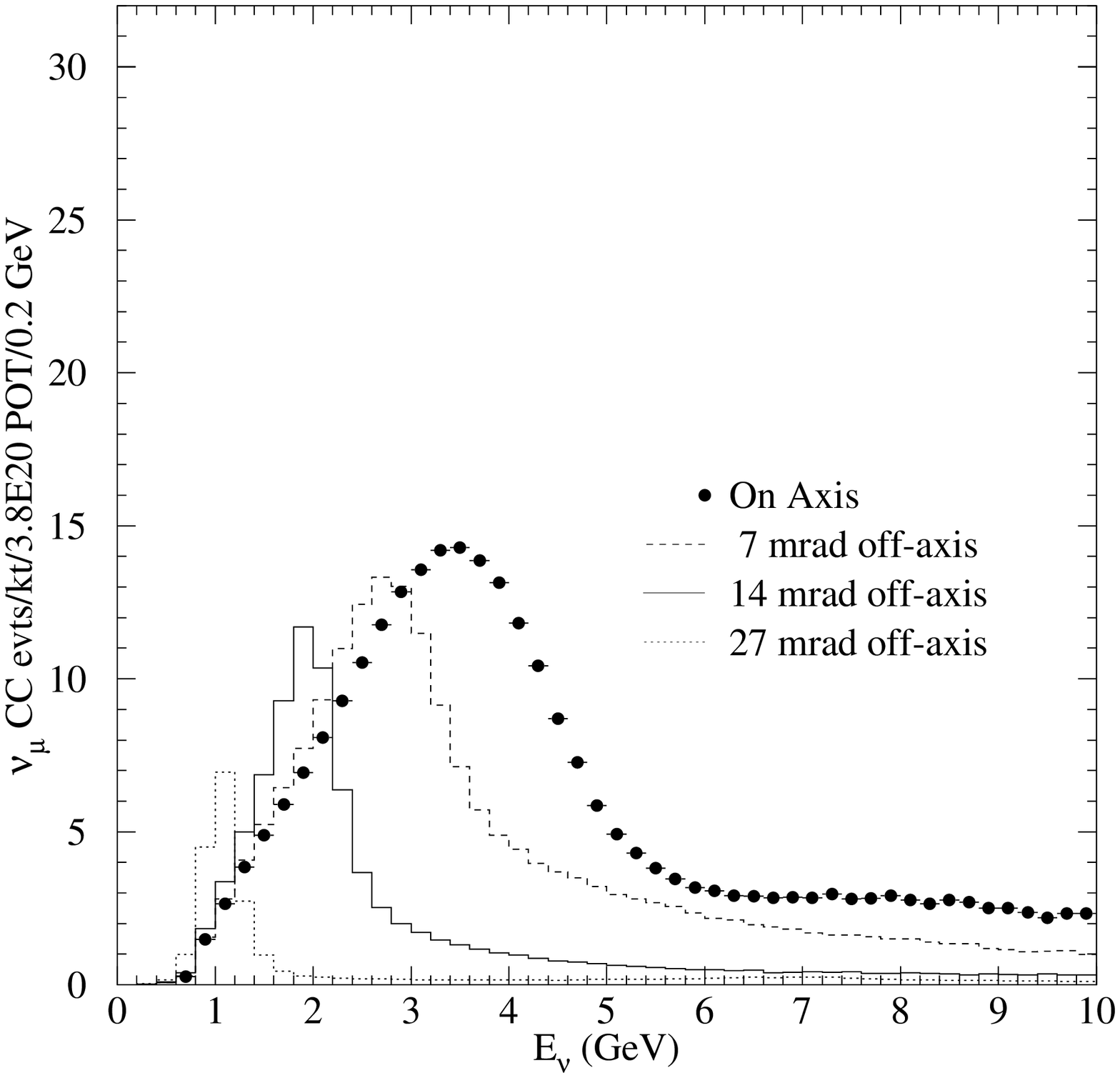,width=2.25in}~
\epsfig{figure=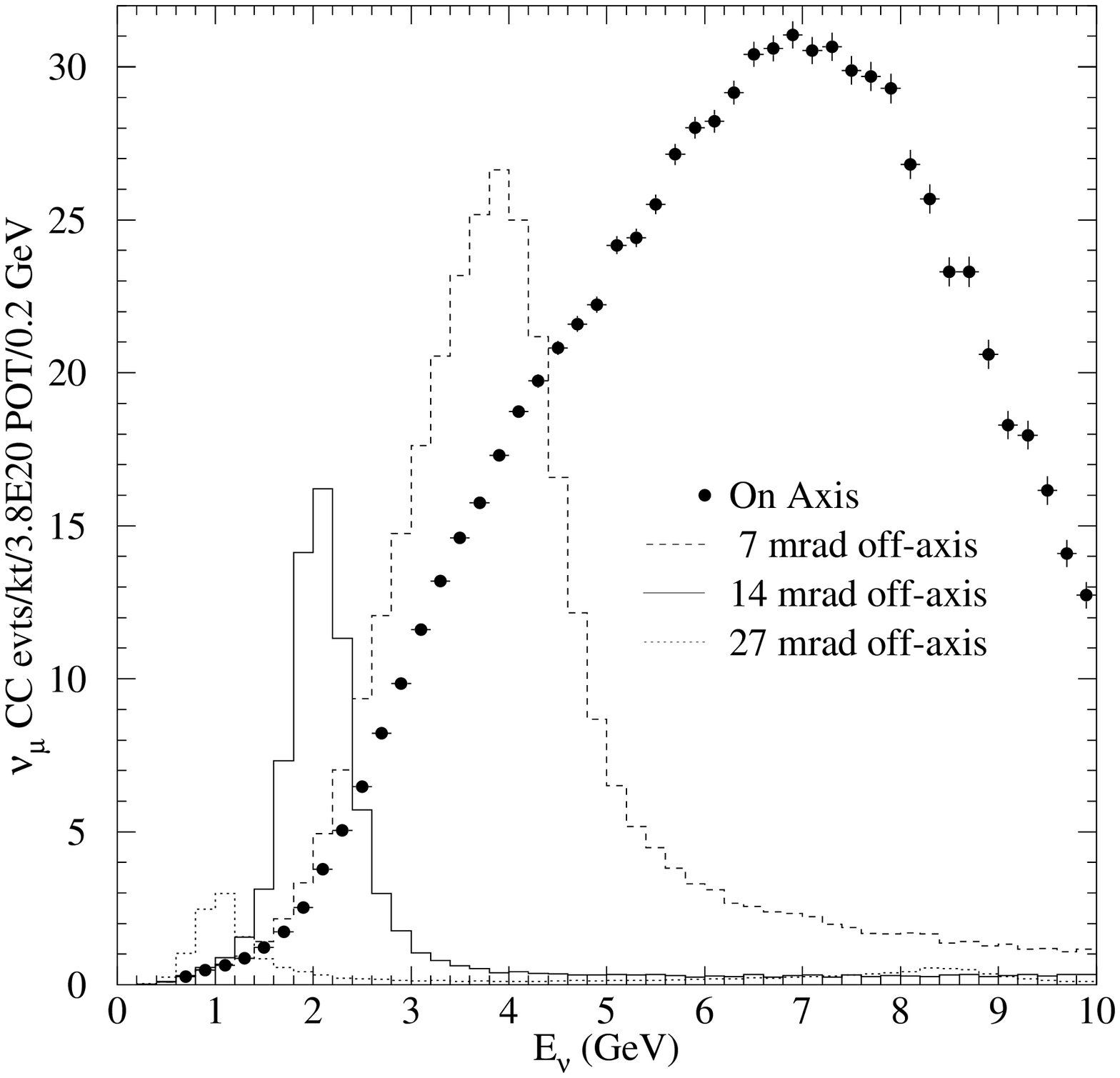,width=2.25in}
\caption{CC muon neutrino events rates expected under `no oscillation' 
hypothesis at a distance of $735~km$ from Fermilab
and at various transverse
locations for the NuMI low-energy beam configuration (left) and
medium-energy beam configuration(right).}
\label{fig:numi-oaspect}
\end{center}
\end{figure}

\section{NuMI Off-Axis Neutrino Fluxes}
Due to the isotropic decays of the focused pions and kaons, the neutrino
beam produced at NuMI will have broad angular distribution. 
For small angles, the flux
and energy of a neutrino produced from the two body decay of a pion in
flight are given in the lab frame by:
\begin{equation}
{\mathcal F} = 
( \frac{2 \gamma}{1+\gamma^2 \theta^2} )^2
\frac{A}{4\pi z^2}, ~
E_\nu = \frac{0.43 E_\pi}{1+\gamma^2\theta^2},
\end{equation}
where $\theta$ is the angle between the pion direction and the
neutrino direction, $E_\pi$ is the energy of the parent pion, and
$m_\pi$ is the mass of the pion and $\gamma=E_\pi/m_\pi$. $A$ and $z$
are the detector cross-sectional area and distance from decay
point. As can be seen in the 
Figure~\ref{fig:oaeandflux}, the neutrino flux
 peaks in the forward direction for all values of $E_\pi$; this 
is why the MINOS detectors are placed on axis. As  the
angle to the beam direction increases, 
however, the relationship between the
pion energy and neutrino energy flattens, with all pions yielding
neutrinos of roughly the same energy. 

It is possible to construct a
nearly mono-energetic neutrino beam by  viewing the NuMI beam
at a location off the beam axis. This is an implementation of a concept 
developed for the proposed experiment E-889 at Brookhaven\cite{BNL_off_axis}.
 Because all pions contribute
neutrinos of roughly the same energy, it is possible to compensate the
reduction in the flux factor. Figure~\ref{fig:numi-oaspect} shows the
neutrino spectra of the NuMI neutrino beam at locations 5, 10, and 20~km 
(corresponding to the decay angles of 7, 14, and 21~mrad) 
off the NuMI beam axis at a
distance of 735~km from the target\cite{off_axis}. As can be seen in 
Fig.~\ref{fig:numi-oaspect}, it
is possible to increase the neutrino yield in the 1-2~GeV region by
placing the detector off the beam axis, while at the 
same time reducing the high-energy component of the neutrino beam. 

The off-axis beam intensity is higher,
the beam energy spread is smaller and the high energy tail is smaller
for the medium energy beam tune than for the low energy one.  The intensity of 
the off-axis beam produced in the high energy beam configuration is
reduced significantly in comparison with that of the medium energy setting
due to the kinematical reduction of the flux.

The low 
energy peak of the off-axis beam spectrum is produced almost exclusively
by decays of pions; hence the prediction of the neutrino flux is quite
insensitive to the experimental uncertainties of the ratio of 
$K/\pi$ production cross sections. At the typical distance of $10~km$ from
the nominal beam axis the resulting spectrum of the neutrino beam is much 
narrower than the expected maximum of the oscillation probability. The
total number of events in the peak region can be predicted on the basis 
of the near MINOS detector with an accuracy better than $2\%$\cite{far_det}.

\begin{figure}
\begin{center}
\epsfig{figure=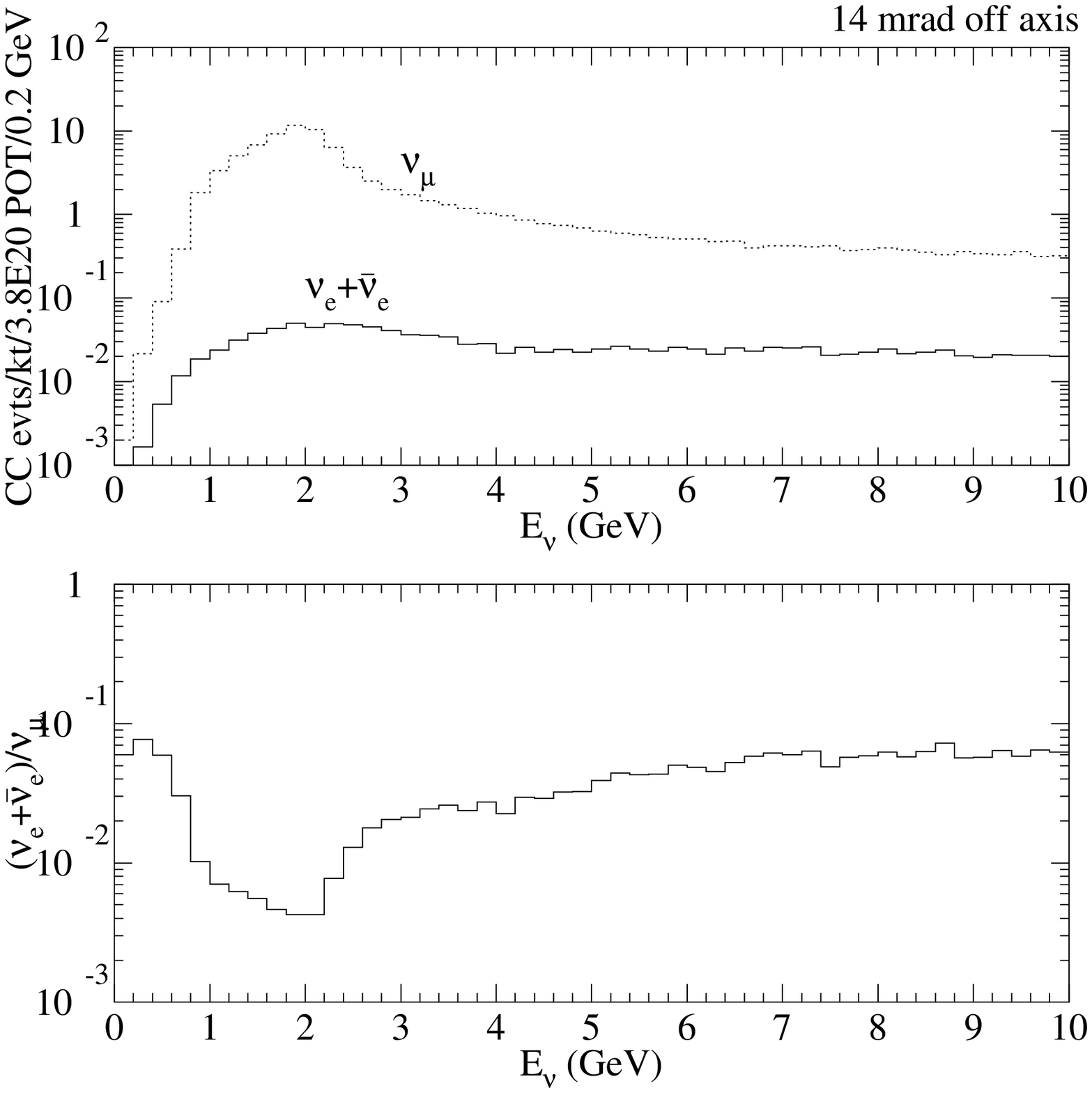,width=2.5in}~
\epsfig{figure=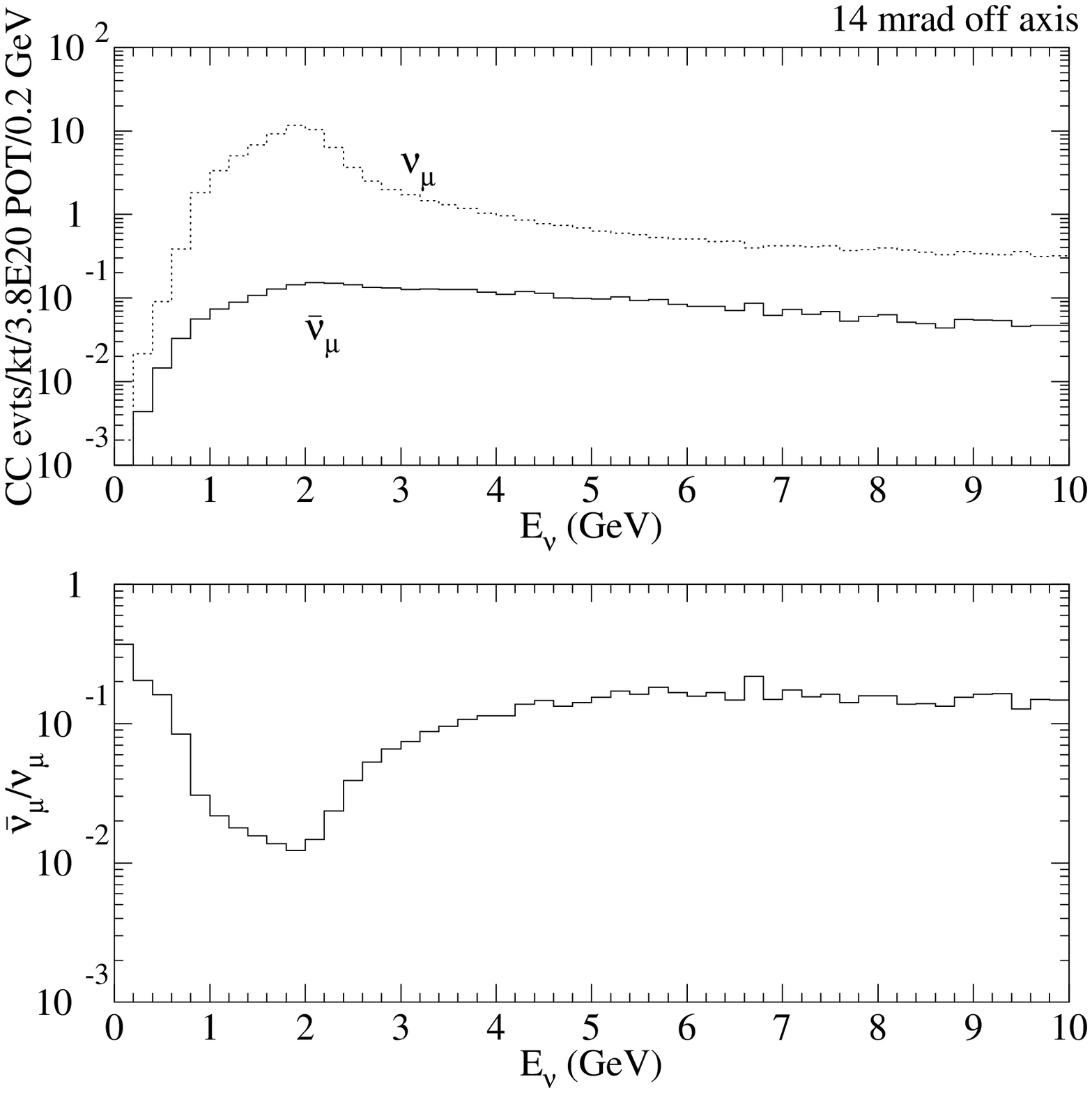,width=2.5in}
\caption{The expected energy distribution of the 
electron neutrino (left) and wrong-sign muon neutrino
(right) components of the NuMI off-axis beams, low energy beam.}
\label{fig:oacomposition-le}
\end{center}
\end{figure}

\begin{figure}
\begin{center}
\epsfig{figure=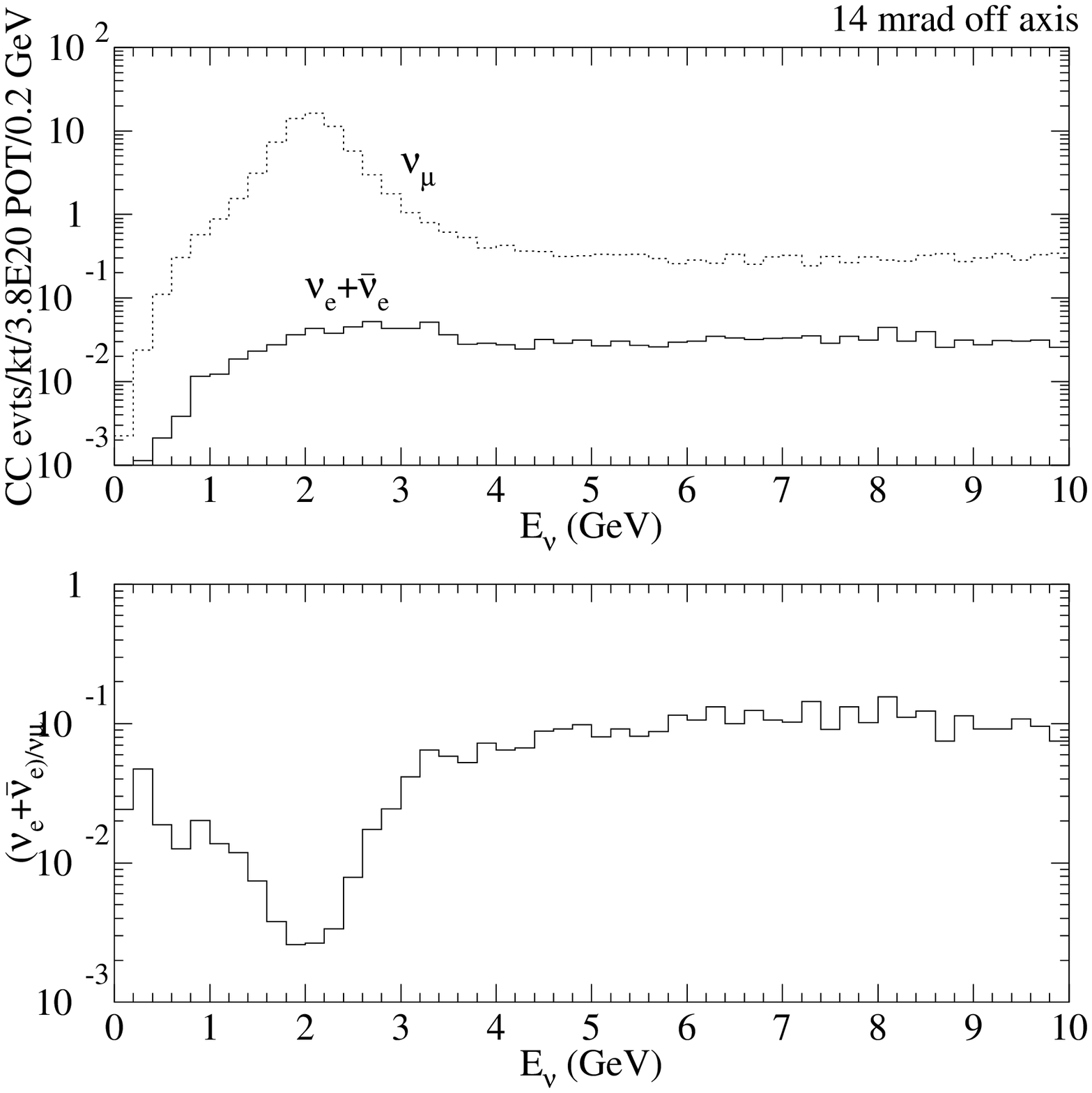,width=2.5in}~
\epsfig{figure=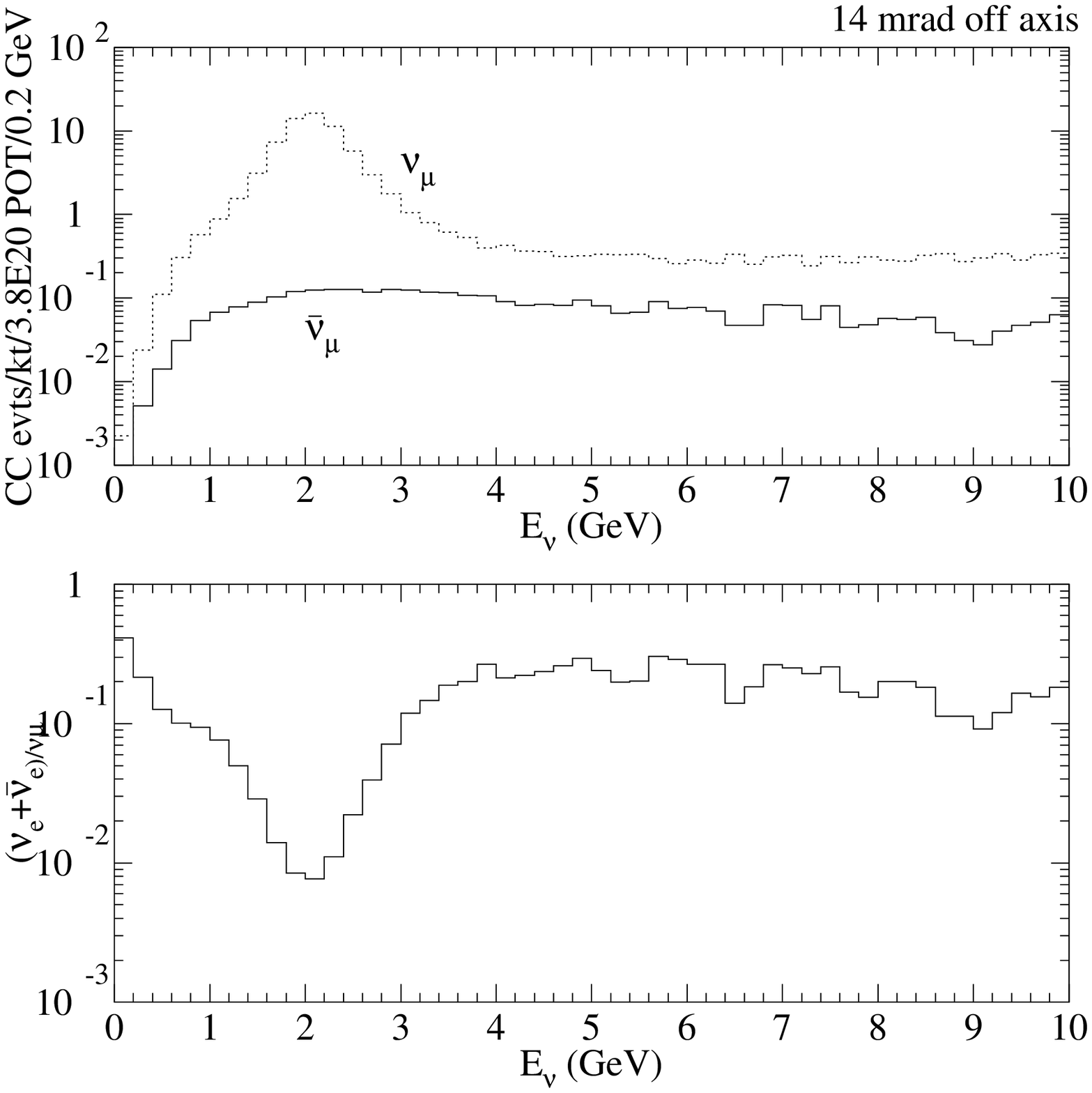,width=2.5in}
\caption{The expected energy distribution of the 
electron neutrino (left) and wrong-sign muon neutrino
(right) components of the NuMI off-axis beams, medium energy beam.}
\label{fig:oacomposition-me}
\end{center}
\end{figure}

\begin{figure}
\begin{center}
\epsfig{figure=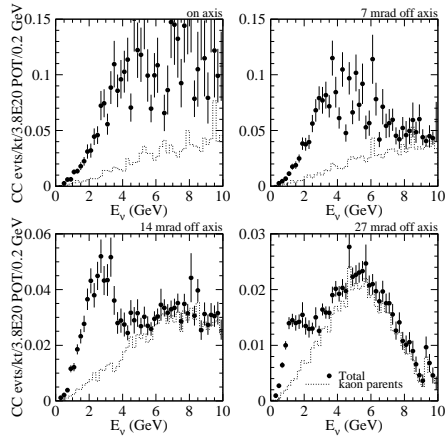,width=2.5in}
\caption{Spectra of the electron neutrino component of the beam at different
off-axis angles. Points represent the total background and a histogram shows 
the component due to $K_{e3}$ decays. }
\label{fig:nuespect}
\end{center}
\end{figure}

\section{Backgrounds}
The neutrino event rates and energy distribution depend on the
detector location,its distance from Fermilab, and the distance from
the nominal beam axis, but their qualitative features are very similar. 
As an example, we show the expected composition of the off-axis 
NuMI neutrino fluxes in Figure~\ref{fig:oacomposition-le} and
\ref{fig:oacomposition-me} at a distance of $L=735~km$.
 The  $\bar{\nu}_\mu$ component of the beam is due to the
decays of high energy $\pi^{-}$ traversing the opening of the magnetic
horn and not being de-focused by the horn system.
 This `wrong' sign beam component contributes roughly $1\%$ of events
at the  peak of the neutrino flux. More
importantly, the component of electron neutrinos is kept to roughly
$0.5\%$ in the peak of the neutrino flux. The $\nue$ component of the beam
is due to $K_{e3}$ decays and to the decays of tertiary muons. 
The $\nue$'s  from the K decays have much higher energies than the main peak
of the $\numu$'s, whereas the neutrinos from the $\mu$ decays populate the
low energy part of the spectrum. 
Fig.~~\ref{fig:nuespect} shows that $\sim80\%$ of the $\nue$ background
at the peak of the $\numu$ flux distribution is due to muon decays.
This component of the background  should be well constrained by
measurements of the muon fluxes and muon neutrino spectra at the
Fermilab site. The systematic error of the extrapolation to the off-axis
detector is of the order of $5-6\%$.

It is worth noting that that the relative contribution of 
backgrounds   is  somewhat smaller for the medium
energy beam setting than for the low energy one.

%
%
%

\begin{figure}
\begin{center}
\epsfig{figure=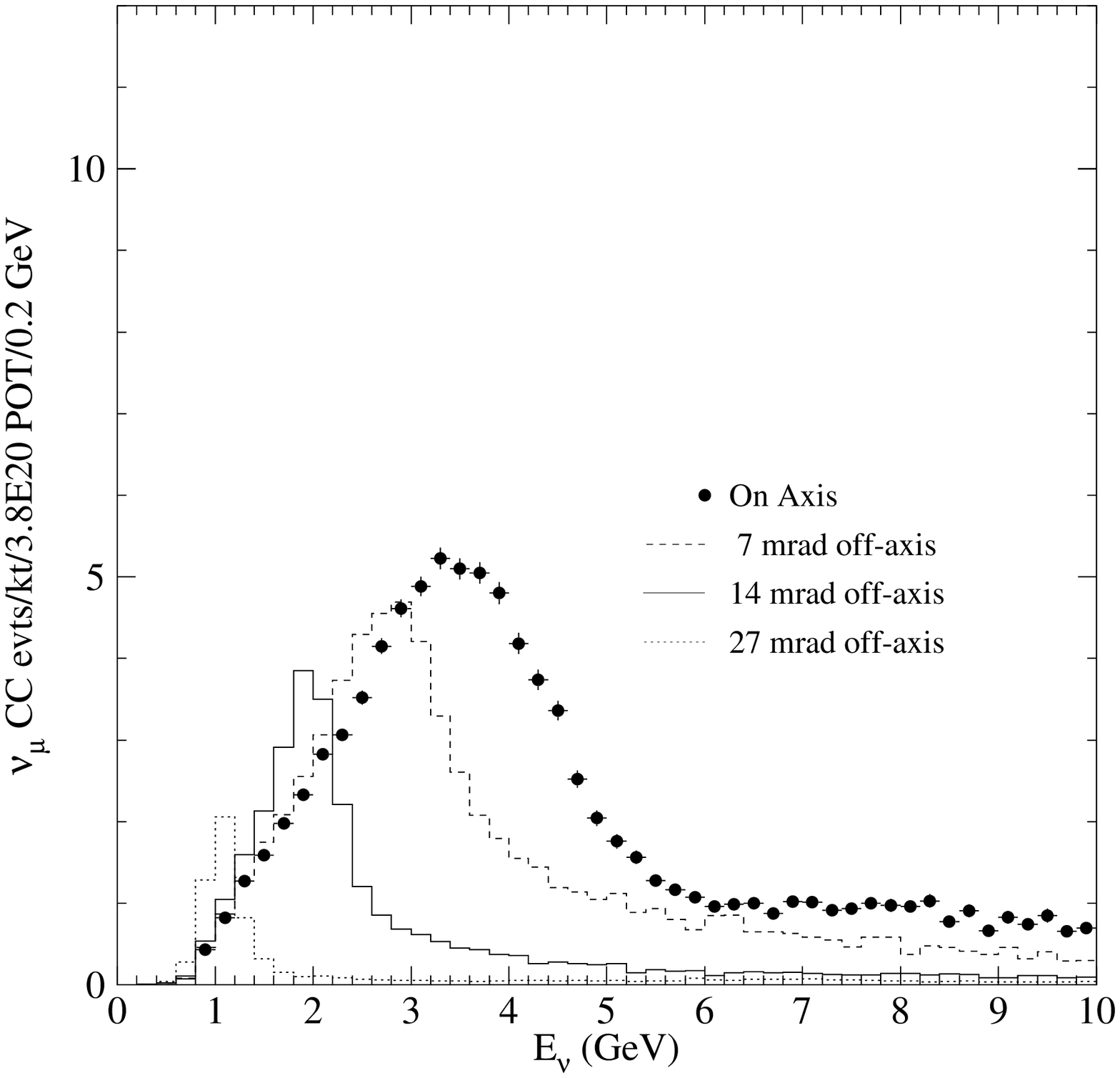,width=2.5in}~
\epsfig{figure=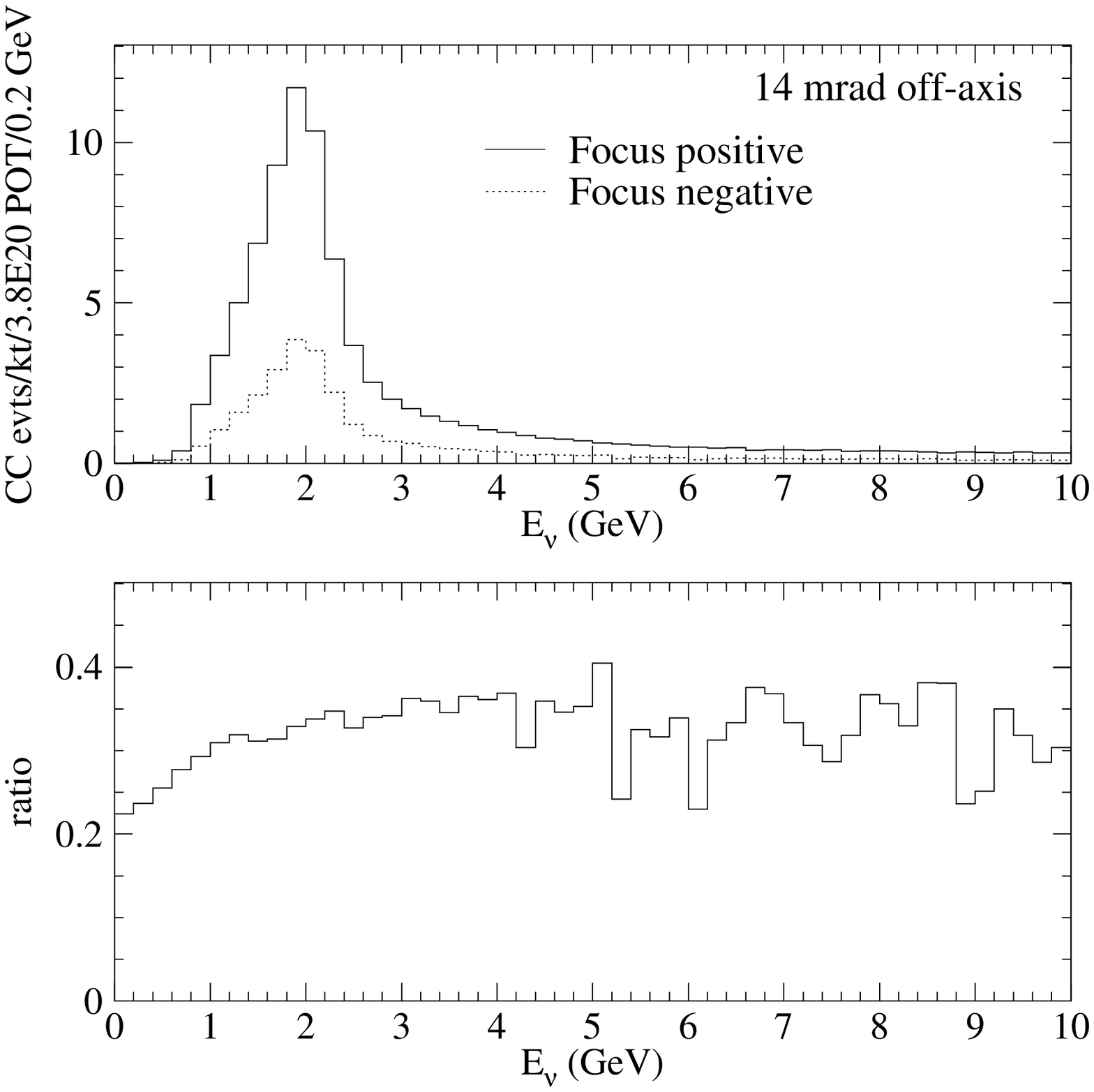,width=2.5in}
\caption{Left: The muon anti-neutrino event rates expected at the
distance $L=735~km$ from Fermilab in the NuMI low energy
beam with reversed horn currents at various transverse
locations. Right: Comparison of the anti-neutrino event rates to those
obtained when running in anti-neutrino mode.}
\label{fig:lerev}
\end{center}
\end{figure}

\begin{figure}
\begin{center}
\epsfig{figure=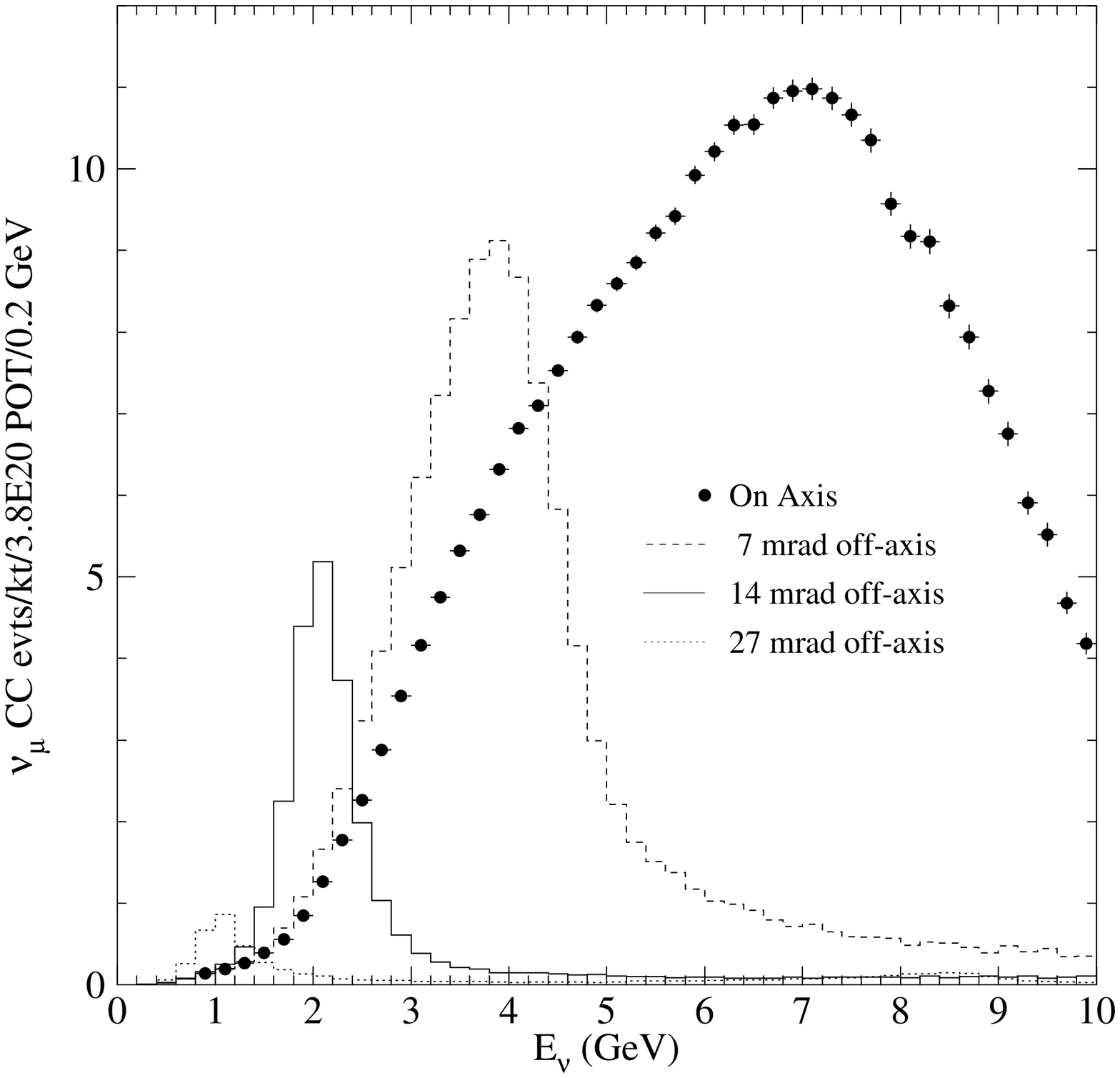,width=2.5in}~
\epsfig{figure=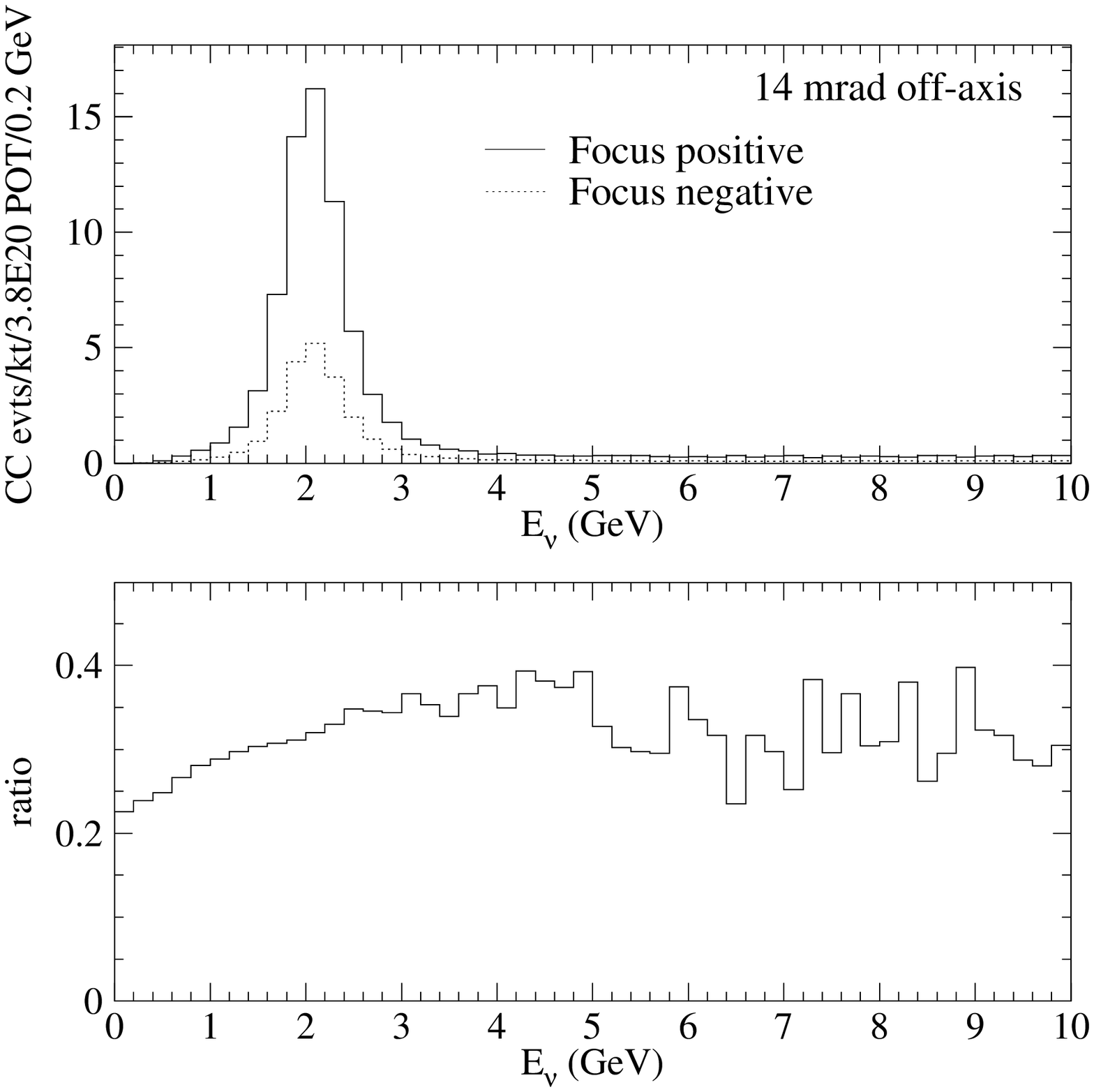,width=2.5in}
\caption{Left: The muon anti-neutrino event rates expected at the
distance $L=735~km$ from Fermilab in the NuMI medium energy
beam with reversed horn currents at various transverse
locations. Right: Comparison of the anti-neutrino event rates to those
obtained when running in anti-neutrino mode.}
\label{fig:merev}
\end{center}
\end{figure}

\begin{figure}
\begin{center}
\epsfig{figure=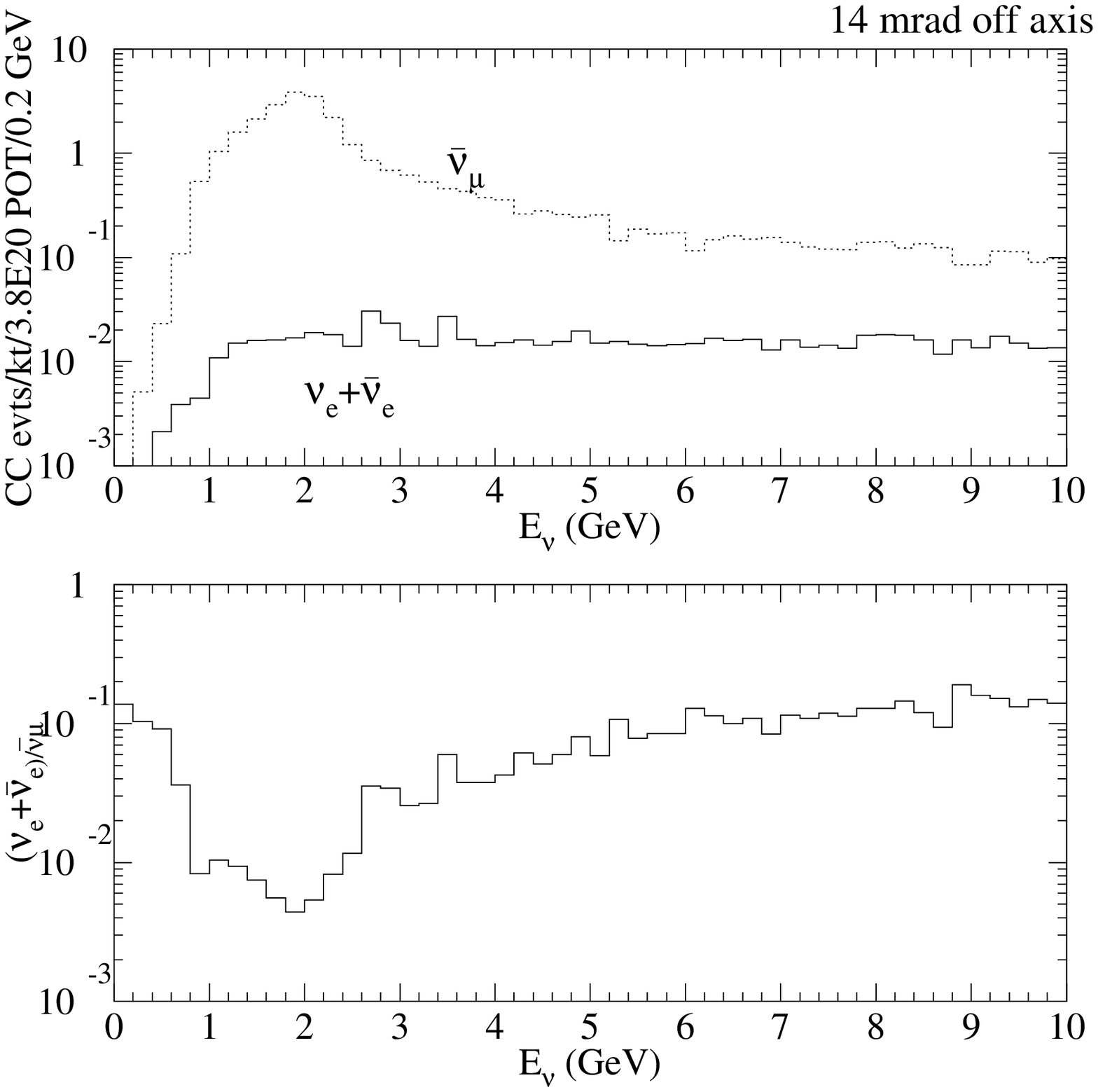,width=2.5in}~
\epsfig{figure=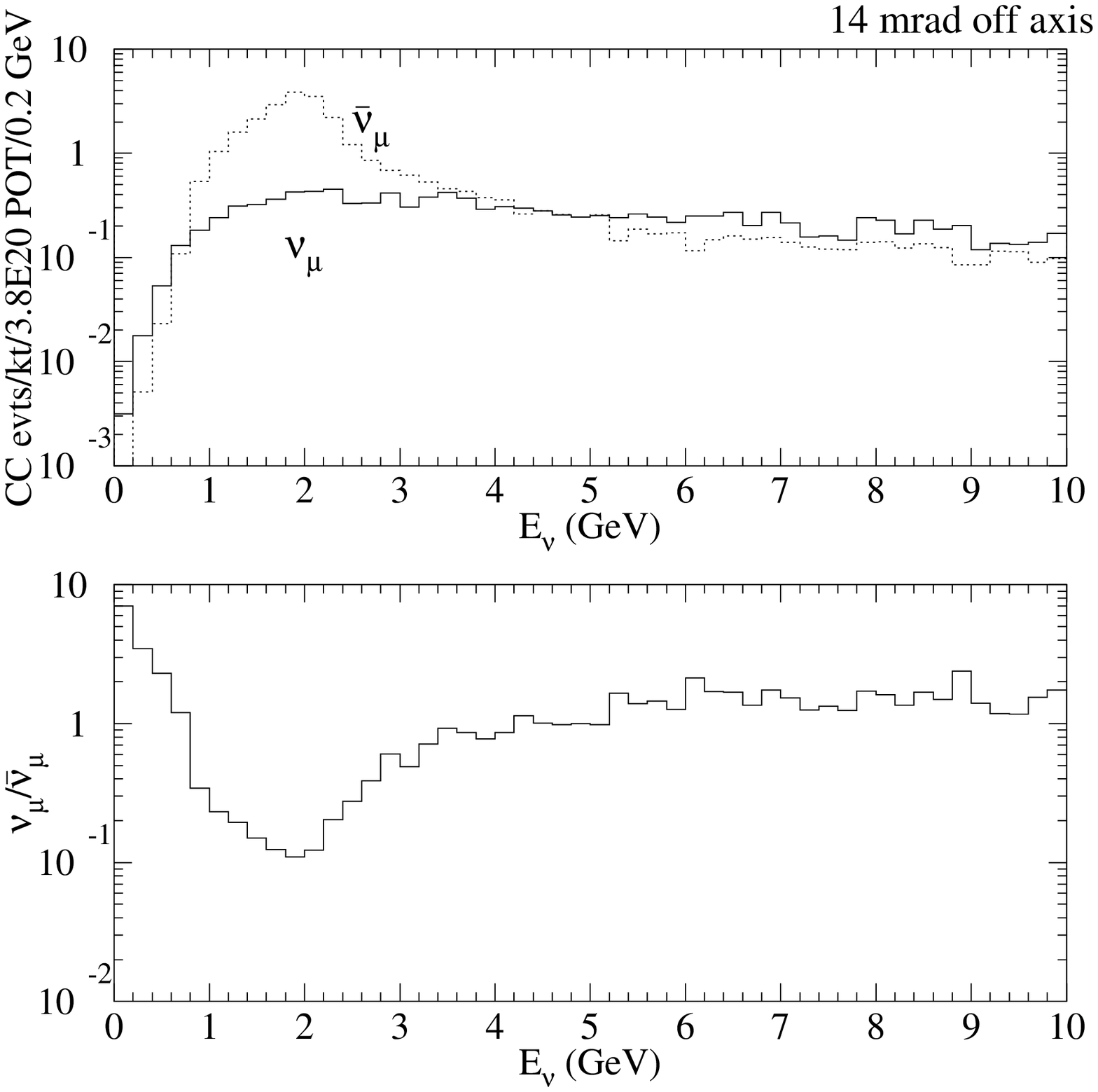,width=2.5in}
\caption{Composition of the off-axis antineutrino beam, low energy
beam configuration.  Left
shows the electron neutrino component. Right shows the wrong-sign muon
component.
}
\label{fig:lerev-composition}
\end{center}
\end{figure}

\begin{figure}
\begin{center}
\epsfig{figure=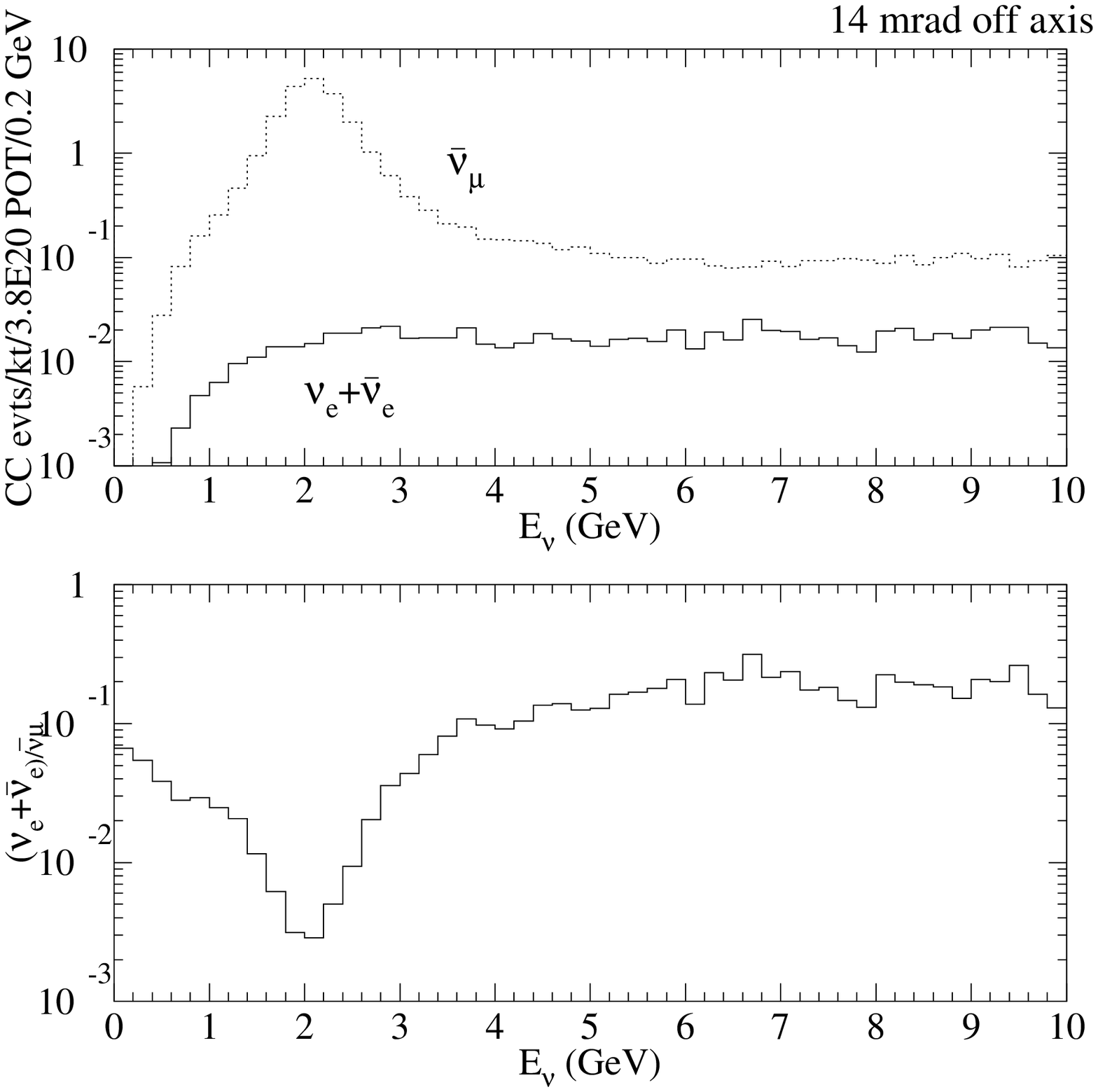,width=2.5in}~
\epsfig{figure=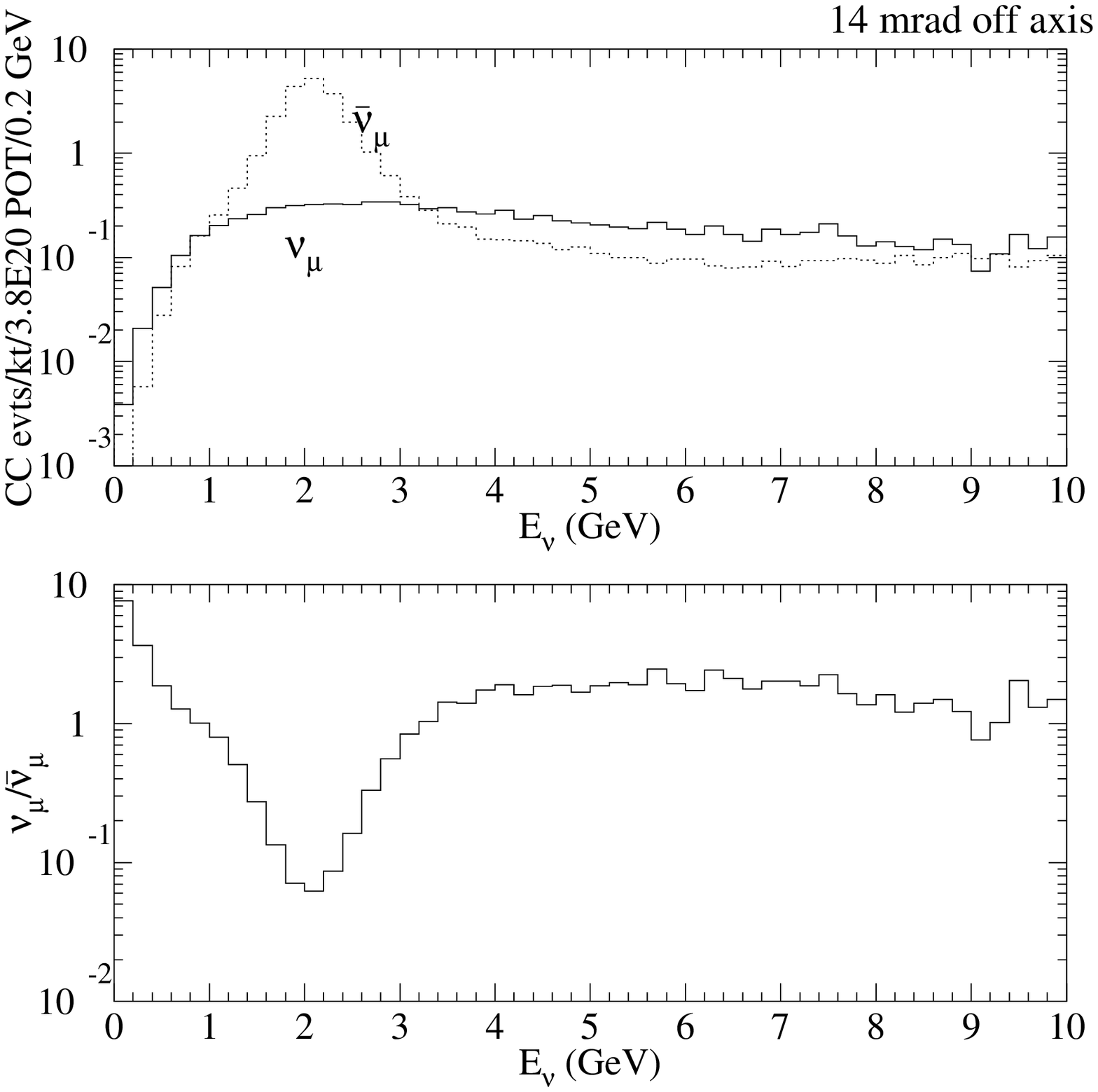,width=2.5in}
\caption{Composition of the off-axis antineutrino beam, medium energy
beam configuration.  Left
shows the electron neutrino component. Right shows the wrong-sign muon
component.
}
\label{fig:merev-composition}
\end{center}
\end{figure}

\section{Antineutrino Beams}

An antineutrino beam, in addition the neutrino beam, is an important
tool for the investigation of the neutrino mass hierarchy and in the
search for CP violation in the neutrino sector.
 
\begin{figure}
\begin{center}
\epsfig{figure=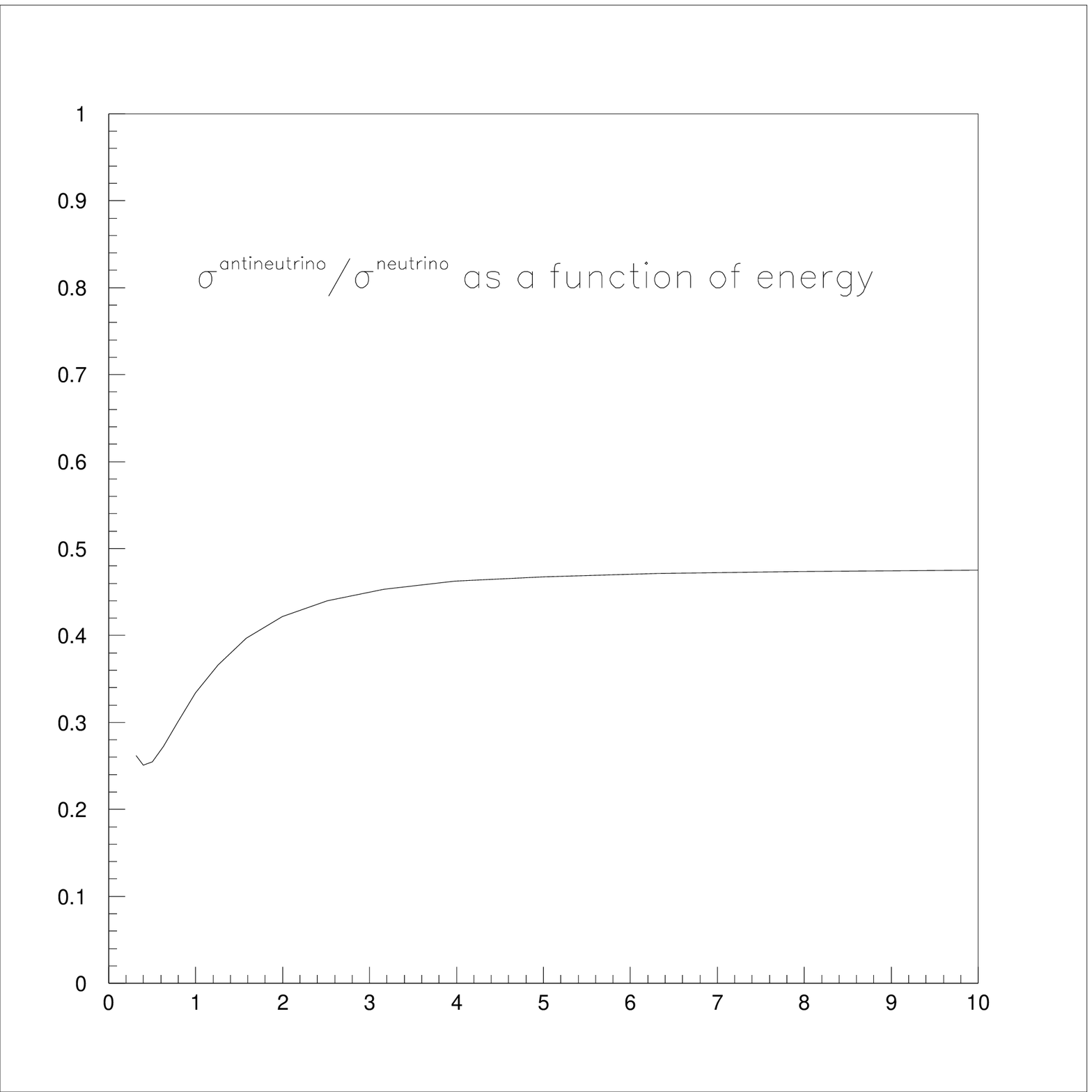,width=2.5in}
\caption{Ratio of antineutrino and neutrino CC cross sections as a function the the neutrino energy}
\label{sigs}
\end{center}
\end{figure}
The direction of the current through the NuMI focusing horns
can be reversed to  produce an
anti-neutrino beam rather than a neutrino beam. The expected
anti-neutrino event rates are shown in Figures~\ref{fig:lerev}
and~\ref{fig:merev}. The events rates in anti-neutrino mode
are $\sim30\%$ of those expected when running in neutrino
mode. Most of the rate reduction is due to the lower anti-neutrino 
cross-sections, as shown in Fig.~\ref{sigs}, and the rest is related to the  lower production of 
$\pi^-$ relative to $\pi^+$ in proton-nucleus
interactions. 

The compositions of the anti-neutrino beams are shown in
Figures~\ref{fig:lerev-composition} and
~\ref{fig:merev-composition}. Note that while the wrong-sign muon
contamination is worse in anti-neutrino mode, the electron neutrino
rates are still roughly $0.3-0.5\%$ at the peak of the neutrino spectrum.

\section{Proton Economics}

In its initial phase of operation NuMI will have to accomodate the $\bar p$
production cycle. Six consecutive batches of 8 GeV protons from the Booster
will be injected into the Main Injector to be accelerated to 120 GeV, with a
cycle time of 1.9~s. One batch will be extracted for $\bar p$ production,
and the remaining five will be extracted onto the NuMI target in about
$8 \mu s$.

At present, with the Booster running routinely at $4.5 \times 10^{12}$
protons/batch, the maximum intensity one can expect  is limited to
$2.3 \times 10^{13}$ protons/cycle.This would provide about $2 \times 10^{20}$
protons/year, assuming 5000~hours/year of operation.

There is no obvious path to increase the total proton flux, but several ideas
are being actively investigated. Both the feasibility of Booster upgrades
to increase the number of protons/batch and the possible gain provided by
proton stacking techniques in the Main Injector (slip-stacking and RF
barrier stacking), to increase the number of protons per Main injector cycle,
are being currently evaluated. Another path would be an upgrade of the
power supplies and the RF system of the Main Injector to allow shorter
operation cycles down to 1 s.

A proton intensity working group has been jointly appointed by the Fermilab
Directorate and the MINOS experiment to produce a roadmap to help maximize
the integrated number of protons to be delivered onto the NuMI target.

\chapter{$\numutonue$ Oscillation Appearance Experiment}

\section{Statement of the challenge}
The challenge for future experiments is to  observe  $\numutonue$ 
oscillations in the atmospheric neutrino $\delmsq$ range down to
the level of a  few parts per mil. 
The CHOOZ experiment gives a limit on $\nue$ disappearance probability 
 of about 0.1. This translates into a limit on the 
$\nue$ appearance probability of 
0.05. MINOS is expected to improve this by a factor of 2-3, setting the 
aforementioned scale of few parts per mil. There are no clear theoretical 
guidelines as to the most likely value of this parameter.

Charged current $\nue$ interactions can be identified by the presence of an 
electron in the final state. The experimental backgrounds to the 
$\numutonue$ oscillation signals arise from two general sources. There are
genuine events with electrons  resulting from 
the intrinsic $\nue$ component in the beam and  
from $\tau$ decays in the  charged current $\nutau$ interactions from 
$\numutotau$ oscillations. In 
addition there are potentially mis-identified NC events or high $y$ $\numu$ CC 
events where one or more $\pi^0$'s in the  final state masquerade as 
an electron.

The intrinsic $\nue$'s in the beam come from $\mu$ decays and $K_{e3}$ 
decays (both 
charged and neutral). They are of the order of $0.5-1.0\%$ of $\numu$'s, 
but can 
be reduced further by an appropriate energy cut. The $\tau$ decay 
contamination 
has a strong energy dependence; the easiest way to reduce it is to have 
a significant fraction of the $\numu$ flux below the $\tau$ production 
threshold. 
$K_{e3}$ contamination is typically of the order of 1/5 of the $\mu$ decay 
contamination in the experiments being considered, as shown in Fig. 
\ref{fig:nuespect}.

The experimental challenge can be separated into two parts: 
\begin{itemize}
\item reducing these 
two backgrounds as much as possible 
\item measuring them as accurately as 
possible so that the principal ultimate uncertainty comes from the 
statistical fluctuations in the event sample of interest
\end{itemize}
 
The background from $\nue$'s can only be reduced by good energy resolution since 
the $\nue$'s from background sources have a broader energy spectrum than the 
potential signal. The NC background, however, can be reduced by a well 
designed  detector. The challenge is to suppress it to a level comparable 
or lower than the intrinsic $\nue$ level without significantly degrading 
the signal detection efficiency.

\section{Possible experimental approaches}
A traditional challenge in high energy physics experiments has been how to 
separate out the electromagnetic component in a hadronic jet from the 
remaining hadrons. In the calorimetric method this was generally achieved by 
having a high Z electromagnetic calorimeter in front of the hadron section of 
the calorimeter. Clearly that technique is not 
suitable for electron/$\pi^0$  separation. The latter has been traditionally 
done in open geometry experiments by using a water Cherenkov , a technique 
 adopted by  
IMB, Kamiokande and SuperKamiokande. The other technology of choice 
 has been  the use of low Z calorimeters (eg CHARM II and BNL 
oscillation experiment) which allow identification of the electron by 
tracking.

In principle, at least, in a highly segmented detector
 electrons can be separated from 
$\pi^0$'s by utilizing several experimental characteristics:
\begin{itemize} 
\item finite 
separation between the vertex and conversion points of the $\gamma$'s from the 
$\pizero$,
\item two electromagnetic showers (for $\pizero$) vs one (for electrons), 
\item double pulse height right after conversion of a $\gamma$. 
\end{itemize}
Success of the 
separation based on these criteria requires fine segmentation: longitudinally,
less than a radiation length, $X_0$; transversely, finer than the spatial 
separation of the two $\gamma$'s from the $\pizero$ decay. The transverse 
segmentation also has to be such that individual tracks in the final state 
can be separated from each other.

Besides the need to distinguish electrons from $\pizero$'s, one must also 
distinguish electrons from hadrons. This is harder in a low Z material and 
 relies on the absence of hadronic interactions (for electrons) and 
a generally broader track from electrons due to accompanying bremsstrahlung.

\begin{figure}
\begin{center}
\includegraphics*{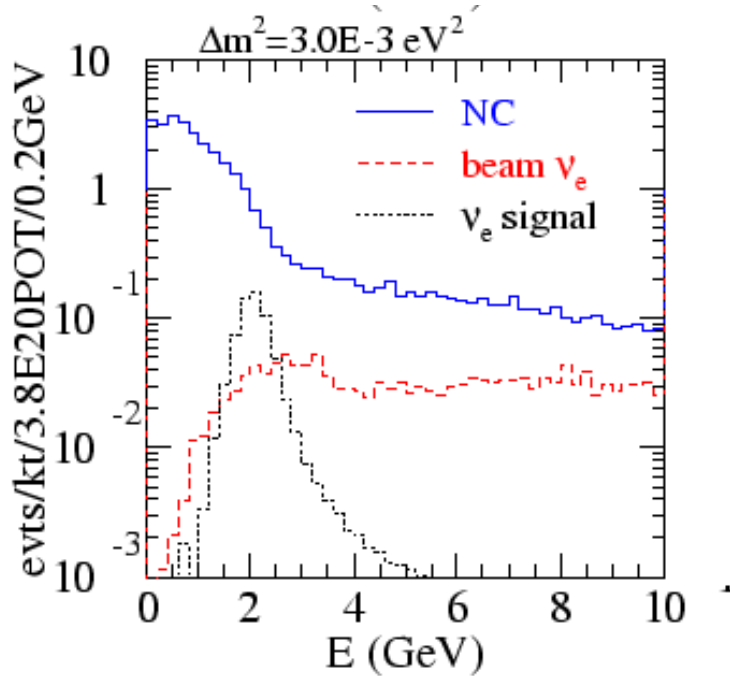}
\caption{Energy distribution of the expected $\nue$ oscillation signal for $\left|U_{e3}\right|^{2}=0.01$  (black
histogram), intrinsic  beam $\nue$ events (red histogram), and observed energy distribution of the NC events before rejection (blue histogram)  }
\label{bckg}
\end{center}
\end{figure}

The other important characteristic of a good $\nue$ detector is its energy 
resolution. The way to reduce the intrinsic $\nue$ background is to 
utilize the 
fact that the  events from $\numutonue$ oscillations will have a  sharp 
energy spectrum 
at a predictable energy in contrast to the backgrounds. 
This is a key feature of the narrow neutrino beam energy spectrum seen
by an off-axis detector.
$\nue$'s from $\mu$ decays will be in roughly the same 
energy 
range as $\numu$'s from $\pi$ decay (and hence oscillated $\nue$'s) but 
have a much 
broader energy distribution. $K_{e3}$ decay will typically give much higher energy 
neutrinos whereas the $\tau$ decay electrons will peak towards low energies. 
The shape and the level of backgrounds and possible signal are shown in 
Fig. \ref{bckg}.

\section{Background measurements and estimates; systematics}

The backgrounds discussed above can be both estimated and measured.
We
discuss each background in turn.
\renewcommand{\theenumi}{\alph{enumi}}
\renewcommand{\labelenumi}{\theenumi)}
\begin{enumerate}
\item $\nue$'s from $\mu$ decay. 

The $\numu$ flux at $0^o$ will be measured in the Near 
Detector located on axis. Muon flux over all angles can be estimated reliably 
from those measurements since only kinematics are involved. 
 The $\nue$ rate and energy distribution from this source can be 
known with an accuracy of $5\%$\cite{nue_bckg}

\item $\nue$'s from K decay. 

This background is about a factor of 5 lower than the 
one from muon decays. It can  be calculated with the basic 
limitation coming from uncertainties of $\pi$/K  production cross section 
by 120 GeV protons. In principle this uncertainty will be reduced by the 
measurements of
the E907 experiment.
Most of this background occurs at high $\nue$ energies, therefore 
 an extrapolation  to the region of interest can be performed.

\item Electrons from $\tau$ decays. 

This background is comparable to but 
somewhat 
smaller than the one from K decays in our energy range.
The $\nutau$ rate can be reliably estimated from oscillation 
parameters and $\numu$ flux. There is some uncertainty in the production cross 
section and angular distributions but, because of the low overall rate,  
these uncertainties will have a negligible contribution to the overall 
systematic uncertainty on the total background.

\item NC events misidentified as $\nue$ interactions. 

An a priori estimate of this background is
 somewhat more difficult than for the cases discussed above and 
 it will need to be measured.  Initial Monte Carlo calculations 
indicate that its contribution can be reduced to a level of the intrinsic
$\nue$ background or better 
while maintaining the efficiency for $\nue$ identification at about $30-40\%$.
 This 
source of background will     probably give the dominant contribution to the 
systematic error.

      A measurement of the NC background can be made using a near off-axis 
detector
 located at an angle similar to the one of the far detector. Such a detector 
is possible in the bypass tunnel downstream of the NuMI absorber where the 
neutrino spectrum would be similar to that in the far detector.

\item  $\numu$ CC events mis-identified as $\nue$ interactions. 

Such a misidentification 
would require a failure to detect the muon, residual hadronic energy in the 
range of the expected oscillation signal, and misidentification of a pion 
as an electron. We note that the rate of the   $\numu$ CC events is 
significantly reduced by the oscillations.
Initial estimates indicate that this background is significantly lower 
than the NC background. 

\item  Hadrons misidentified as electrons. 

There is now a significant amount of 
data on electron-muon (or pion) separation in water Cherenkov detectors with 
the conclusion that this background is insignificant for energies 
 above few hundred 
MeV. A low Z detector would have to be exposed to hadrons and electrons in 
a test beam. Such an exposure will guide the design of the detector and allow 
one to develop an optimum algorithm for e/hadron separation.
\end{enumerate}
In summary, the Monte Carlo studies to date indicate that for a detector in 
the tens of kton range the systematics associated with imperfect knowledge of 
the background levels can be kept sufficiently low so that the dominant 
uncertainty
will be due to statistical fluctuations. More detailed studies with the 
detector of choice are needed, however, to confirm up this initial impression. 

\section{Physics Potential of the NuMI Off-axis Beam}

At the present time there are a large number of unanswered questions in 
neutrino physics. Some of the most important ones that are relevant to the 
proposed program are:
\begin{itemize}
\item Is the LSND effect related to neutrino oscillations?
\item Is the LMA solution to the solar neutrino problem the correct one?
\item What is the value of $\theta_{13}$?
\item What is the sign of the mass hierarchy?
\item Is CP conserved in the neutrino sector?
\item Is CPT conserved in the neutrino sector?
\end{itemize}

It is highly likely that the first two questions will be answered by the time 
the experimental program described here will begin. The last 4 topics are the 
focus of the proposed program.

Our discussion below tacitly assumes that the answer to the first question is 
in the negative. If that assumption is incorrect, the neutrino community will 
have to rethink seriously the future program. 

Regarding the second question, KamLAND should provide a definitive answer 
here should they see depletion of events. If they do not, and future NC data 
from SNO reinforce the support for the LMA solution, then very likely new and 
unexpected physics lurks in the neutrino sector. The argument for new 
experimental efforts becomes even stronger.

The experimental program being proposed here takes into account the need to  
maintain maximum flexibility as answers to the last four questions become 
slowly elucidated. 
We divide our program into two phases: the second phase will 
build on the results from Phase I as well as on the information provided by 
the potential JHF program. We can, however, start planning Phase I now.

At the present time two generic site possibilities appear to be front runners 
for the Phase I of the proposed experiment. 
The first one is at 712 km, near the current Soudan 
experiment. The second site possibility is in the region in Canada between 
the southern and northern branches of the Trans Canadian Highway, between 
850 and 980 km. The near-Soudan site gives good sensitivity to $\nue$ 
appearance 
signal, on the assumption of  $ \dmsq23 = 3 \times 10^{-3}~eV^2 $. 
The further sites are better 
if $\delmsq$ is lower and 
offer larger matter and CP violation effects and thus 
 a chance to obtain at least the first glance at these two phenomena.

\begin{figure}
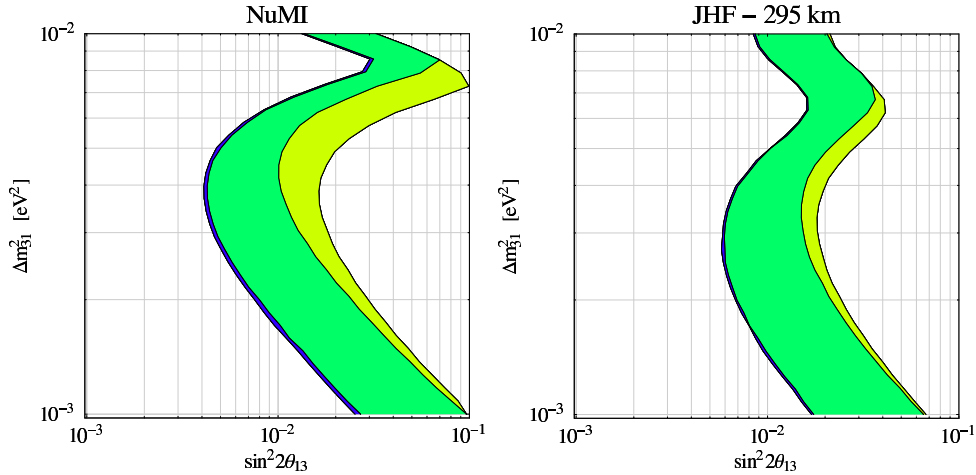

\begin{center}
\epsfig{figure=NUMI.eps,width=2.5in}~
\epsfig{figure=jhf.eps,width=2.5in}
\caption{Interpretation of the observed $\numutonue$ transition probability 
in terms of $\ssq2t13$ for the NuMI(712 km), left,  and the JHF (right). The
left edge is the limit implied by the statistical errors. The right edge of the
 blue region corresponds to a $10\%$ error on the background subtraction.
The green band represents uncertainty due to the value of $\delta$. The yellow
band shows the impact of degeneracy between $\theta_{23} < \pi/4$ and 
$\theta_{23} > \pi/4$ solutions\cite{huber}.  }
\label{degeneracies}
\end{center}
\end{figure}

   The sensitivity of Phase I to observation of $\nue$'s is hard to quantify
because the size of the expected signal depends not only on the value of
$\th13$ but also on the pattern of mass hierarchy (through matter
effects) and the value of the CP violation phase as shown in 
Fig.~\ref{degeneracies}. At low values of $\ssq2t13$ there is an additional
potentially important contribution to the uncertainties in the interpretation
of the experimental results due to the uncertainty of the value of
$\Delta m_{12}^2$.
We try to give a
general estimate by making a rough comparison with the JHF\cite{JHF} 
proposal (see
below, Tables~\ref{JHF_table}  and~\ref{NUMI_table}). 
 In Table~\ref{NUMI_table} we have assumed a 20 kt detector, 5 yr
experiment, $85\%$ fiducial volume, NC background equal to the beam $\nue$
background and $4\times 10^{20}$ protons delivered per year and no matter or CP
effects. Because the matter effects are much more
important at NuMI distances, NuMI sensitivity could increase by up to
$30\%$  at $712~ km$ for normal mass hierarchy, or more at $900~km$.

\begin{figure}
\begin{center}
\epsfig{figure=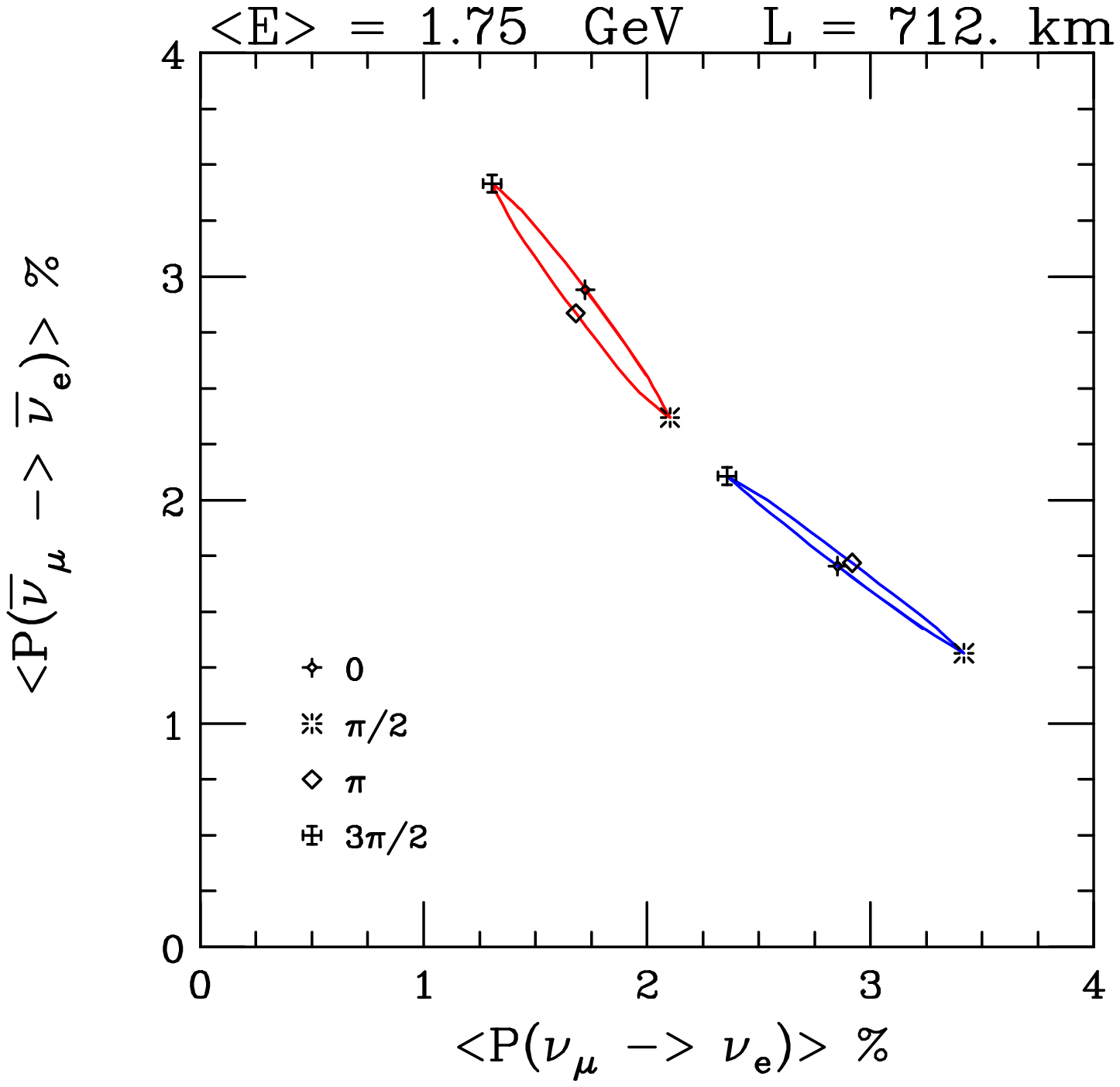,width=1.5in}~
\epsfig{figure=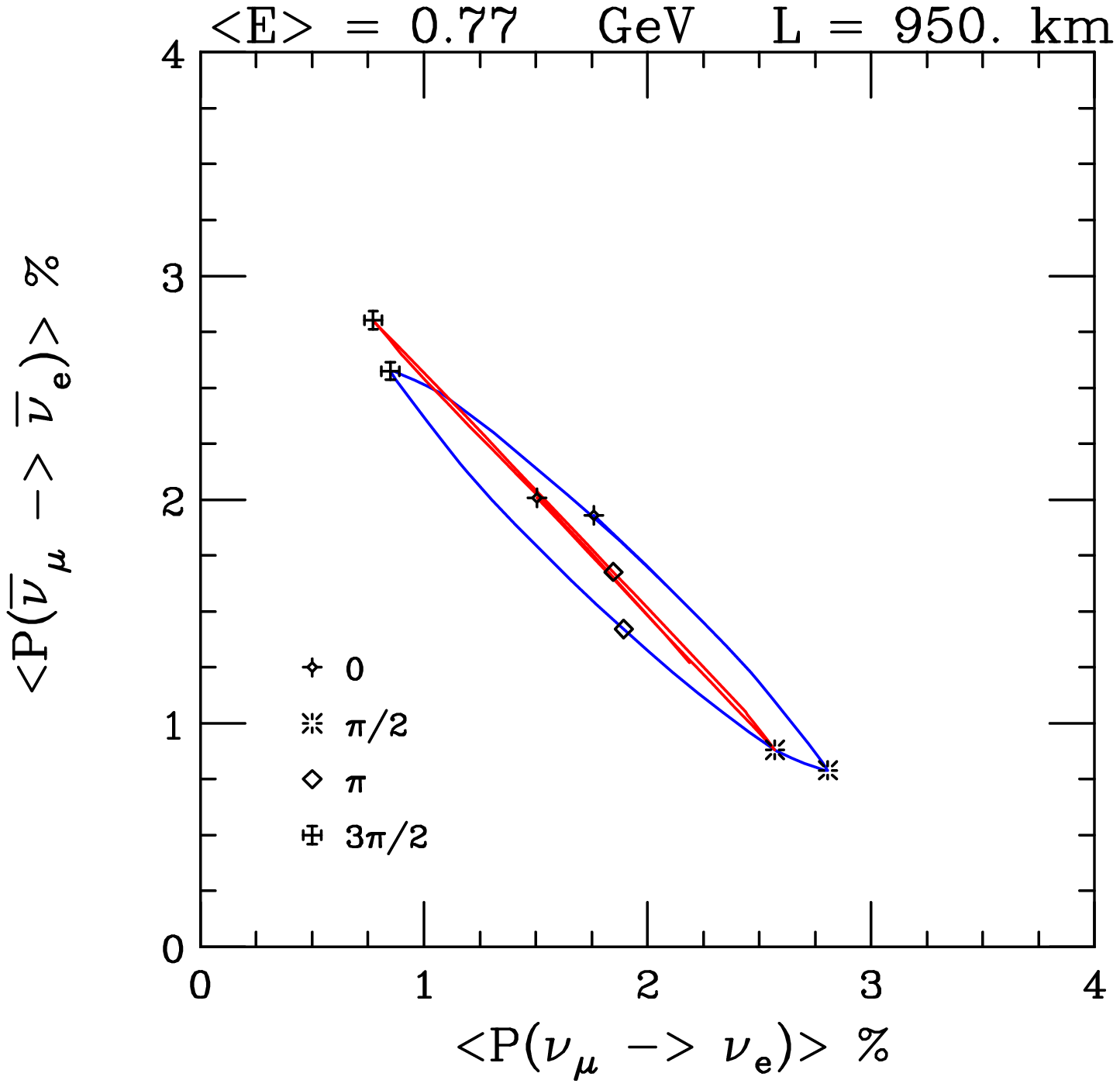,width=1.5in}~
\epsfig{figure=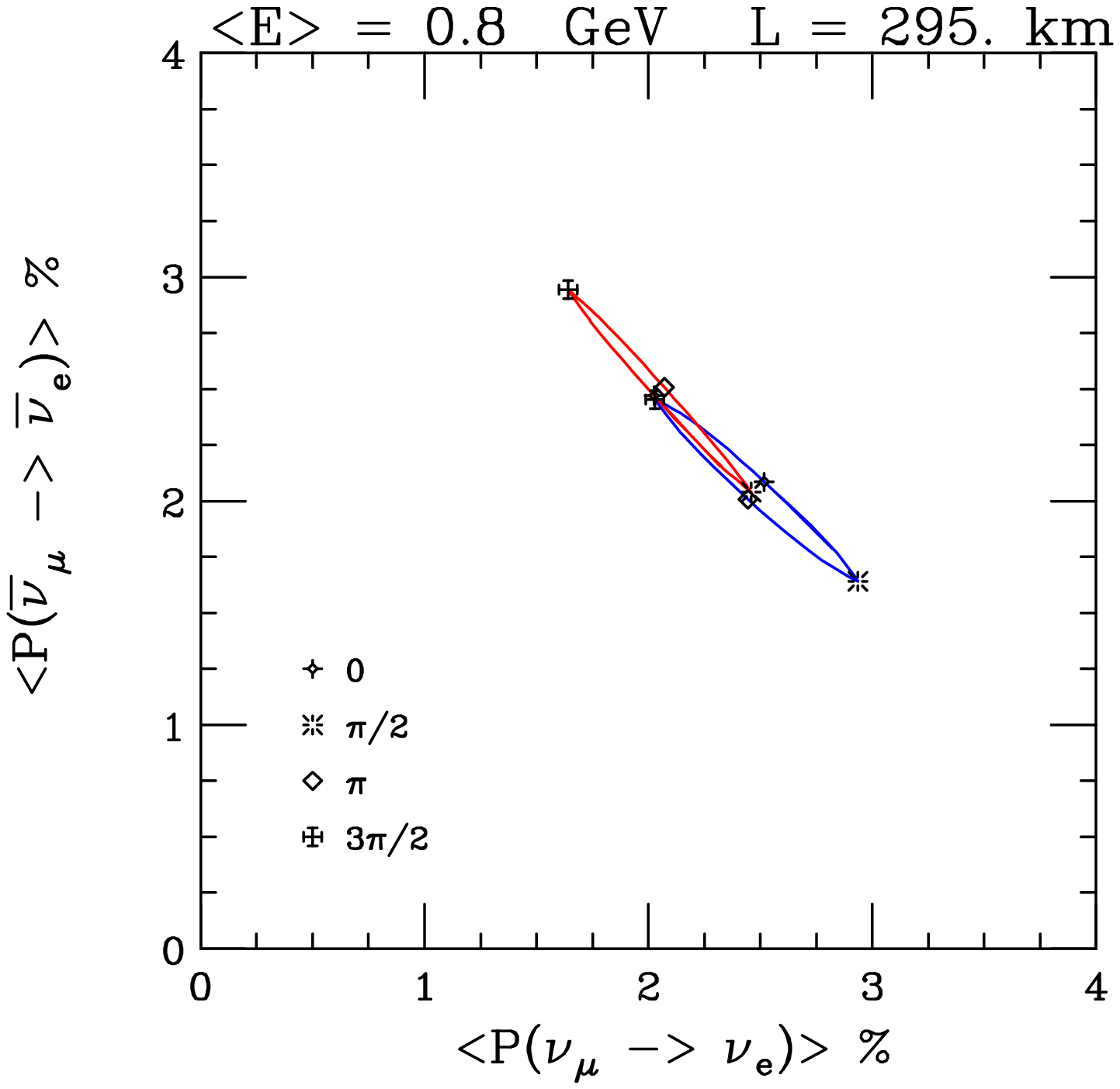,width=1.5in}\\
\epsfig{figure=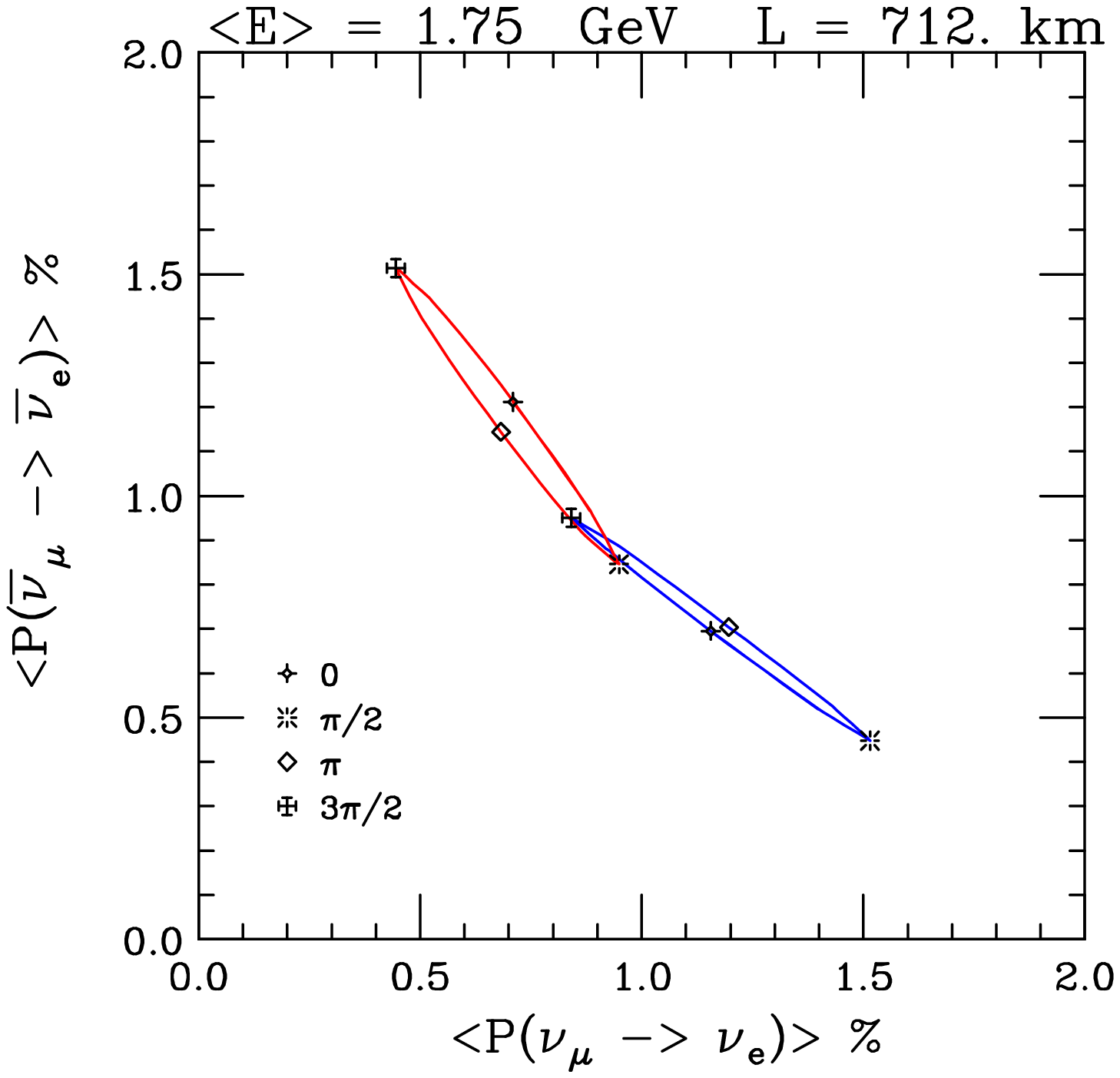,width=1.5in}~
\epsfig{figure=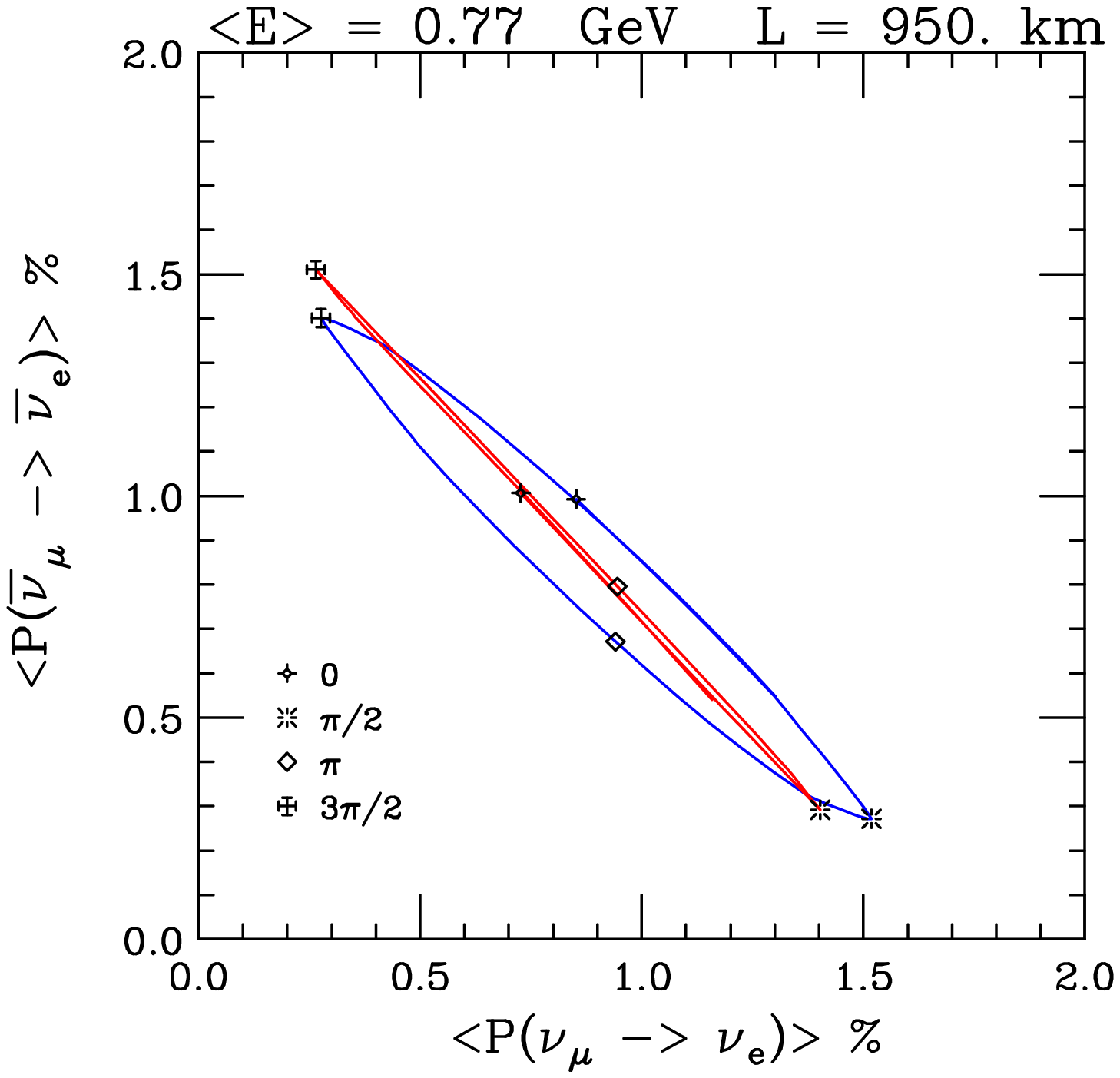,width=1.5in}~
\epsfig{figure=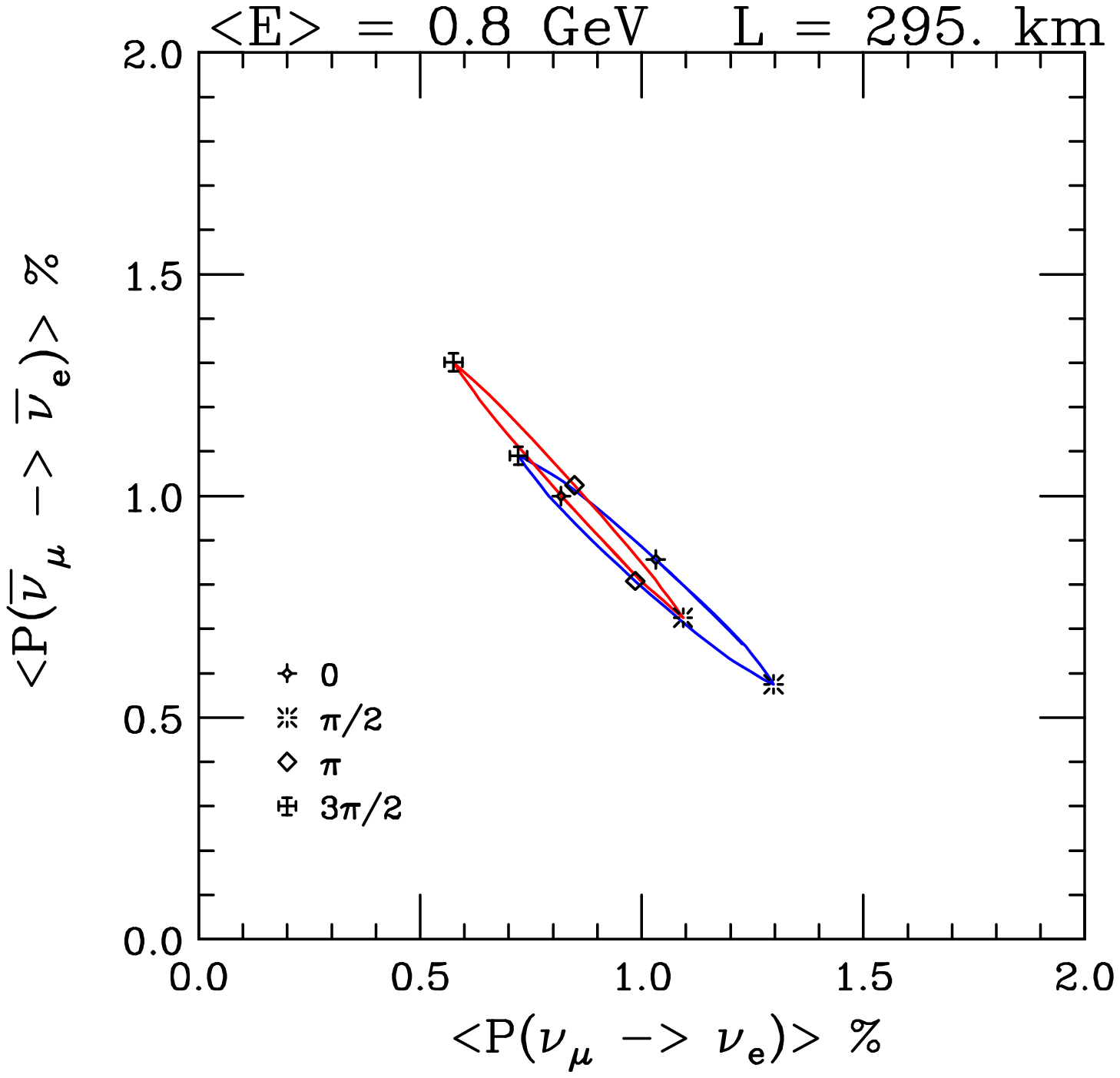,width=1.5in}
\caption{CP trajectory contours in the bi-probability space for 
$\nue$ and $\nuebar$ appearance at $295~km$, $712~km$ and $950~km$. Top row 
is for $\ssq2t13=0.05$, bottom row is for $\ssq2t13=0.02$. Symbols denote the 
expected values of probabilities for specific values of the CP phase.}
\label{CPtraj}
\end{center}
\end{figure}

Fig.~\ref{CPtraj}  shows the CP trajectory contours in the bi-probability space for 
$\nue$ and $\nuebar$ appearance at $712~km$ and $950~km$ using optimized 
energies at 
those distances. We use the oscillation parameters of the solar neutrinos
${\sin^{2}2\theta_{12}=0.8}$, $\Delta m^{2}_{12}=5\times 10^{-5}~eV^{2}$  and values of $\ssq2t13$
corresponding to an oscillation probability equal to 1/2 and 1/5 of the 
current CHOOZ limit. It is clear that given sufficient statistics the 
proposed experimental program would be very well positioned to identify the 
sign of the matter effect and obtain some information on CP violation, 
especially when one combines information from neutrino and antineutrino 
running. For comparison we show also the same contour plots, for identical 
oscillation parameters, for the proposed JHF program with $L=295~km$ and 
$E_{\nu} = 0.8~ GeV$.
 
Studies of energy dependence of the disappearance of $\numu$ CC events can 
provide information about equivalence of $\dmsq23$ for $\numu$ and $\numubar$ 
oscillations and thus validity of CPT in the neutrino sector. This comparative
 measurement will probably be dominated by systematic effects; no quantitative
 estimates of how well one can do have been performed to date but $1\%$ 
accuracy 
does appear possible.

\section{Comparison with JHF Phase I}

The phase I of the contemplated JHF program would utilize the proposed 
neutrino beam from the JHF accelerator, currently under construction, and 
the SuperKamiokande detector, to be reconstructed after the end of the K2K 
experiment. The source to the detector distance would be $295~km$ and the mean 
energy for the currently favored $2^o$ off-axis beam is $0.8~GeV$, thus the 
L/E ratio would be quite comparable to the value in the proposed NuMI program.

Similarly to the NuMI program it plans to achieve better than a factor of 
10 improvement in $\ssq2t13$ over the CHOOZ limit at a value of 
$\dmsq23 = 3\times 10^{-3}$, somewhat better at lower values and 
somewhat worse at higher 
values. The matter effects are much smaller than in NuMI case and it 
probably will 
not be possible to obtain any significant information on CP violation in this 
Phase. The most recent proposal does not indicate any plans to run in the 
$\numubar$ mode.

The sensitivity of the proposed JHF PhaseI to matter effects and CP violation 
is quantified in Fig.~\ref{CPtraj}.  The relative sensitivity of the 
two proposed 
programs to the $\nue$ appearance signal are compared in 
Tables~\ref{JHF_table}~and~\ref{NUMI_table}.
We use 
the same parameters ($\dmsq23 = 3 \times 10^{-3}$ and $\ssq2t13 = 0.1$, i.e. 
at the CHOOZ limit, and a 5 yr exposure for both proposed programs). We assume 
$4\times 10^{20}$ POT/yr for NuMI and a 20 kt detector with $85\%$ 
fiducial volume. 
We use numbers from  Table 2 of the JHF proposal.

\begin{table}
\begin{center}
\begin{tabular}{|c|c|c|c|c|}
\hline
 & $\numu$ CC & NC & Beam $\nue$ & Signal $\nue$ \\ \hline
all        & 10714 & 4080 & 292 & 302 \\ \hline
after cuts &   1.8 &  9.3 &  11 & 123 \\ \hline
\end{tabular}
\end{center}
\caption{Anticipated number of signal and background events in JHF
OAB $2\deg$ beam}
\label{JHF_table}
\end{table}

\begin{table}
\begin{center}
\begin{tabular}{|c|c|c|c|c|}
\hline
 & $\numu$ CC & NC & Beam $\nue$ & Signal $\nue$ \\ \hline
all        & 12104  &  5696    & 295.4  &  293 \\ \hline
after cuts &    & 10.2   & 10.2  &  85.5 \\ \hline
\end{tabular}
\end{center}
\caption{Anticipated number of signal and background events 
in the detector located $9~km$ off-axis at a distance
of $712~km$ from Fermilab}
\label{NUMI_table}
\end{table}

\chapter{Future Evolution of the Off-Axis
Neutrino Program}

In addition to searching for unexpected
phenomena, such as CPT violation, the off-axis
program strives to measure \sinsqthot, the
order of the mass hierarchy, and the
CP-violating phase $\delta$.  It is rather
unlikely that the Phase I program described in this
proposal will be able to complete these objectives.  In
almost all scenarios there will be
a strong argument to enhance the program by increasing
the flux times detector mass and possibly by moving
detectors or building additional detectors to obtain
different values of $L$ and $E$.  

There are three reasons for expecting the need to
eventually expand the program:
\begin{enumerate}
\item The effects, particularly those due to CP violation,
are quite likely be small, requiring greatly increased
statistics.  A reasonable expectation is that one would
like to eventually increase the proton flux by a factor
of four, as envisioned in the Proton Driver study, and to
increase the detector mass by a factor of five.
\item Since the parameters scale differently with
$L$ and $E$, different detector locations will be
advantageous to measure them.
\item Different detector locations might be
necessary to resolve ambiguities that can occur with a
single detector location.
\end{enumerate}

We will give a few examples to illustrate these
points.  The first goal of the Phase I program is to
detect a positive signal for $\nu_\mu\rightarrow\nu_e$
oscillations.  Such a signal should be seen if
\sinsqthot\ is greater than 0.01.  If Phase I is
successful in this, then we can proceed to try to
determine the parameters.  If no significant signal is
seen, then increased flux times detector mass will be
necessary to continue the search for a non-zero value
of \sinsqthot.  Systematic errors need to be considered
carefully to determine what additional subsidiary
experiments are needed to control them.   

Detection of a signal for $\nu_\mu\rightarrow\nu_e$
appearance will not in itself provide a measurement of
\sinsqthot\ since relatively large matter effects
introduce a two-fold ambiguity in the measurement of
\sinsqthot, in the absence of CP-violating effects.  In
the presence of CP-violating effects, there will be
additional ambiguities caused by the value of $\delta$,
unless the value of \sinsqthot\ is relatively large,
greater than about 0.02.  

To understand how we proceed to resolve ambiguities
and measure the parameters, it is useful to review how
these effects scale with distance, energy, and the small
parameters.  Since both effects are maximal at the peaks
of the
\dmsqoth
$\approx$ \dmsqtt\ oscillation, we assume that
measurements will be done at odd-integer values of the
oscillation phase.  The practical values of $n\pi/2$ are
the first peak, $n=1$, and the second peak, $n=3$.  This
implies the a scaling in $E$ is equivalent to a scaling
in $L/n$.
In the NuMI region, the matter effects scale as
$\theta_{13}^2E$ or $\theta_{13}^2L/n$.  CP violating
effects scale as $\theta_{13}\Delta m_{12}^2n$.  

Consider the case of
relatively large values of $\theta_{13}$ first.   The
matter effects, which scale as $\theta_{13}^2$ dominate
the CP-violating effects, which scale as $\theta_{13}$. 
In this regime, the mass hierarchy can be determined
without ambiguity by the combination of neutrino and
antineutrino runs.  (See Fig.
\ref{probsfig}) For NuMI baselines, the matter effects are
large.  The double ratio
$$\biggl[{\sigma(\nu_\mu\rightarrow\nu_e)\over
\sigma(\nubar_\mu\rightarrow\nubar_e)}\biggr]_{\rm
normal}\biggl/\biggl[{\sigma(\nu_\mu\rightarrow\nu_e)\over
\sigma(\nubar_\mu\rightarrow\nubar_e)}\biggr]_{\rm
inverted}$$
is 1.7 for $L=712$ km, 1.8 for $L=850$ km, and 2.0 for
$L=985$ km.  Thus, once electron appearance has been
seen, the value of $\theta_{13}$ and the mass hierarchy
can be determined simultaneously.   However, the necessary
antineutrino running requires increased integrated flux
times detector mass, since fewer antineutrinos are
produced and their cross section is lower.

If \sinsqthot\ $< 0.02$, then there can be ambiguities
caused by the CP-violating phase $\delta$.  To resolve
these ambiguities, and to make a measurement of $\delta$
regardless of the value of \sinsqthot, it may be
desirable to make measurements at the second oscillation
maximum,
$n=3$.  The scaling laws indicate that at the second
maximum the matter effects decrease by a factor of three
and the CP-violating effects increase by a factor of
three.  (There is however, a decrease in both the flux
(see Fig.
\ref{fig:numi-oaspect} and the cross section, due to the
reduced neutrino energy.)  Since, in the off-axis beam,
each detector location represents a single $L$ and a
single $E$, this would require an additional far
detector site.  Water Cherenkov detectors might be more
appropriate for the low energies (about 500 MeV) of the
second maximum.  They have the great advantage that the
cost of detectors scales as the area of the detector
rather than as the mass.

Finally, it should be noted that there is a high degree
of complementarity between a JHF program at 295 km, and
NuMI off-axis detectors at 712 km (first maximum) and 985
km (second maximum) as shown in the following table.

\begin{center}
\begin{tabular}{|c|c|c|c|c|} \hline
Detector & $L$ & $E$ & Relative & Relative \\
         & (km) & (GeV) & matter effect & CP effect \\
\hline\hline
JHF & 295  & 0.6 & 1.0 & 1.0 \\ \hline
NuMI Phase I & 712 & 1.4 & 2.9 & 1.0 \\ \hline
NuMI Phase II & 985 & 0.7 & 1.1 & 3.0 \\ \hline
\end{tabular}
\end{center}

Measurements at both a 300 km and a 700 to 1000 km
baseline will further aid in resolving ambiguities and
checking  consistency with the theory.

In summary, we expect that the investigation of the
neutrino mass hierarchy and the parameters of the MNS
matrix will be an extended and fruitful program.  Arguably it is
as or more important than the efforts now underway to
measure the parameters and check the consistency of the
CKM matrix, and should receive, with time, at least a
similar level of support.  We expect that it will be a
major component of the Fermilab program in the LHC era.

\chapter {Detector}

An off-axis NuMI neutrino beam offers an unique opportunity to study 
$\numutonue$ oscillations. There will be a very large number of $\numu$'s
oscillating away. Most of the resulting $\nu_{\tau}$'s will be below the 
kinematical threshold for $\tau$ production hence a small  admixture
of  $\nue$'s should be detectable with as small background as 
possible. 

To take full advantage of this opportunity it is necessary to construct a new 
detector capable of the detection and identification of the $\nue$ charged 
current interactions. 
Such a detector must meet several challenges:
\begin{itemize}
\item it must have fine granularity in order to identify the final state 
electrons
\item it must have very large mass to provide maximal 
sensitivity  to the  oscillation amplitude
\item it must have an acceptable cost per unit mass
\item it must be able to operate on surface or under a small overburden, as 
there are no convenient underground locations 
\end{itemize}
The detector should be optimized for the neutrino energy range of $1 - 3~GeV$. 

The ultimate sensitivity of the experiment, at the  $90\%~C.L.$ will be 
determined by the condition 
$S\geq1.26\sqrt{B}$, where $S=\epsilon\times\#~of~signal~events$ and 
$B=\epsilon\times\#~of~\nu_{e}+\eta\times NC$ with $\epsilon$ being the 
efficiency for the detection and identification of the $\nue$ interactions and
$\eta $ being the mis-identification probability for  NC interactions.
The sensitivity of the experiment improves with the square root of the exposure
time until  it becomes limited by the the systematic error of the 
background. The latter is likely to be dominated by the error on the NC
component of the background.
 
There appear to be three classes of a possible new detector:
\begin{itemize}
\item water Cherenkov. It is likely to offer the biggest advantage in terms of
the mass of the detector but in the energy regime of interest at NuMI it
appears to be dominated by the NC background, i.e. 
$\eta\times NC \gg\epsilon\times\nue$. The efficiency for the signal events
is likely to be around $25\%$.
\item fine grained calorimeter. This class of detectors is characterized by
the signal detection efficiency of the order of $35-40\%$ while maintaining
a balance between the intrinsic background and the NC-induced one,
 i.e. 
$\eta\times NC \sim\epsilon\times\nue$
\item Liquid Argon TPC. This is probably the most challenging, but the most 
powerful detector, offering the 
highest signal efficiency $\epsilon\sim90\%$ and background level at the
intrinsic $\nue$ component, $\eta\times NC \ll\epsilon\times\nue$
\end{itemize}
     
The NuMI neutrino beam will start operations in the early 2005. It is highly 
desirable that the Phase I of the proposed program is implemented as early as 
it is practically possible to yield the initial information on the $\nue$
appearance and to provide guidance for the future phases of the program.
Such an accelerated time scale leaves no time for the extensive detector R\&D 
program, but fortunately none is necessary. Forty years of experience in 
constructing neutrino detectors combined with the impressive progress in the 
detector technology in the past decade offers several possible implementations
for a detector of the 20 kton class.

 Phase II of the program may require 
much large detector, perhaps in the range of 100-200 ktons. It may well be 
that the the future detector may be required to operate at the energies below
$1~GeV$, if the second oscillation maximum will offer the best sensitivity.
It is not clear, at the present time, if the detector technology for the 
Phase II will be necessarily the same as the Phase I detector.

Studies of  neutrino oscillations, especially in  Phase  II may require
optimization of the detector location. Given the cost of the initial investment
it may be advantageous to have a detector design allowing the relocation of
the detector from one location to another, or, perhaps, reconfiguration of the detector into two or more smaller detectors located in different positions.
  
In  Appendix A we describe several alternatives for a possible new detector 
to demonstrate that there are realistic conservative designs of affordable
detectors as well as there are possible new alternatives.

\chapter{Detector Sites, Physics  Considerations}
 
 Progress in neutrino physics worldwide will be optimized if the  
future NuMI program is comparable in quality and as complementary as possible 
to the planned JHF program. The interpretation of any observation of $\nue$'s 
due
 to oscillations will be difficult because of simultaneous contributions of 
$\th13$, matter effects, and CP phase, $\delta$. Several measurements with 
different conditions will be necessary to unravel the situation.

For a given baseline distance, $L$, different contributions to the $\numutonue$
oscillation probability are principally a 
function of $L/E$.  
In our region of parameter space a contribution of  matter effects grows with 
the distance $L$.  The baseline 
in the JHF program is constrained to be relatively short, i.e. $295~ km$. 
Thus the 
natural choice for NuMI would be to have as long a baseline as is consistent 
with the energies of Main Injector beams and availability of satisfactory 
sites.

It is convenient to discuss optimization of the detector position
 in terms of the oscillation phase 
$\phi =1.27 L \dmsq23 / E$. Generally, unless there are strong cancellations 
from different contributions, the maximum $\nue$ appearance  signal will be 
obtained when  $\phi = \pi/2$. Due to $1/L^2$ dependence of the signal $S$ and 
the background $B$, the optimum 
sensitivity, taken as a ratio of $S/\sqrt{B}$, will correspond to a 
distance L about $25\%$ shorter. 
For $\dmsq23 = 0.3~ eV^2$, 
this translates to an optimum transverse distance away from the beam axis of 
about $9~ km$ (independent of $L$). This distance 
scales inversely with $\dmsq23$.

The maximum matter effects occur at a larger value of the phase which 
translates into an equivalent  transverse distance of $13-15~ km$. 
In addition, 
as mentioned above, they grow with L. CP violation effects also generally 
increase with increasing $\phi$.  The NuMI neutrino beam comes to the surface 
at the distance $L=747~km$ from Fermilab. 
A transverse displacement from the beam axis can be achieved either
 by going  further or closer, along the nominal beam direction,  
 effectively taking the off-axis beam in the
 vertical plane or by going sideways at a baseline corresponding to the 
nominal  $0^o$ distance, or a combination of both. 

An important issue is the change in sensitivity as we move away from the 
optimum distance. This defines how precisely we need to know $\dmsq23$ before 
choosing the site. This change is relatively small if we try to optimize the 
size of the total appearance signal, and a transverse displacement from the 
optimum (at a fixed L) can be as large as $\pm 2~ km$.The sensitivity is greater 
(by roughly a factor of 2) if one tries to maximize a contribution of the 
matter effects. These 
comments should be viewed as qualitative statements since due to the coherent 
nature of all contributions, a more rigorous statement can be made only in the
 context of specifying values of all the relevant parameters.

Another issue which needs to be considered is the variation of total $\numu$ 
flux as a function of distance $L$ and mean beam energy, controlled by the 
distance from the beam axis. The flux will fall off with $L$ as $1/L^2$. 
For 120 GeV proton energy and off-axis angles corresponding to neutrino energy
 range 1 - 2 GeV, the flux times cross section drops a little faster than 
$E^2$. Thus we gain slightly in event rate, for the same phase $\phi$, 
as we go to 
larger distances, but that effect is sufficiently small so as not to be very 
important.

To give an idea of the relative magnitude of potential contributions due to 
matter effects and CP violation in different sites, we present those 
contributions, in parts per thousand addition to the $\nue$ appearance 
probability,  in Table~\ref{osc_contr}.
 We show them for two values of $\th13$, 0.07 and 
0.1, corresponding roughly to a factor  of 5 and 2.5 below the current CHOOZ 
limit, and for $\dmsq23=2.5\times10^{-3}~eV^2$,
 the currently best fit to the K2K and SuperKamiokande data.

\begin{table}
\begin{center}
\begin{tabular}{|c|c|c|c|c|}
\hline
Distance &  0.07 - matter & 0.07 CPV & 0.1 - matter & 0.1 CPV \\ \hline
295 km   &       0.56     &  2.87    &   1.14       &   4.08  \\ \hline
712 km   &       2.45     &  4.92    &   4.97       &   6.99  \\ \hline
848 km   &       3.01     &  5.08    &   6.07       &   7.21  \\ \hline
985 km   &       4.50     &  5.76    &   9.12       &   8.18  \\ \hline
\end{tabular}
\caption{ Potential contributions, in parts per thousands, 
to oscillation probability from matter 
effects and CP violation}
\label{osc_contr}
\end{center}
\end{table}

\chapter{Cost and Schedule, Milestones}

\section{Cost}

Given the status of the project it is not possible to have a credible cost
estimate. It appears, though, that the main cost driver will be the 
construction of the active detectors.
There are several possible detector designs
using mature and well understood technologies, thus making a cost of the
active detectors reasonably well predictable. These costs are in the range 
of $~\$15-20M$ 
for a 20 kton class detector. Engineering issues associated with construction 
of a very large and massive detector are very challenging. Costs associated 
with the construction of the absorber, support structures, experimental halls
etc.  cannot be  very well known before the  engineering studies are  
completed. We aim 
at the design of the Phase I of the experiment for a cost not exceeding 
$\$50-100M$.

The second phase of the program may involve upgrades to the accelerator 
complex of Fermilab. These upgrades are subject of a dedicated 
studies \cite{proton_driver}. It is likely that the Phase II detectors
will require a significant increase in the fiducial mass (a factor of 5 or so)
and, perhaps, in the detector capabilities. It is not possible to give a 
reliable cost estimate of these detectors, but it appears likely that
this phase will require investment of the order of $\$250-300M$.
     
\section{Schedule}
   
NuMI beam line will start delivering neutrinos, on- and off-axis, at the 
beginning of the calendar year 2005. While it is not 
possible to complete a construction of a significant fraction of a large 
detector in such a short time frame, it is important to point out that a timely
construction of a new detector would maximize the scientific output of the
NuMI facility. It is important to note that a modular detector design may 
enable a start of the data taking long before the coentruction is completed.

Proposed Phase I detectors utilize  off-the-shelf technology, not requiring
time-consuming R\&D process. Detailed engineering studies are necessary, 
though, before the final approval of the experiment. Modular design of the
experiment may enable parallel construction at several locations, it may
simplify the final assembly as well.

It appears, therefore, that the schedule of a possible experiment will be 
driven by the approval process and by the availability of the construction 
funds.  We hope for an approval process to proceed efficiently enough to
provide construction funds starting in FY2005. We also hope that the 
construction  of the detector may start earlier through forward-funding from
universities and/or from foreign sources. 

\section{Milestones}

\begin{description}

\item[2002 Summer]  submission of a Letter of Intent

\item[2002 Fall/winter]   detector design work, measurement of the cosmic ray
background

\item[2003 Spring] experimental verification of the level of cosmic ray-induced   background

\item[2003 Summer]   Proposal to the Fermilab PAC , NSF

\item[2003-2004] construction of prototypes, 

\item[2003 Fall] Far site selection

\item[2004 Summer] electron and hadron test beam studies of the detector 
prototypes 

\item[2004]  Reviews, validation, 

\item[2004 Summer] final approval  

\item[2004 Fall] start construction of the far detector building and facility

\item[2005] Construction and installation of the Near Off-Axis detector

\item[2005 Summer] Far detector building complete

\item[2005 Fall] Near off-axis detector starts taking data
 
\item[2005-2006-2007] construction of the Far Off-axis detector

\item[2006 Summer] start taking data with the initial 5 kton detector 

\item[2007-2012] Phase I of the program

\end{description}

We envisage that the Phase II of the program will require an increased 
useful  mass of the neutrino detector by a factor of five or so. It can be 
achieved through construction of more detector modules of the same design
as in the Phase I or by a construction of a diferent  detector with much 
higher detection efficiency (like Liquid Argon TPC) or much larger mass 
(like water Cherenkov). To be in a position to select the most apropriate 
strategy for the Phase II it is imperative that adequate funding for the
required R\&D process is available as early as possible.

\vspace{2cm}
\begin{center}
{\Large \bf Acknowledgments}
\end{center}

Many people have contributed to the ideas and concepts presenetd here.
Their contributions are gratefully acknowledged. 

Special thanks go to
Friederich Dydak for his contribution to the chapter on the evolution of 
neutrino physics. 

Achim Geisser,  Tommaso Tabarelli de Fatis and Carlo Gustavino were the source
of  inspiration for the LoDen detector concept.

Dan Marlow was a source of advice on the RPC chambers and has contributed 
a section on the gas and HV systems.

Marc Mengel made it possible to overcome all annoying problems with forcing
Postscript files into a Latex document. 

\newpage
\appendix
\renewcommand{\thepage}{\Alph{chapter}-\arabic{page}}
\setcounter{chapter}{0}
\setcounter{page}{1}
\chapter{Possible Detector Technologies}
\section{Low Z Tracking Calorimeter}

Identification of the final state electron in a calorimetric detector requires
that the sampling frequency is high, of the order of $1/4-1/3$ of the radiation
length $X_0$. Neutrino detectors must serve as a target and as a detector at
the same time, hence their mass must be maximized. These two requirements lead 
to a conclusion that the absorber should be made out of a low Z material to 
maximize the mass of the detector while maintaining good sampling frequency.
Low Z absorber will lead to a minimal number of the active detector planes, for
a given total mass of a detector, hence it will minimize the cost of the
detector.

A required transverse granularity of the detector is related to the 
local particles density on one hand and to the Moliere radius on the other 
hand. Hadron and electron showers develop over large volumes in a low density
detectors, hence the requirements on the transverse granularity of the detector
will be relatively modest.

A direct consequence of the large spatial extend of events is a necessity
of a fiducial cut of the order of 1 meter away from the edges of the detector.
This, in turns requires that the transverse size of the detector is as large
as it is practically possible. As an example, with the $1~ meter~from~edges$ 
fiducial cut only $81\%$ of a total mass of a $20\times20~m^2$ detector
is in the fiducial volume, for  the $12\times12~m^2$ detector it is only 
$69\%$. 

\subsection{Low energy neutrino interactions in low Z calorimeter}

Neutrino interactions in the region of $1-3~GeV$ involve very low multiplicity
 reactions: quasi-elastic scattering, resonance production, single pion
production.  Deep inelastic scattering sets on above $2~GeV$ although 
multiplicities of the hadronic final states are still very low. $\nue$ charged
current events are recognizable by the presence of an electron in the final 
state. The electron, on the other hand, will be characterized by the presence
of additional hits, due to the onset of the electromagnetic shower, along 
its trajectory.

\begin{figure}
\begin{center}
\epsfig{figure=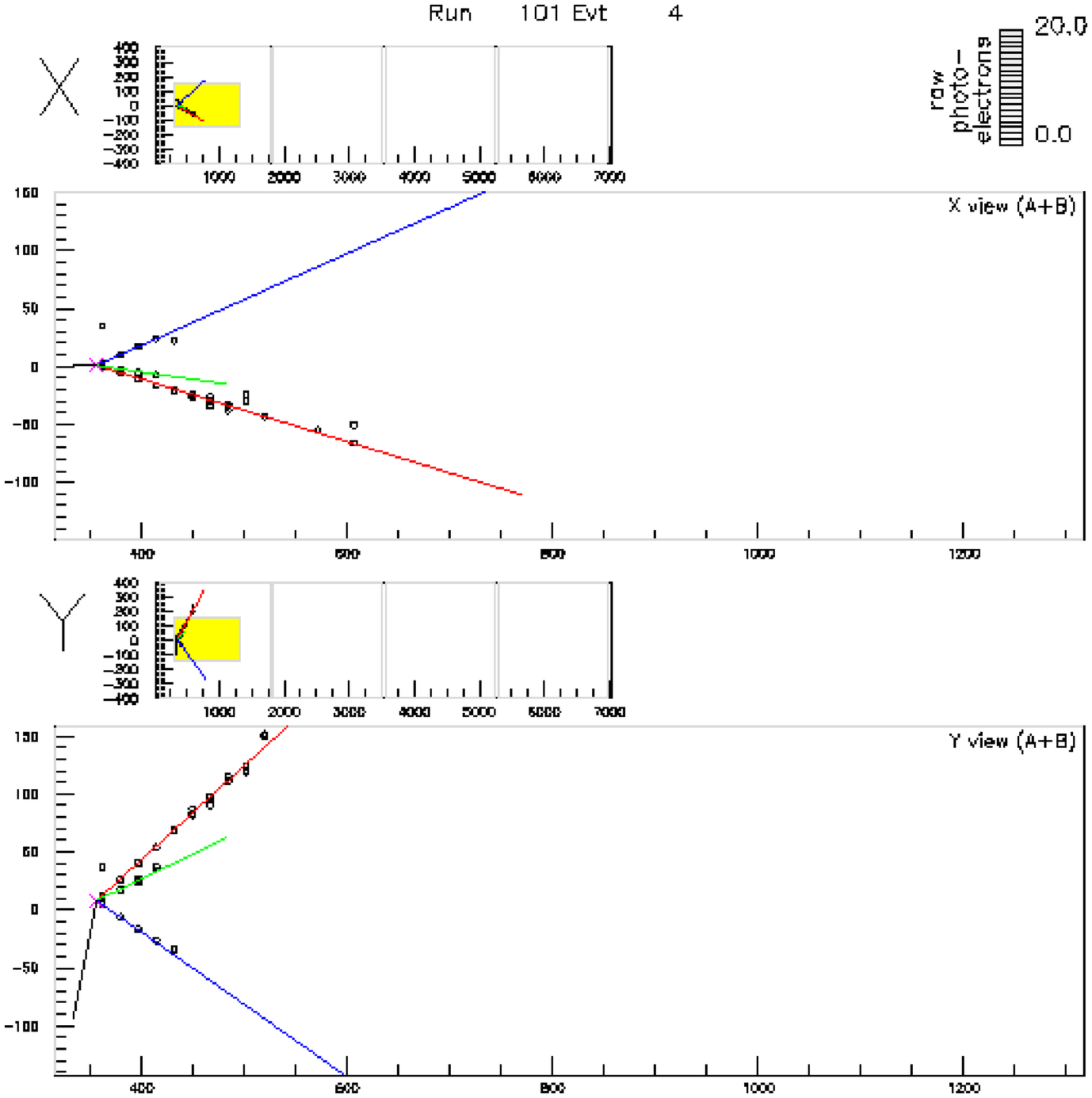,width=2.5in}~
\epsfig{figure=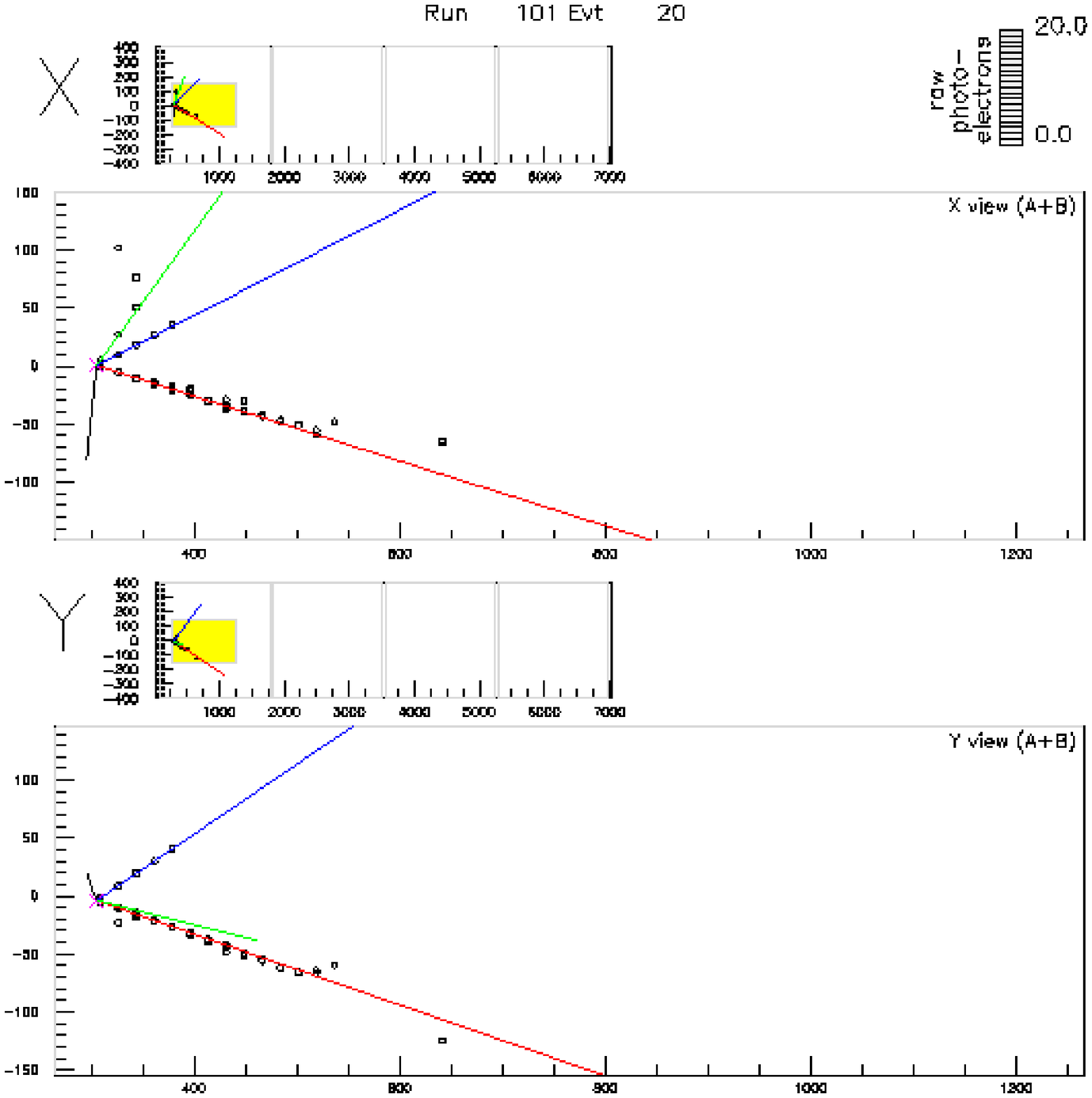,width=2.5in}

\epsfig{figure=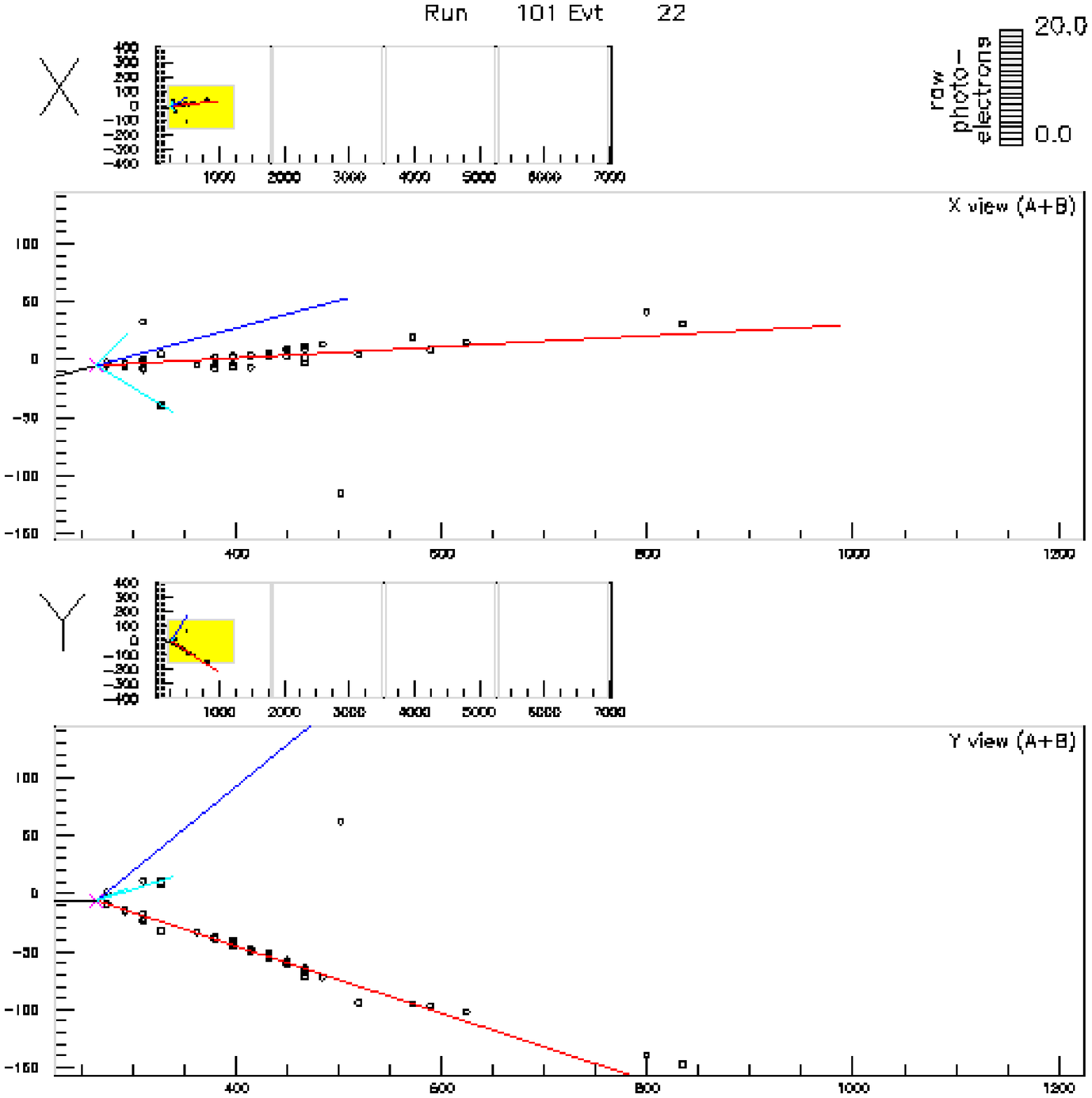,width=2.5in}~
\epsfig{figure=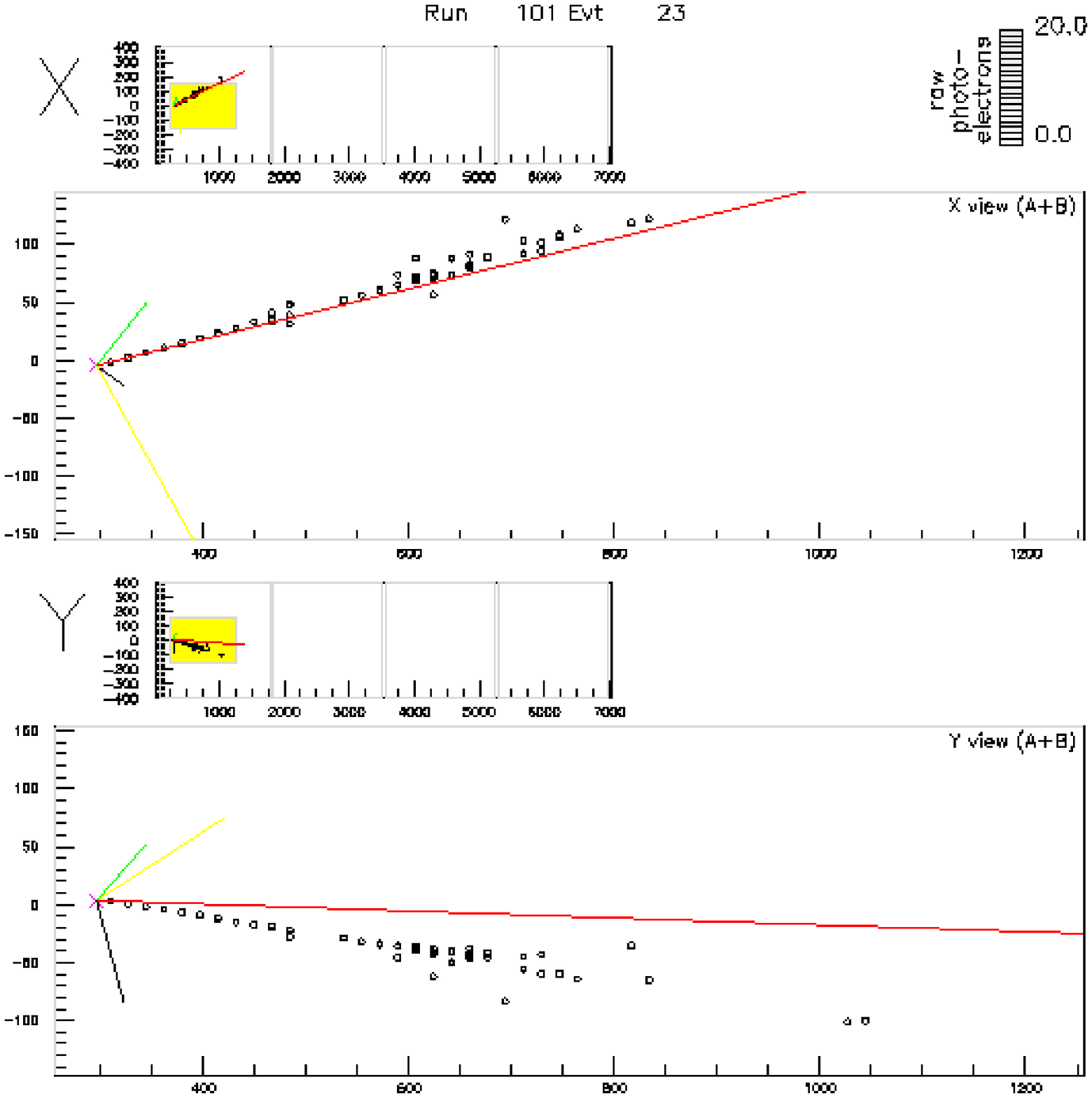,width=2.5in}

\caption{Typical charged current $\nue$ interactions with energies $1-3~GeV$
as detected in a fine grained calorimeter with the longitudinal sampling of
$1/3~X_0$ and the transverse sampling of $3~cm$.}
\label{signal_events}
\end{center}
\end{figure}
      
  In such a low multiplicity environment a low Z tracking calorimeter will
provide a valuable information on the topological properties of the event.
This potential is illustrated in Fig.~\ref{signal_events} showing several 
examples of low energy $\nue$ CC events are registered in a plastic calorimeter
with $1/3~X_0$ sampling and with the transverse granularity of $3~cm$. Black 
circles denote the hit detector strips. Colored lines are indicating the 
initial directions of final state particles: red - electron, blue - proton,
green - charged pion, yellow - neutron and cyan - gamma.

\begin{figure}
\begin{center}
\epsfig{figure=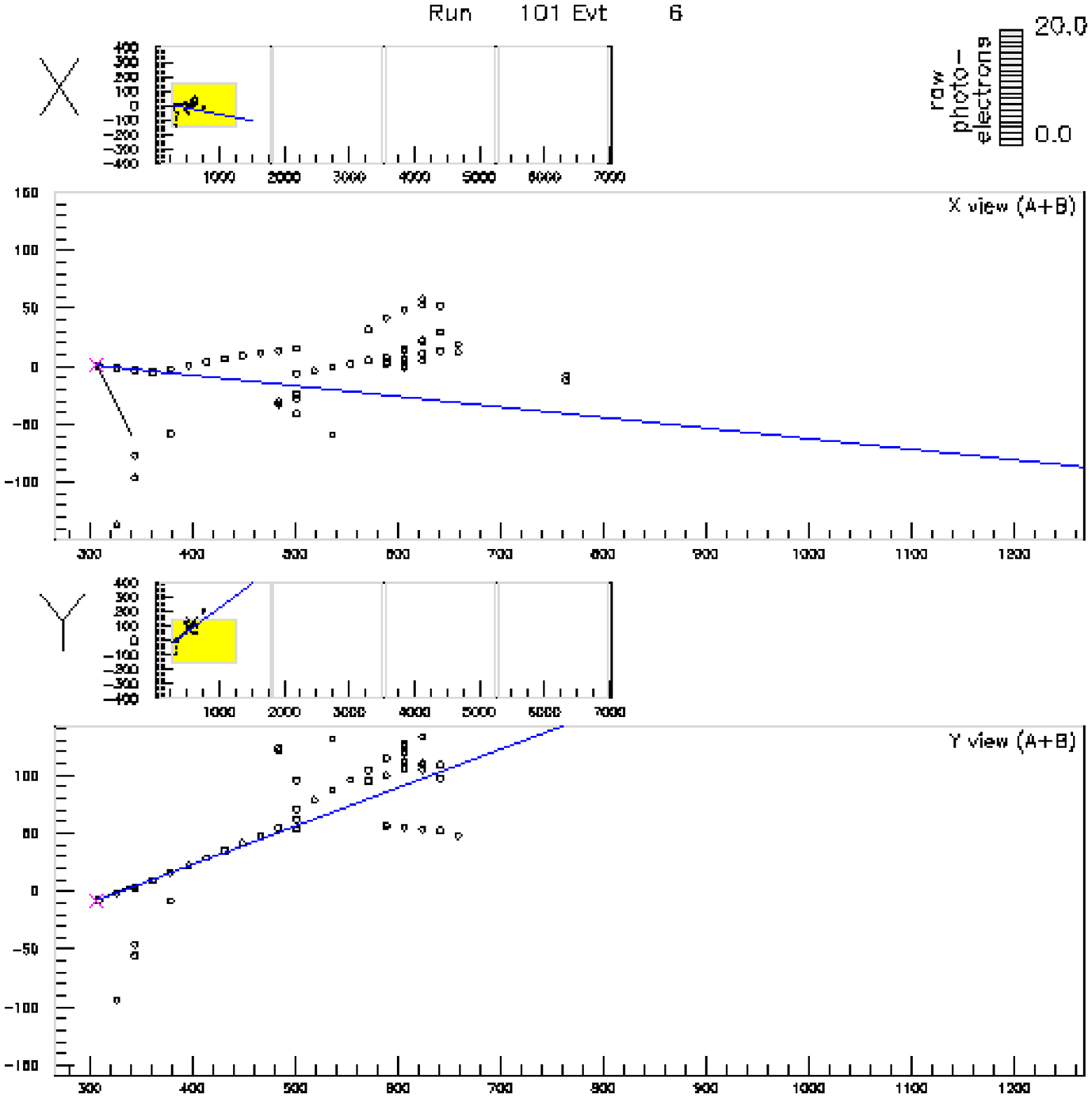,width=2.5in}~
\epsfig{figure=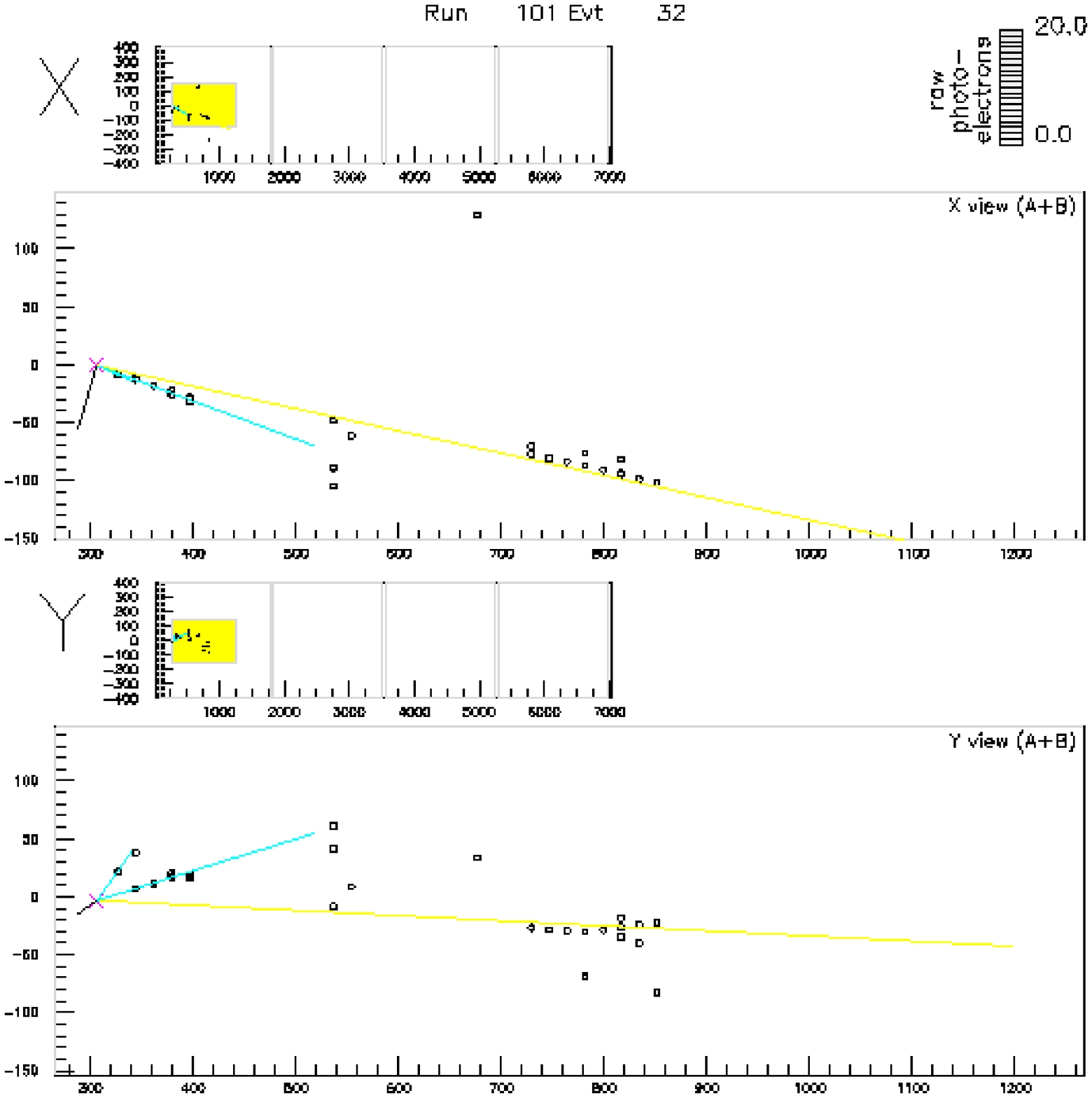,width=2.5in}

\epsfig{figure=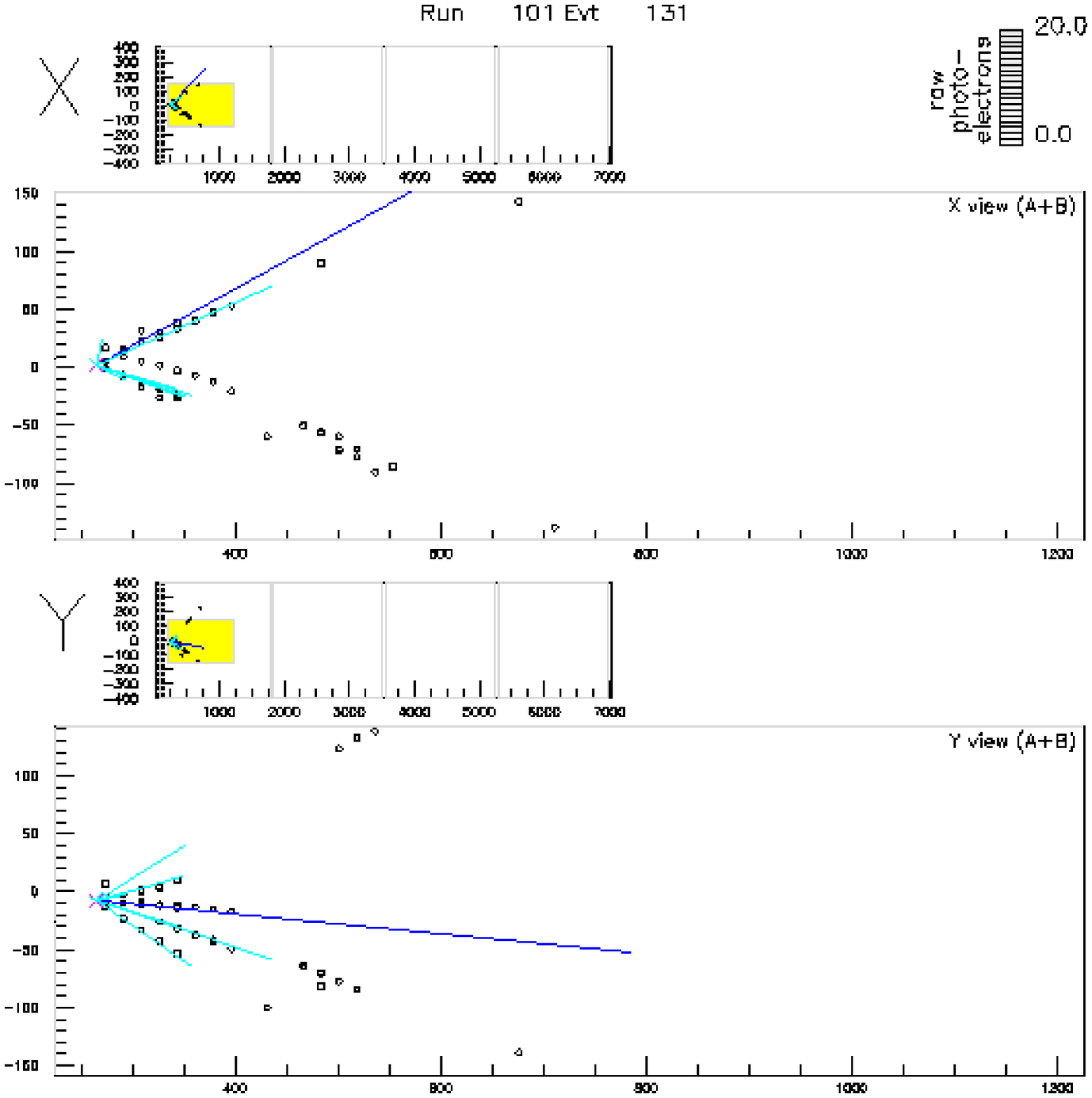,width=2.5in}~
\epsfig{figure=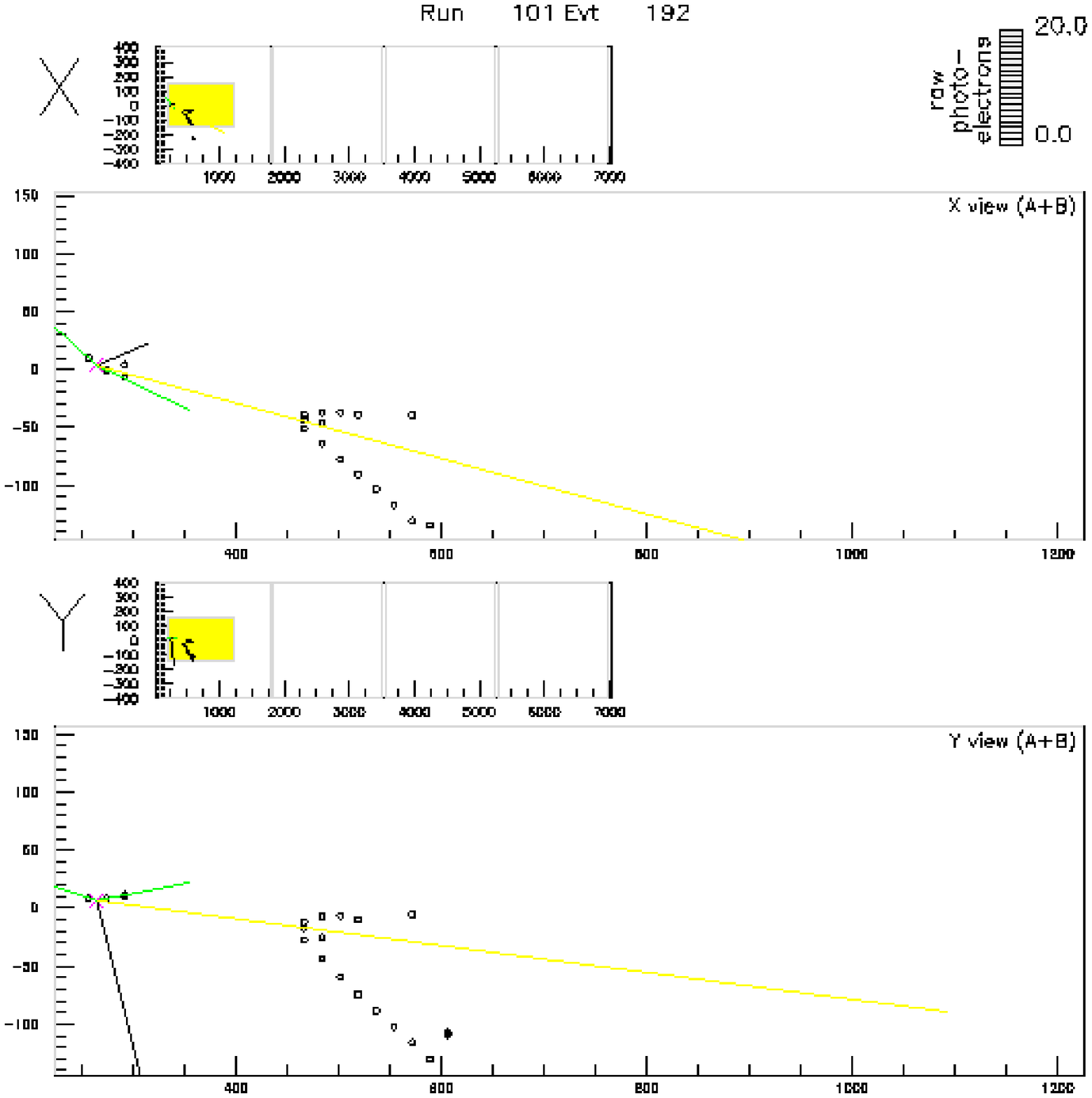,width=2.5in}
\caption{Typical neutral current  interactions with energies $1-3~GeV$
as detected in a fine grained calorimeter with the longitudinal sampling of
$1/3~X_0$ and the transverse sampling of $3~cm$.}
\label{bckg_events}

\end{center}
\end{figure}

\begin{figure}
\begin{center}
\epsfig{figure=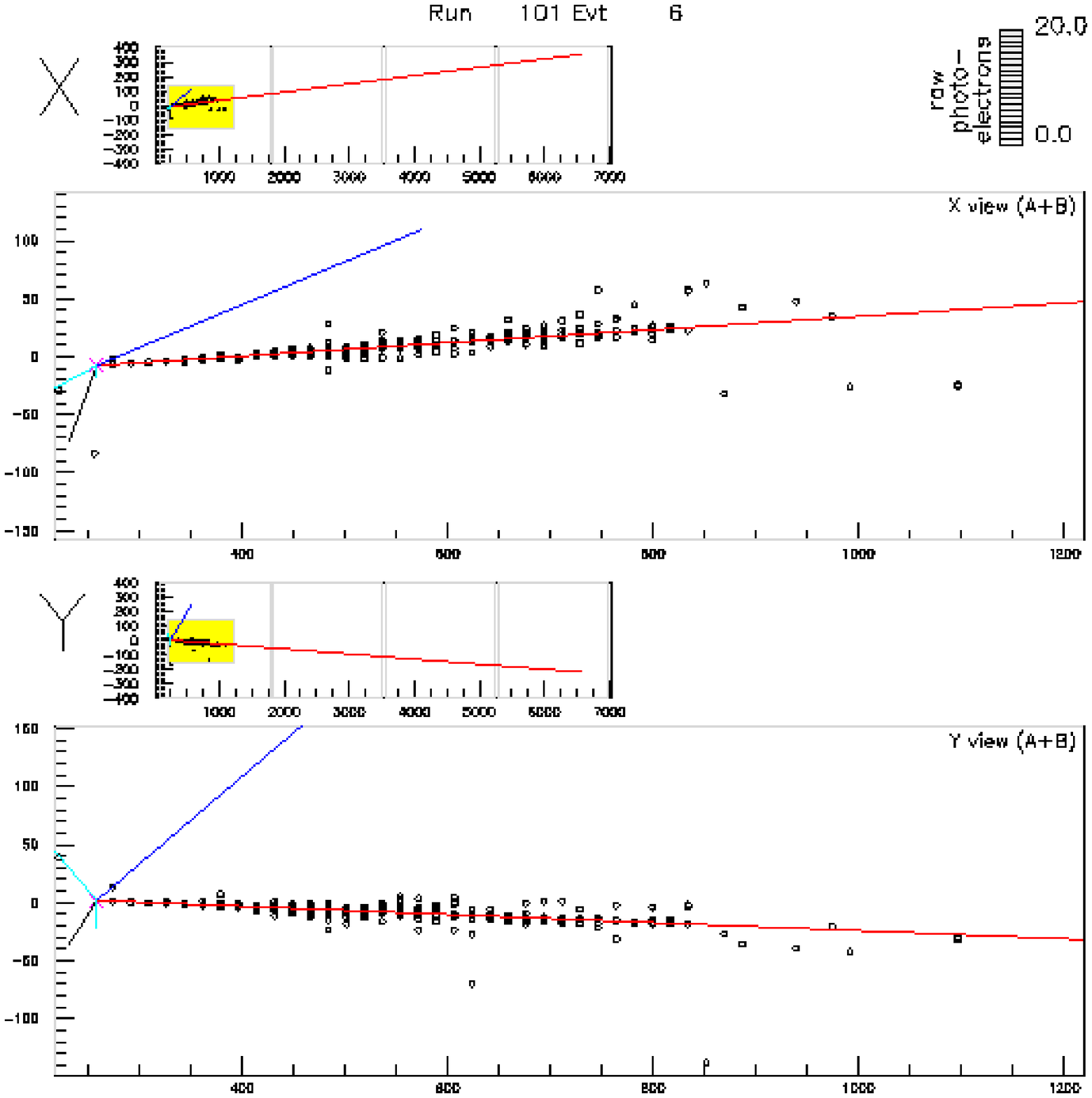,width=2.5in}~
\epsfig{figure=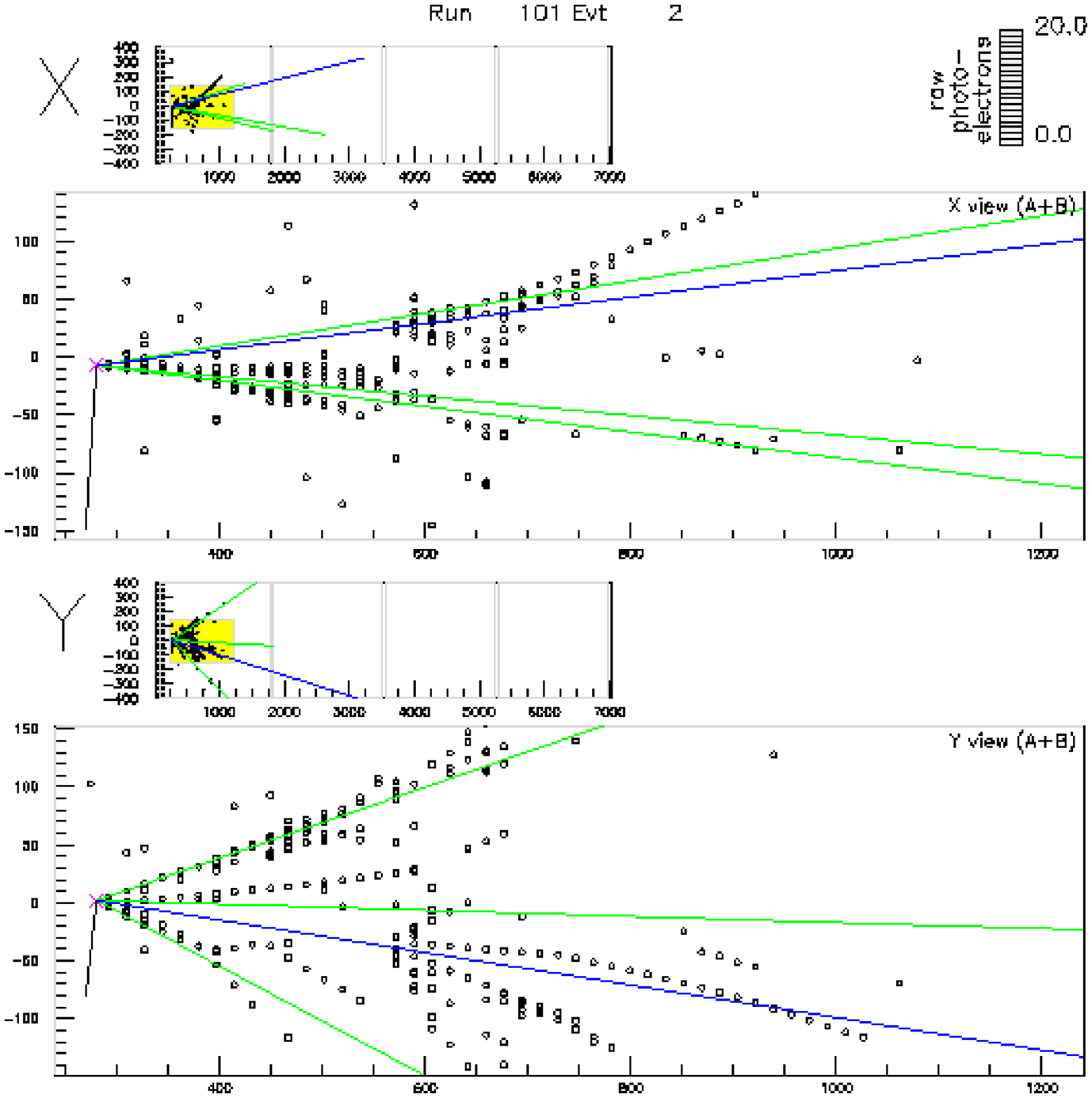,width=2.5in}

\caption{Typical charged current and neutral current $\nue$ interactions with 
energies $\sim 5~GeV$
as detected in a fine grained calorimeter with the longitudinal sampling of
$1/3~X_0$ and the transverse sampling of $3~cm$.}

\label{he_events}
\end{center}
\end{figure}

Typical NC events are not very spectacular - they have just few hits from low 
energy final state hadrons. More important are the $high~y$ events as they 
constitute potential background for the $\nue$ sample. Some example of such
events are shown in Fig.~\ref{bckg_events}. Most of them do not represent a 
serious source of background, it is only the small fraction with relatively 
energetic $\pi^0$ which are problematic. To illustrate the qualitative
 difference between neutrino interactions at very low energies, in $2~GeV$ 
region and at higher energies we show examples of $\nue$ CC event and a typical
NC event for neutrino energies of the order of $5~GeV$ in Fig.~\ref{he_events}.

\subsection{Electron identification}

Charged current $\nue$ interactions are identified by the presence of an
electron among the final state particles. There are several characteristic 
features of these events which can be used to differentiate them from the NC
and $\numu$ CC events:
\begin{itemize}
\item they do contain a long track
\item the track is accompanied by additional hits due to the onset of an 
electromagnetic shower
\item track energy is a significant fraction of the total energy of the event
\item track direction is at a small or moderate angle to the neutrino beam
direction
\item the energy deposition at the very beginning of the track is consistent
with a minimum ionizing particle
\end{itemize}

\begin{figure}
\begin{center}
\epsfig{figure=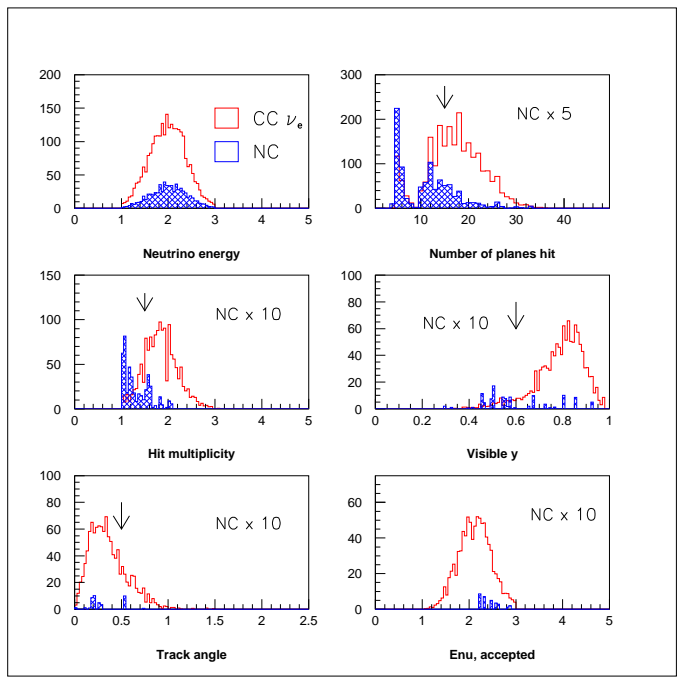,width=3.5in}

\caption{Typical evolution of the events sample: open histogram for
$\nue$ CC events, shaded histogram is for the NC sample. From top left to
bottom right: Neutrino energy,  number of planes in the longest track of the 
event, average number of hits (per plane) in a road around the longest track,
fraction of the event energy contained in the longest track, track angle with
respect to the beam direction, neutrino energy of the events passing the cuts. 
Cut values are indicated by arrows.}

\label{nue_vs_nc}
\end{center}
\end{figure}

Several independent studies involving full GEANT simulation, realistic 
reconstruction and analysis of events in the fine grained calorimeter 
indicate that such a detector enables identification of the $\nue$ CC 
interactions with a typical efficiency $\epsilon=0.35-0.4$ while keeping
the contribution   of NC contamination at the level of the intrinsic $\nue$
component of the beam. An example of the analysis chain is shown in 
Fig.~\ref{nue_vs_nc}.
 These results serve as a basis of the estimates of the 
physics potential of the off-axis experiment presented in this letter.

\subsection{Energy resolution}

Energy resolution of the detector is an important parameter. It provides 
a significant rejection against the backgrounds, as the oscillation signal 
events are concentrated in a narrow range determined by the beam spectrum, 
whereas the backgrounds are characterized by a broad distribution of the
observed energy.

Good sampling frequency ensures that the sampling fluctuations will be small, 
hence the energy resolution will be good. This is true even if the readout
is purely digital, i.e. the energy is determined by a number of the
hit detectors strips, irrespective of the deposited energy per strip.

\begin{figure}
\begin{center}
\epsfig{figure=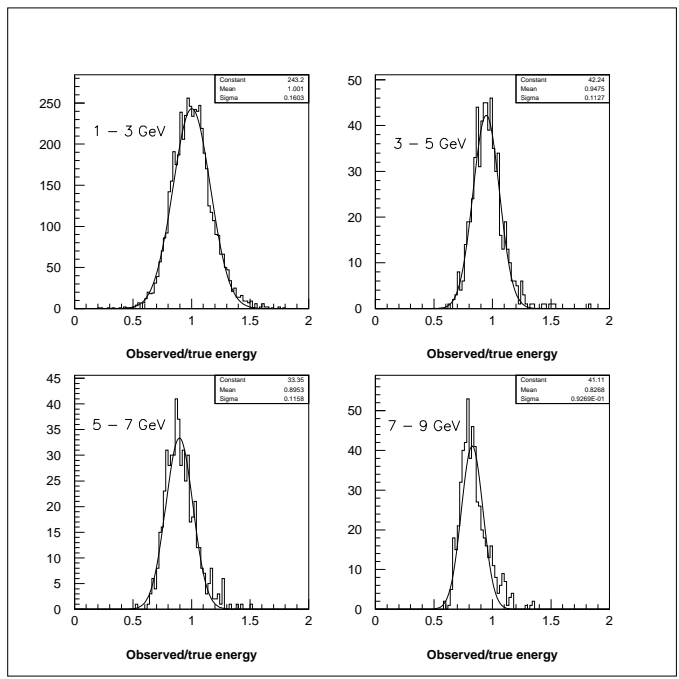,width=3.5in}

\caption{Total neutrino energy resolution for low y ($y<0.5$) $\nue$ CC 
interactions for different neutrino energies} 

\label{loden_energy}
\end{center}
\end{figure}
 
The resulting energy resolution, $\Delta E/E\sim16\%$ is well matched to the
off-axis energy spread. Purely digital readout leads to a non-linearity of the
response of the detector,  reaching some $18\%$ at $E_\nu = 8~GeV$. This
feature, although irrelevant for the neutrino oscillations experiment,
can be avoided by having an analog readout, for example in the liquid 
scintillator design.

\subsection{A specific detector example: LoDen - LOw DENsity calorimeter}

Neutrino experiments do not present significant challenges to the 
active  detectors. The low cost and long term stability of operations are 
among the most important characteristics.

The principal challenge of the $\nue$ detector will be therefore in the
engineering area: how to construct as large as possible and as cheap as
possible low Z calorimeter. 

The Loden detector consists of the alternating absorber planes, $1/3~X_0$
in thickness and Glass Resistive Plate Chambers. One detector plane represents 
a mass of about $70~tons$ hence 300 planes are necessary to attain the 
mass of $20~kton$. Total area of the active detectors is $120,000~m^2$.

The choice of the Glass RPC as a an active detector is primarily dictated by
the low unit cost of such detectors. These detectors have several other
attractive features: they are robust and reliable. They allow physical 
decoupling of the active elements (spark gaps) from the readout elements 
(pick-up strips) making it relatively easy to construct a very large
area detector with relatively few readout channels out of a large number
of smaller, hence manageable, individual detectors. 
Glass RPC are read out via pick-up strips providing $x$ and $y$ coordinate
in every plane.
Optimal transverse granularity of the active detectors needs to be studies 
in details, but the initial investigations indicate that a strip readout with
a strip width of $3~cm$ is adequate. Such a granularity leads to some 
$\sim1400$ readout channels per plane, or $400,000$ channels in total.

Practical design of the detector needs to be validated by the detailed 
engineering studies. Preliminary calculations indicated that the absorber
planes of dimensions approaching $20\times20~m^2$ can be constructed in 
a self-supporting manner, not requiring external support structures.
To prevent potential buckling the absorber planes will be bolted around
the periphery and along the center line into modules involving 20 planes.

\begin{figure}
\begin{center}
\epsfig{figure=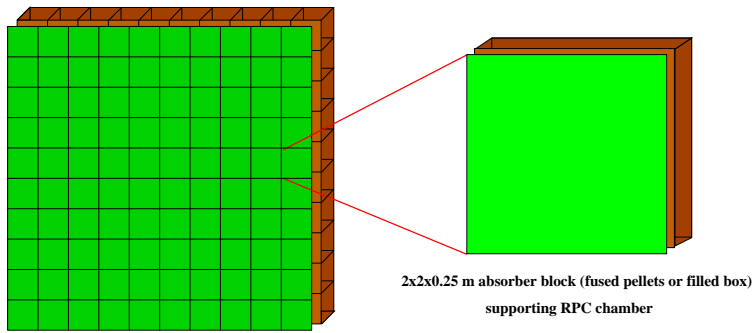,width=4.0in}

\caption{Modular detector design: $20\times20~m^2$ planes constructed by
stacking absorber modules, active detectors supported by the absorber.} 

\label{loden}
\end{center}
\end{figure}

Practical aspects of constructing very large but relatively thin detector 
walls 
must not be underestimated. It would be very desirable to develop a detector
design based on a 'human' scale modules, say $2\times2~m^2$ which could be
stacked to construct w final large detector plane, as illustrated 
in Fig.~\ref{loden}.

\subsection{Glass RPC Chambers}

  Glass RPC chambers belong to a family of the Resistive Plate Chambers,
but they utilize inexpensive commercial float glass of high resistivity 
($10^{12}~\Omega cm$) instead of the traditional bakelite.

\begin{figure}
\begin{center}
\epsfig{figure=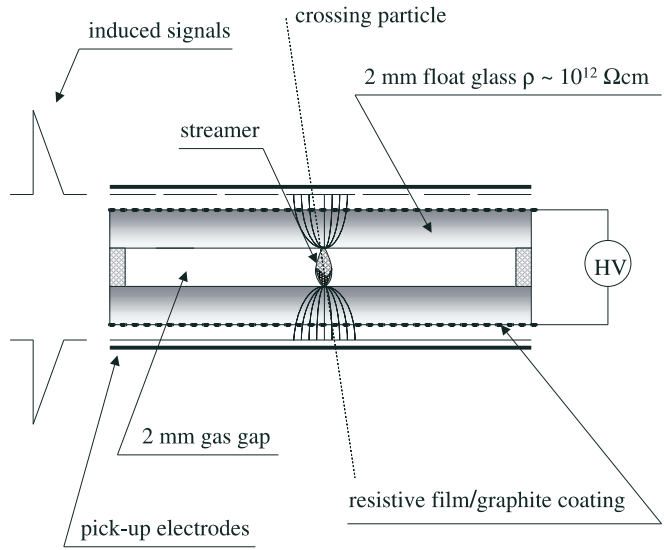,width=4.0in}

\caption{Glass RPC detector principle}

\label{RPC_fig}
\end{center}
\end{figure}

The detector, shown in Fig.~\ref{RPC_fig}, is composed of two parallel 
glass electrodes, $2~mm$ thick, kept
$2~mm$ apart by appropriate spacers. The gap between electrodes is filled 
with a suitable non-flammable gas mixture. The resistive coating on the outer
surfaces of the glass connected to the HV power supply creates a strong
electric field $\sim4\div5~kV/mm$ across the gap. An ionizing particle 
initiates a local discharge which induces a signal on external pickup strips.
The induced pulses are 
typically $100-300~ mV/50\Omega$  with $\sim ns$ time resolution.
Large signals of $100-200~pC$ allow for the possibility
 of a variety of cost saving options in the readout electronics.

This type of detector has been 
used successfully for muon and K-long detection in the BELLE experiment at 
KEK \cite{BELLE}. 
A single detector plane has an efficiency of typically $90-95\%$.  Double 
planes in the BELLE experiment were used to provide $99\%$ efficiency.  For 
this experiment we expect single planes to be adequate.

 The high resistivity of the glass and the quenching properties of the gas 
limit the discharge to a small area.   This type of detector does not have 
the rate capability necessary for hadron colliders but, as demonstrated by 
BELLE, is more than adequate for this neutrino experiment. They are ideal 
for large area applications since the pickup strips can be configured as 
transmission lines covering many meters in length.  

 These detectors are relatively simple and inexpensive to construct. The 
typical cost of the detectors is in the range of $\$100-150 /m^2$ of the detector \cite{gustavino,caren}.
They have been shown to be stable and reliable in more than four years of 
operation at KEK.

  These low cost detectors which can  be configured to cover a large area.
As an example, the Monolith \cite{Monolith} group has developed a design 
for the RPC planes as large as $15\times 30~m^2$. This group also developed several
possible industrial production techniques whicch may allow further cost 
savings.

\subsection{Readout Electronics}

When operated in streamer mode, RPC detectors produce a large pulse in 
response to the ionization within the gas.  The signals are large so that 
a significant voltage (100 mV or more) can be developed across a 50 or 100 
ohm resistor.  Because the measurement of events requires only the observance 
of hits in the detector, it is sufficient to use a simple discriminator as 
the front-end electronics, without the need for additional amplification or 
signal processing.

\begin{figure}[t]

\begin{center}
\scalebox{0.8}{\includegraphics*{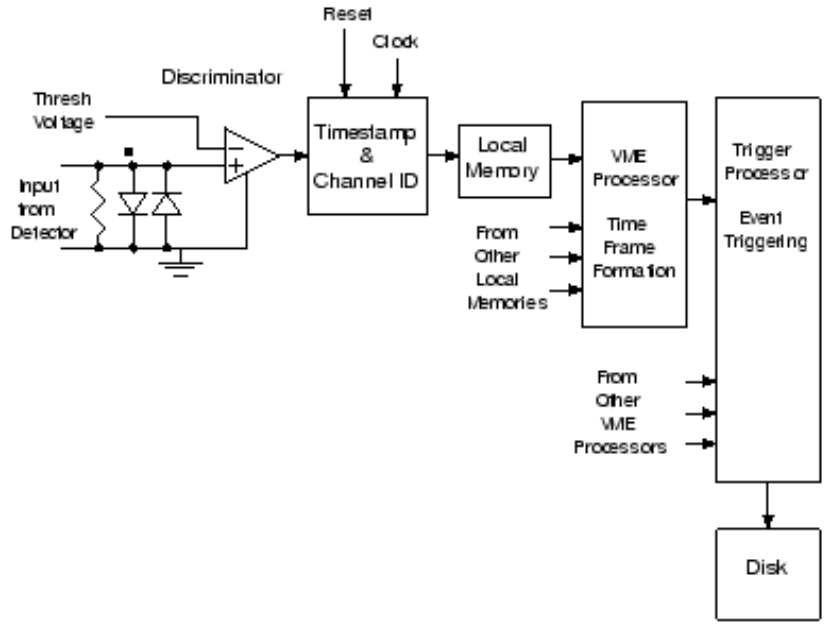}}
\end{center}
\caption{Block diagram of the readout electronics }
\label{block}
\end{figure}

To facilitate event reconstruction, the output of each discriminator would 
latch a time-stamp for the event.  The time-stamp 
is formed using local counters, which receive a common clock and counter reset
 signal from a global timing system. In this way, all time-stamp counters 
across the detector are synchronized.  When an RPC channel is hit, the data 
that is recorded is the value of the time-stamp counter.  The time-stamp hits 
are stored in a local memory, and read out later.  A block diagram is shown 
in Fig.~\ref{block}.  The resolution of the time-stamp is determined by the 
clock speed,
 which might be 100 ns (10 MHz.)  The number of bits in the counter is 
determined by the frequency of the counter reset, which might be 0.5 Hz, 
corresponding to frequency of the NuMI beam spill.

Because the data rate is low, it is envisaged that no trigger hardware is
needed.  Instead, the formation of a trigger and the analysis of events are 
done using a series of processors.  This is similar to the data acquisition 
system of MINOS\cite{MINOS}.  The initial sorting of hits by 
time-stamps is done using a VME-based processor in the front-end crate.  The 
processor would form "time frames" using the time-sorted data.  The time 
frames are then sent to a trigger processor, which receives time frames from 
the entire detector.  The trigger processor runs algorithms that look for 
tracks and discard noise hits.  Those events that pass are either written to 
disk, or passed to another processor for further analysis.  
\begin{figure}[t]
\scalebox{1.0}{\includegraphics*{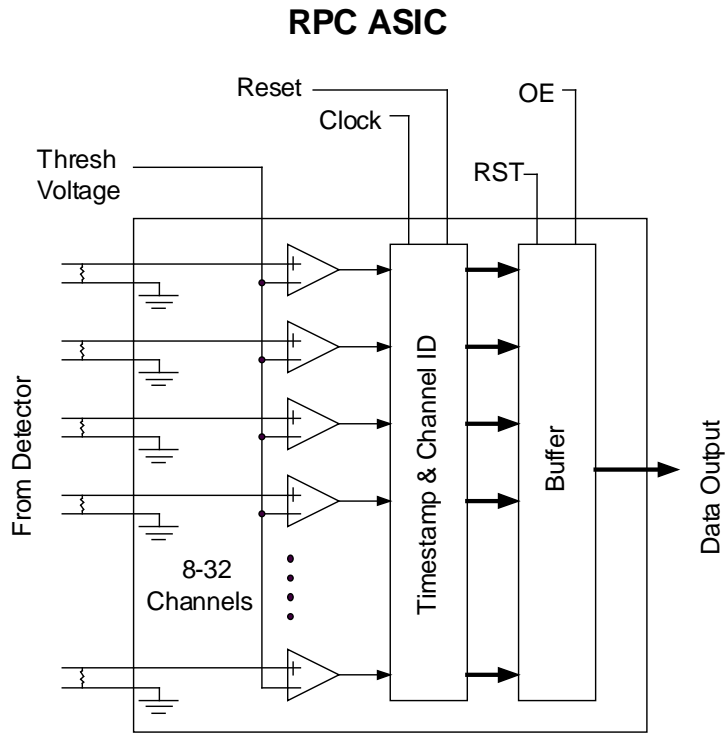}}
\caption[]{ASIC for the front-end readout electronics }
\label{asic}
\end{figure}

A goal in the electronics development is to reduce the cost.  Given the high 
channel count and relatively simple front-end configuration, it is practical 
to implement the functionality in a custom ASIC.  The comparator and time-stamp
 functions are easily realized in silicon.  The cost per channel for the ASIC 
would be $~\$1.50$ 
 for a 16-channel device.  A block diagram of the chip is shown
 in Fig.~\ref{asic}.  

\begin{figure}
\begin{center}
\scalebox{0.7}{\includegraphics*{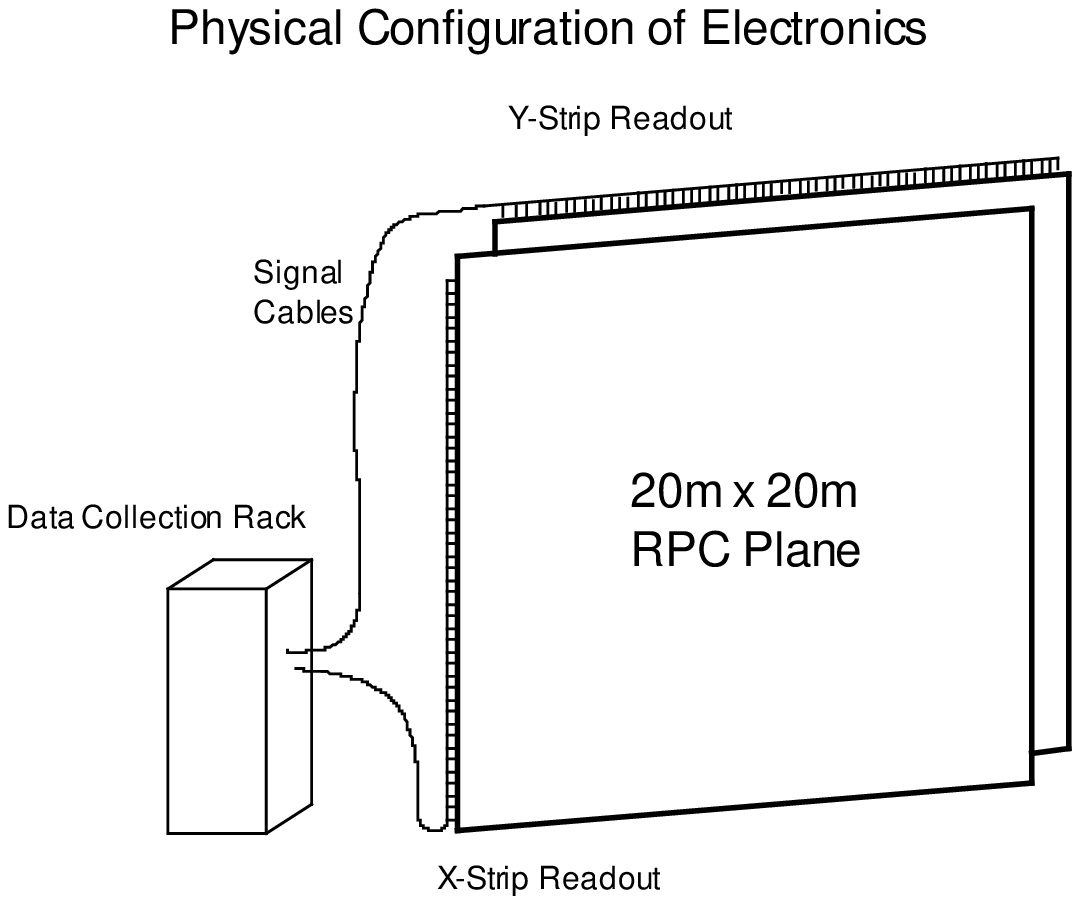}}
\end{center}
\caption[]{Physical layout of the readout electronics  }
\label{det}
\end{figure}

\begin{figure}
\scalebox{0.7}{\includegraphics*{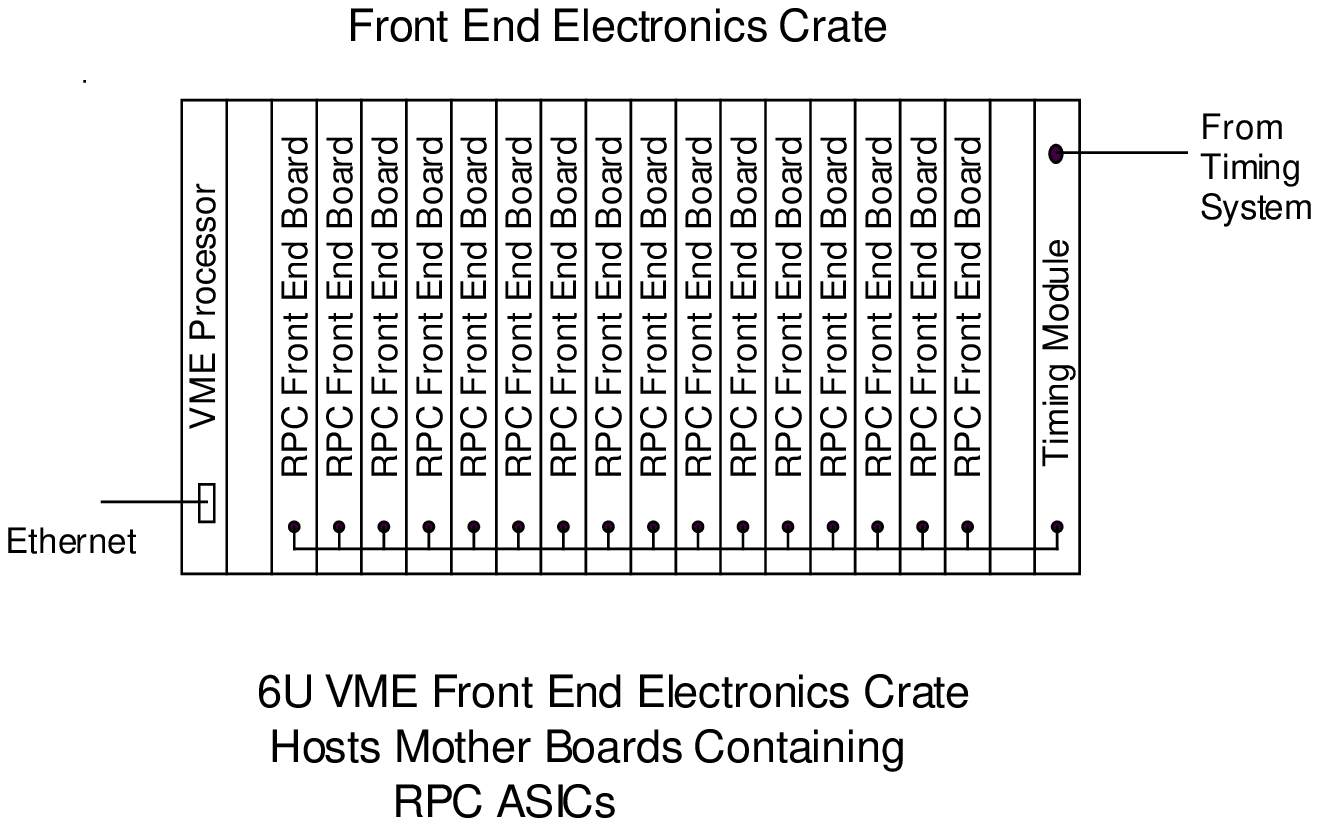}}
\caption[]{Front end electronics crate }
\label{crate}
\end{figure}

The front end electronics would reside in racks situated close to the 
detector, as shown in Fig.~\ref{det}.  The ASICs would be mounted on VME boards, 
providing easy access of the data to the front-end processor.  Each front-end 
board would host 16 chips, servicing 256 detector channels.  A VME crate would
 hold 16 such cards, servicing 4096 channels, as shown in Fig.~\ref{crate}.  

A total of
 100 front crates would be needed for 400K channels in the detector.  

\begin{figure}
\scalebox{1.0}{\includegraphics*{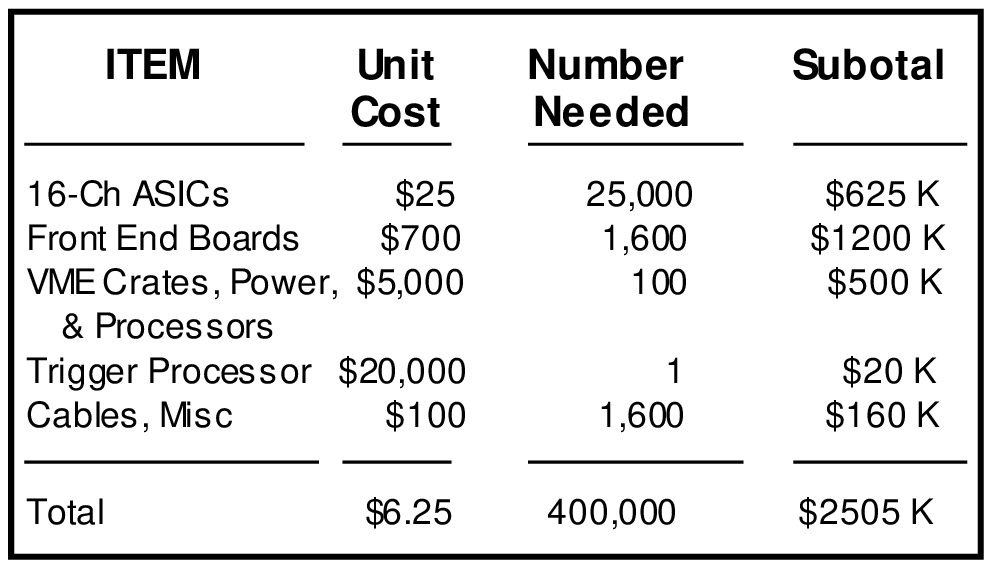}}
\caption[]{Cost estimate of the front-end electronics}
\label{cost}
\end{figure}

\clearpage
A preliminary cost estimate is shown in Fig.~\ref{cost}.   The total cost of the
electronics is estimated to be $\$2.5 M$ , or $~\$6/channel$.
\clearpage

\subsection{Ancillary Systems}

Assuming that the RPC mechanics and electronic
readout are engineered to 
minimize the cost of their fabrication, it will be
important to take care in the design of the ancillary
systems, namely the gas mixing and distribution system
and the HV system.   

\subsubsection{Gas System}

The design of the gas system must address a number of
challenges: 
\begin{itemize}
\item the number of distribution channels will
be large; 
\item the gas entering the chambers must be
very dry; 
\item safety; 
\item cost of operation.   
\end{itemize}

Although
it may be possible to feed the gas in series through multiple
gas volumes, there will be limits on this since the accumulated
effect of the flow resistance of the chambers will eventually
result in a pressure that exceeds the limit of the glue joints
(the geometry of the RPCs is by its nature very weak with
respect to internal overpressure).   Moreover, experience
with series connection of the Belle RPCs showed that the chambers
near the end of the line were slow to achieve stable operation.
Thus it will be important to minimize the per-channel cost
of the distribution system and to develop a design that provides
a balanced flow between channels without the need for extensive
adjustment.  To solve this problem, Belle used ``flow resistors,''
which are 10 cm stainless-steel tubes drilled down the center
with sub-mm bores.  These tubes are series inserted in the output
of each channel and set the flow impedance of the channel.  This
provides a simple and economical way to balance the flow from
channel to channel without need for tedious adjustment.

Keeping the gas dry (water vapor at the 100 ppm level or less)
is very important to avoid formation of hydrofloric acid,
which etches the glass, ultimately destroying the RPCs
(Belle came very close to having this happen).   Achieving
the requisite dryness is reasonably simple provided copper
tubing is used (plastic tubing is too permeable).

The importance of safety and economy are clear and will
no doubt be the subject of extensive design studies.

\subsubsection{HV}

Perhaps the most challenging aspect of the RPC high voltage
system is the voltage across the gap must be in the neighborhood
of 8~kV.   Although it is possible to achieve this with
a single ended supply, supplies capable of sourcing 
more than 5~kV tend to be expensive on a per-channel basis.
For that reason the Belle experimented employed a differential
scheme, wherein the total bias of 8~kV was achieved by
using a pair of supplies operating at opposite polarities.
A single 0-6kV supply was used on the positive side.  Since
these units were somewhat pricey, several RPCS were 
ganged onto a single supply.   For the negative bias,
independent (actually only quasi-independent) supplies were 
used on each channel.   This allowed Belle to monitor the
current on individual RPCs, which  proved to be very
important in the operation of the system, as discussed
below.

Experience with previous RPC HV distribution systems indicates
that it is essential to monitor the current drawn by the
RPCs on a per channel basis (the dark current of an RPC
is an excellent diagnostic of its overall health).   The
ability to switch a single chamber on or off is also useful,
although perhaps not essential (one could always resort
to simply unplugging the cables of misbehaving chambers).
Still less important is the ability to adjust the HV on
individual chambers.    

Once again safety and cost will be important practical
considerations.

\subsection{Modular Detector Design}

	One possibility for the detector assembly is to use pre-engineered 
intermodal containers as the basic modular unit and then replicate the unit 
many times to create a giant structure.  Dry Freight containers are used for 
international exchange of over $90\%$ of the world's manufactured goods and 
can be transported easily by truck, rail, and sea (therefore termed 
"intermodal").  These objects are in widespread use throughout the world 
and the existing mass production infrastructure offers a low cost solution 
for a low-density neutrino detector.

	The basic intermodal container is a steel box 8 ft wide by 8 ft 6 
inches high by 20 ft long with an empty weight of about 2,300 kg and a 
payload capacity of 22,000 - 28,000 kg depending on the manufacturer.  These 
containers have internal volumes of $31 - 36 m^3$ and therefore can hold 
cargo with an average density of about $0.7 - 0.8~ g/cc$.  The box has rails 
around the edges, a set of steel floor supports, a pair of doors at one end, 
and is typically covered with corrugated steel.  Such a container is called 
a "Twenty-foot Equivalent Unit" or TEU.   The International Standards 
Organization specification calls for the TEU box to be based around 4 
corner posts that can each take a load of 86,400 kg.  This ISO specification 
means that the boxes can be stacked 9 high when fully loaded and they are 
routinely stacked this high in large ocean-going container ships.  The cost 
of a TEU (in quantity) is in the range $\$1500 - \$2500$.  Millions of TEUs 
exist and hundreds of thousands of new ones are built each year.

	A modular neutrino detector, 18 m by 20 m by 80 m, could be assembled 
from about 800 such TEUs at a cost of approximately $\$ 1.6 M$.  
The resulting structure would be self-supporting and weather-tight.  The 
empty containers by themselves would constitute about 1600 metric tons of 
steel.  Off-the-shelf pre-engineered gantry cranes, lifting fixtures and 
many other attachments like box-to-box clamps and casters are available.

	The small size of the basic TEU would allow construction of detector 
modules at many institutions around the country.  The intermodal feature 
would allow easy shipment between institutions for different steps in the 
assembly process and the completed TEUs could easily be shipped to any site 
along the NuMI beamline.  The modular structure  allows for mobility of the 
detector if a different off-axis position becomes desirable part way through 
data collection, or if a different neutrino beam eventually becomes available 
from a superbeam or from a muon storage ring.

	The concept does have two drawbacks.  First, temperature control is 
not a built in feature of containers, but this can likely be solved using 
excess load capacity of the stacked TEUs as the supporting structure for an 
insulating shell.  Secondly, when the TEU boxes are stacked, the floor 
supports carrying the load in each individual TEU mean that the assembled 
detector will have an un-instrumented area with an active detector to active 
detector vertical crack size of about 20 cm.  Both of these problems require 
further study.     

\subsection{Absorber  Construction and Composition}

Typical low Z absorber
materials have a density of $0.7~g/cm^3$, and a radiation length of about 
$64~cm$ (similar to a mixture of $70\%$ polyethylene and $30\%$ air) leading 
to absorber  elements $16-20~ cm$ thick. 

Modular construction techniques include: standardized shipping containers 
filled with alternating layers of target and active detector, layers of 
target and active detector supported by an external framework, and 
free-standing structures of the target material with regularly spaced gaps 
for active detectors.
 
\subsubsection{Containers and Loose Fill Materials}

Standard shipping containers are inexpensive and strong enough to be stacked 
at least 9 containers high.  The effects of unsampled target and voids in 
a container array are unlikely to cause serious problems but must be 
investigated. Using aluminum (instead of steel) containers may reduce the 
unsampled energy loss, and the inactive volume can used for service and 
signal connections between containers.
\begin{figure}[t!]
\begin{center}
\includegraphics[width=7cm]{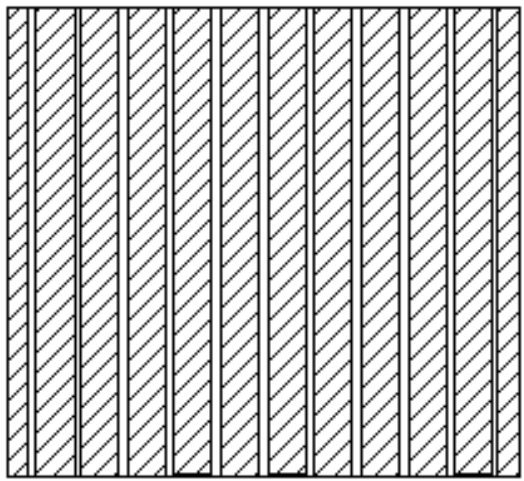}
\vspace{0.5cm}
\caption[]{Container $2.1\times2.3~m^2$ cross section filled with target slabs
creating 10 active gaps, $5.4~cm$ wide.}
\label{absorber}
\end{center}
\end{figure}
The interior of a standard shipping container has the 
dimensions $(h\times w\times l)~ 2.1\times 2.3\times 5.8~ m^3$. The interior 
230 cm width can be divided, as shown in  Fig.~\ref{absorber}, into 12 layers 
of target material $(10\times 16~ cm + 2\times 8~ cm)$ with $10\times 5.4~ cm$
 slots for active detectors. Assuming a target density of $0.7~ g/cm^3$, the 
container holds $15450~ kg$ of target mass, well below the load limit.

	Target layers can use a loose fill if contained in a cavity formed 
by (wood) walls. To prevent bulging, however, the cavity might require an 
internal bracing. Low Z, loose fill materials include recycled petroleum 
based or agricultural products. Some typical loose fill materials are listed 
in Table~\ref{fill}.

\begin{table}
\begin{center}
\begin{tabular}{|c|c|c|}
\hline
Loose Fill Material & Density $(g/cm^3)$ & Price $(\$/lbs.)$ \\ \hline
Plastics (recycled) & 0.6 - 1.0 & 0.10 - 0.30 \\ \hline
Shredded Tires      & 0.4 - 0.8 &   0.15   \\ \hline
Walnut Shells       & 0.6 - 0.7 &   0.1    \\ \hline
Cracked Corn        & 0.7       &   0.05   \\ \hline
\end{tabular}
\label{fill}
\caption{Typical Low Z Fill Materials}
\end{center}
\end{table}

Recycled fill materials include plastics that are clean but can obtained only 
in relatively small batches with large variations in properties, and 
shredded tires which are quite stable and can be had in various particle 
sizes. The least expensive loose fill materials are agricultural products 
that include nutshells and cracked (and dried) corn. However, a polypropylene 
bag might be required to package these materials. Better yet, loose fill 
material can be molded into precast blocks, but add significantly to the cost.
 
\subsubsection{Molded Particleboard}

A standard building material called particleboard, e.g., 
Georgia Pacific's (GP) MicroFine Novoply, is likely the cheapest 
(about $\$0.123/lbs$.) 
commercially available low-Z molded product. Particle board has very 
good mechanical properties and  can produced in  8' x 28' (8.3 m)  sheets 
up to 1.44" thick. The boards can be laminated into slabs of the appropriate 
thickness, and installed in the containers, as shown in Fig.~\ref{absorber}.
This would form smooth and straight gaps for the active elements. 

 Particle board for a 20 kT 
detector would  cost $5.5 M\$$.
\subsubsection{External Framework and Self Supporting Structures}

Particleboard, laminated into target slabs, $8$ or $16~ cm$ thick, can be 
bolted together and then supported by an external framework to generate 
a large array of target and active elements. However, particleboard might 
be strong enough to create a stable (container sized) module 
$(2.1 \times 2.3 \times 5.8 m^3)$ with, as described earlier, gaps for the 
active elements. Bonding or bolting a steel skin to four sides (top, bottom, 
and both ends) would stabilize the module against flexing. For active elements
 that must run the full height of the target, e.g., liquid scintillator filled
 tubes, internal bracing between the target slabs (instead of an external steel
 skin) can create continuous vertical slots.

\subsection{Liquid Scintillator Option}

Liquid scintillator makes an ideal active detector for a large inexpensive 
tracking calorimeter.  It is a proven technology that has been used in large 
quantities in detectors over long periods of time.  It gives good energy 
resolution and charged particle tracking efficiency.  Liquid scintillator 
allows for a flexible geometry and segmentation that can be optimized for 
the expected events.  Because it is a liquid, it can be added to the detector 
after it is assembled to minimize assembly cost.  It can even be removed if 
repair is necessary or if the detector needs to be disassembled and moved.   
Modern off-the-shelf photonics can be used to readout the detector so that 
additional electronics is minimal.  Scintillator segmented into cells gives 
both pulse height information and tracking.  
   Since this type of detector was 
investigated as a candidate for MINOS, very little additional R\&D and 
engineering is required to construct a 20 kTon detector from this technology. 

\subsubsection{Support Structure}

As a specific example, we present a conceptual design of a detector
 based on liquid 
scintillator interspersed between planes of water approximately 1/3 of a 
radiation length thick.   Both the liquid scintillator and the water would 
be contained in modules of extruded PVC plastic colored by titanium dioxide 
for good reflection.  Although this is a conceptual design, its structure 
has been investigated previously and those investigations documented 
\cite{border}.

 The design we consider here is based on modules that contain 6 layers of 
30 cells each with the dimensions of each cell 3.3 cm x 3.3 cm x 12 m long.  
The outer walls are 1 mm thick while the inner webbing of the extrusion is 
0.5 mm thick.  One of the layers would hold liquid scintillator and the 
others would hold water.  Although the cell sizes are chosen here for ease of 
calculation and would be optimized by Monte Carlo studies of neutrino events 
together with structural studies of the plastic, they are close to those 
required for a final detector and are structurally sound.  

For each module, 
the cells are sealed at one end by a single plug of PVC glued in place.  This 
construction has been tested to be reliable with no leaks to a pressure of 
over 6.5 atmospheres.  If the detector consists of scintillator planes 
separated by 1/3 radiation length of water as the target mass, a 
12m x 12m x 300 m detector will have a 20kT fiducial mass (assuming a 1 m 
fiducial cut around the edge.)  Alternate planes of scintillator/water 
modules would have perpendicular cells to give an X - Y readout.  This 
geometry would give 1500 planes of liquid scintillator.  

\subsubsection{Signal Collection}
Light would be collected using a 1 mm diameter wavelength shifting fiber 
similar to that used in MINOS.  Each cell would have a 12.5 m long fiber 
placed in it.  Our previous studies have shown that the amount of light 
collected by the fiber is not sensitive to the position of the fiber in the 
tube.  The end of the fiber will dipped in white paint and then in epoxy to 
provide $30\%$ reflectivity at the end.  Tests have shown a minimum ionizing 
particle this geometry with BC517L scintillator will give 40 photons from 
the end of the fiber.  The fibers from module would be gathered in a manifold 
to an optical connector similar in design to that of MINOS.  The design of the
 PVC manifold would also include fittings to fill and, if necessary empty the 
cells of scintillator.   Chemical activity tests  
show \cite{border} there will be no measurable effect of the scintillator on 
either the 
fiber or the PVC extrusion over the lifetime of the experiment.

\subsubsection{Support}

The construction of the detector is simplified because the light plastic 
extrusions would be stacked to form the detector before the liquid was added.
  Mounting the plastic extrusions with alternating planes at $45^o$ to the 
horizontal and $45^o$ to the vertical would allow the detector to be read out 
and filled from the top.  A possible mounting structure would be a V shaped 
trench.  The sloping walls could be supported by earth in a manner similar 
to culvert wall construction.  The longitudinal direction the trench would 
be sloped at about $1^o$ for ease of construction and to allow drainage for 
water seepage from the ground.   

The stacked extrusions would be supported 
primarily by the sides of the trench and at the ends of the trench by a 
bookend structure.  Most of the weight of the detector is supported by 
compression on the floor of the trench carried by the entire length of the 
extrusion.  In this configuration, all mechanical stresses are well below the 
tested strength of the extruded structure.  We envision that the detector 
would be covered with a roof, possibly of Quonset hut design, which would then
 be covered by about 3 meters of earth.  This would ensure a stable operating 
temperature for the detector and eliminate the soft component of the cosmic 
ray flux.  

\subsubsection{Readout}
Image intensifiers provide a low cost readout well matched to the rates of 
cosmic rays through the detector.  Reading out fibers into a standard 25 mm 
image intensifier will require one image intensifier for every plane of the 
detector.  Standard image intensifiers have a quantum efficiency of better 
than $10\%$ in the green and a gain of at least $10^5$.  An image 
intensifier would 
be read out by a video camera.  The CCD in a modern off-the-shelf video camera
 can be gated to have an exposure of as little as 10 microseconds and can be 
read out at 30 Hz.  

For this experiment we propose an exposure gate of 20 
microseconds.  Even on the surface, the occupancy rate of a single cell would 
then be about $1\%$.  The 360 cells of each plane of the detector could be 
read 
out by a single image intensifier and video camera.  Standard firewire readout
 would be done into processor for each camera, the processors would be 
sparsified and read into a PC.  If necessary, several PCs would alternate 
spills to assure adequate readout time.

\subsubsection{Cost Estimate}

 The costs of constructing the detector described above is based on quotes 
and engineering estimates of the liquid 
scintillation detector proposed for MINOS\cite{border}.
It is expected that the cost of the $20~kton$ detector, excluding support
structures and building, is of the order of $\$40M$.

\subsection{ Backgrounds in a surface detector}

\subsubsection{  Cosmic ray rates in the detector}

At the surface of the earth, the products of the extensive air showers
initiated by primary cosmic rays high in the atmosphere provide a
background flux to any detector. We present some preliminary estimates of these 
rates.
 
The principle component is muons with a
flux of 110 cos$^2\theta /m^2/sr/s$\cite{PDG}. The net flux on a detector of
area 3000 $m^2$ is then about $7 \times 10^5$ muons per second. The
average energy of these muons is 4 $GeV$ and about 30\% will stop inside a
detector of average thickness 1200 $g/cm^2$.

The live-time of the off-axis detector is given by the product of the
20$\mu s$ spill length and the $10^7$ spills per year  = 200 seconds.
There are about $1.4 \times 10^8$ muons passing through the detector in
this time. In any given 20$\mu s$ spill there are an average of 14 muons
distributed throughout the detector. 

Accompanying these muons are low energy electrons and photons with net
flux about 30\% of the muons and with energies in the range $\sim$tens of MeV.
In addition, there is a small component of low energy neutrons and
protons (the proton flux is $\sim$0.1 the neutron flux) with net vertical
flux approximately $3 \times E^{-1.5}/m^2/sr/s$ ($E$ in $GeV$) \cite{Hess},
i.e. the flux of neutrons with energy above 2 $GeV$ (the onset of
inelastic pion production) is about 1.0\% of the muon flux. A small
overburden is sufficient to effectively remove the soft electromagnetic
component but it requires about 3 m of rock or earth ($\sim$10 interaction
lengths) to attenuate the neutron flux so that there will be $\sim$50
inelastic neutron interactions per year. Pattern recognition plus a
moderate fiducial cut will be enough to make this small remnant
negligible.

\subsubsection{  Tertiary particle production by muons interacting in the environment
outside the detector}

There will be a small flux of electrons and photons produced by muon
bremsstrahlung in the overburden directly above the detector.  This is a
very characteristic signal that is trivially accounted for by a small
fiducial cut.  A neutral hadronic component of such interactions is
potentially more serious.

An estimate of the rate of hadronic interactions of the muons passing
through the detector can be made by finding the equivalent spectrum of
photons using the Weizsacker-Williams formalism coupled with
cross-sections for real photons.  This has been done by Perkins with the
result
 $dN/dE \approx (0.61/E) \times 10^{-6}/g/cm^2$ where $E$ 
is in $GeV$\cite{Perkins}.
Integrating this over the muon energy spectrum we find that the
probability of any single muon depositing more than $2 GeV$ hadronic
energy in one interaction length ($\sim 100 g/cm^2$ of water or other 
low-Z material) is $\sim 5 \times
10^{-5}$.  The fraction of these interactions giving a final state
neutron with energy above $2 GeV$ is $\sim 10^{-3}$\cite{Ruddick}. 
Estimating the
relevant muon flux as 
$10^8$ year and taking the angular  acceptance into account, the
corresponding flux of energetic neutrons into the detector is
$\sim$1/year.
\vspace{.25in}

\subsubsection{  Particle production by unseen muons inside the detector}

The proposed detectors contain planes of inert material.  A cosmic ray
muon can pass through one of these regions undetected but might produce
a hadronic or electromagnetic shower which is detected. The probability
that a muon will pass through one of these spaces is approximately {\it
w/L} where {\it w} is the width of the inert space and {\it L} the
transverse size of the detector. We will assume that this probability is
1\%. Any such interaction will have an energy flow exactly orthogonal to
the events of interest. Simple pattern recognition should make any
bremsstrahlung-induced background negligible.

Based on the previous discussion, the probability of a hadronic
interaction ($>~2~GeV$) in the full $12 m$ thickness of the detector is 
$\sim 6\times 10^{-4}$. Thus the net  background comes from a potential 
$1.4 \times 10^8 \times .01 \times 6 \times 10^{-4} \approx $ 1000 muon
hadronic interactions/yr. The highest energy pions in these interactions
are produced in the forward direction (vector dominance) which is again
exactly orthogonal to the events of interest. Large angle pion
production occurs at a rate comparable to the large angle nucleon
production and will be accompanied by significant nuclear break-up.
Pattern recognition must reduce this background to a negligible level.
\vspace{.25in}

\subsubsection{  Gamma rays}

There will be a gamma ray flux around the periphery of the detector
coming from decay of $^{40}$K, and the U and Th decay chains in the rock.
These are low energy gammas which will give very small signals in the
detectors; the net rate will be similar to that in the MINOS detector,
i.e. 1 - 2 kHz/detector plane.
\vspace{.25in}

\subsubsection{  Measurement of cosmic ray backgrounds}

Very accurate background measurements will  be made in a relatively small test
setup, as the effective lifetime of the neutrino detector is of the order of
$200~s$ per year of operation. These measurements will validate the 
correctness of the above estimates. 

During the duration of the experiment the background rates will be monitored by 
triggering the detector during the beam-off time.  

\section{Water Cherenkov option}

Large volume water Cherenkov detectors provide a proven technology that must
be considered as an off-axis neutrino beam far detector. The Super-Kamiokande
detector is already planned as the far detector for the JHF neutrino
beam\cite{JHF}. Whether or not water Cherenkov is appropriate for
the NuMI off-axis beam depends critically on the difference between the 0.8
GeV peak neutrino energy planned for JHF and the 2 GeV peak neutrino energy
anticipated for the NuMI project. At the low energy of KEK/JHF beams
 the 22.5 kton
Super-Kamiokande has proven capability for separating $\nue$ CC events from
 NC $\pi^0$ background
due to:
\begin{itemize}
\item quasi-elastic kinematic constraints, 
\item good $\pi^0$
reconstruction efficiency due to a larger opening angle between the gamma
showers, 
\item low intrinsic NC $\pi^0$ background due to the low
neutrino energy and small cross section for heavy resonance channels such as
the delta. 
\end{itemize}

These advantages can be recovered in the NuMI project by going
further off-axis to utilize the 2nd oscillatory maximum; this will be briefly
discussed at the end of this section.

The advantages of water Cherenkov for 2 GeV off-axis experiment are:
\begin{itemize}
\item it is a mature and proven technology, 
\item the cost of such a detector can be
accurately estimated, 
\item the cost per kiloton is low, no higher than
\$2M/kton \footnote{Estimated in round-numbers from the \$100M cost of the
Super-K.}, 
\item the energy resolution is good, for example at Super-K one
finds for electromagnetic showers that 
$\sigma / E  = 0.5\% + 2.5\% / \sqrt{E}$ , 
\end{itemize}

\begin{figure}
\begin{center}
\includegraphics[width=4.5in]{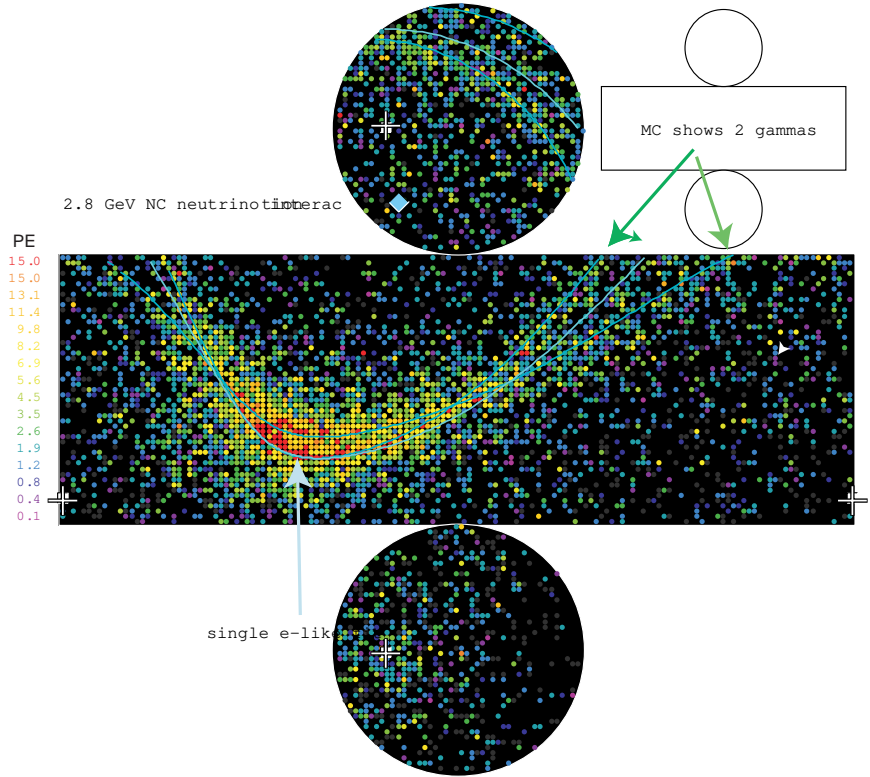}
\caption{An event display showing a typical neutral current
$\pi^0$ event for the NuMI off-axis spectrum. The Super-K reconstruction
software finds a single e-like ring.}
\label{sk-ncevent}
\end{center}
\end{figure}

The experimental challenge facing this detector option is accurately counting
a small sample of CC $\nu_e$ interactions over a background of NC
interactions with an associated $\pi^0$. In Super-K, the current $\pi^0$
analysis is based on ring-finding satisfactory for proton decay signatures
(pion momentum 0.5 GeV/c or lower). The efficiency for identifying two
separate showers, using the current algorithms, is approximately 10\% at 0.8
GeV\cite{mauger-nuint} and nearly zero at higher energies. This is mostly due
to events with a small opening angle between the gamma showers, typically 4
degrees for a 2 GeV $\pi^0$.  Figure~\ref{sk-ncevent} shows an event display
of a simulated event in the Super-K detector. The event is a 2.5 GeV neutral
current $\pi^0$ where one Cherenkov ring was found. The Monte Carlo
simulation shows two gamma rings, as does a reasonably trained human eye
scan.

The problem in separating the two showers becomes more difficult
up to about 0.8 GeV, but then should not get worse as one goes
to higher energies. This is because the vertex separation becomes
dominated by the photon attenuation length rather than the angle
between the two decay gammas. Firstly, the mean of the angle ($\theta$) 
between the two gammas is given by:\\

\begin{equation}
\overline{cos{\theta}} = 1-\frac{2}{\beta \gamma^{2}}ln[\gamma (1+\beta)]
\end{equation}

At higher energies the second term becomes small, so that the mean
separation angle goes as:\\

\begin{equation}
\overline{\theta} \sim \frac{2}{\gamma}\sqrt{2\gamma}
=\frac{2 m_{\pi}}{E_{\pi}}\sqrt{2\gamma}
\end{equation}

\begin{figure}
\centerline{\epsfig{file=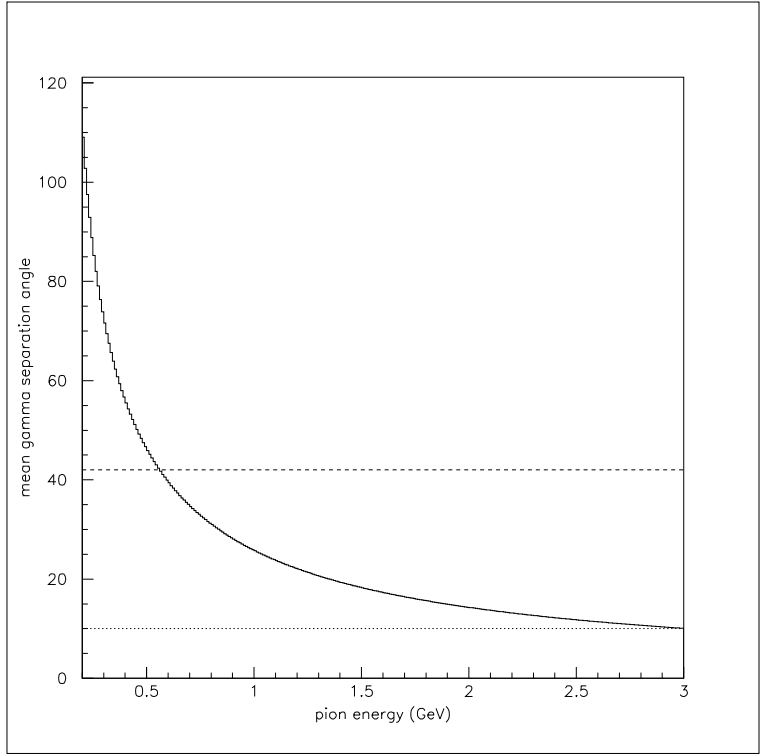,height=10.cm,width=10.cm}}
\caption{The cosine of the mean separation angle between $\gamma$'s
from $\pi^{0}$ decay as a function of $E_{\pi}$.}
\label{F:gamangle}
\end{figure}

When the mean angle becomes less than the intrinsic detector
resolution, angular separation becomes less important as a means
to recognize $\pi^{0}$'s from $e$'s
Figure~\ref{F:gamangle} shows the mean separation angle
as a function of pion energy. This becomes equal to the Cherenkov
light in water (dashed line) at about 0.6 GeV. At this energy, the
rings from the shower begin to overlap and so algorithms must
decide which PMT hit goes with which gamma. For a NuMI off-axis
experiment, good separation to better than 10 degrees (dotted line)
is required.\\

 Until now, experiments
have used standard TDC/ADC technology to record PMT data. It is clear
that multi-hit waveform digitizers might prove to be very useful in
better separating light from individual rings. This is because light
from the two rings is likely to be separated in time even though overlapping
in postion. Figure~ref{F:gamdist} shows the mean distance between the first
interaction point of the two gammas. At energies below 1 GeV this is
influenced by the opening angle between the gammas so the difference
is near a meter. At higher energies, however, the mean distance saturates due
to the slow change of the photon attenuation length in water and
so even near 3 GeV the average distance is more than half a meter. This
might be detectable with reasonably fast PMT's and standard 
waveform digitizers. Such studies are now in progress.\\

\begin{figure}
\centerline{\epsfig{file=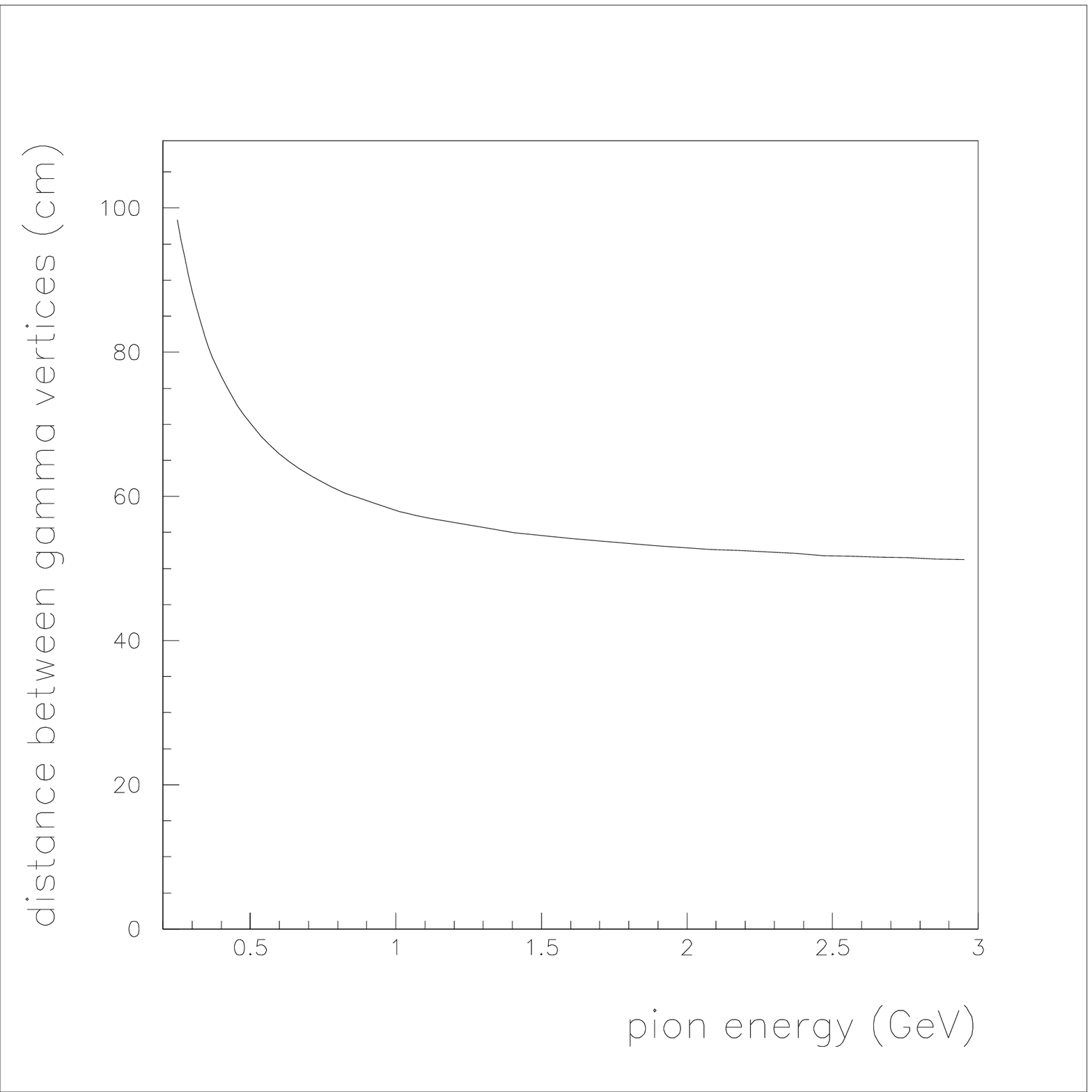,height=10.cm,width=10.cm}}
\caption{The mean distance between the initial interaction point of
the two $\gamma$'s as a function of $E_{\pi}$.}
\label{F:gamdist}
\end{figure}

It is hopeful that improved and specialized reconstruction algorithms may
have a significant impact on such high energy $\pi^0$ reconstruction. It is
already the case that a specialized likelihood fitter has been employed to
distinguish between single and double Cherenkov rings. This technique was
employed in the studies for section 5.2 of the JHF neutrino proposal
\cite{JHF}.

Using a prelimary version of a specialized $\pi^0$ fitter developed for
Super-K, a study was performed of the hypothetical case of transplanting
Super-K into the NuMI off-axis beam\cite{messier_workshop}. The
standard Super-K atmospheric Monte Carlo was used, reweighting each event to
the NuMI beam spectrum using the Monte Carlo truth information. Nine event
variables, such as decay electron tag, angle to the beam, the ratio of light
in and outside of the single ring fit, and the value of the $\pi^0$ fit
quality were combined into a likelihood to distinguish NC from $\nu_e$-CC
interactions. The analysis was roughly 20-30\% efficient for accepting
$\nu_e$-CC signal, with a signal-to-background ratio of 0.7-to-1.

There are further possibilities to be explored, including adding new
transient recording electronics that can reconstruct multiple close in-time
hits on a single PMT, to recover some events where one of the two gammas
converts at a significantly displaced vertex.

With further study, the sensitivity of the analysis for $\nu_e$ appearance
could be clarified using the comparably mature Monte Carlo and reconstruction
software used for Super-K and IMB. At this time, it seems like a water
Cherenkov is a difficult but but potentially viable detector for the off-axis
experiment.

There remains an interesting alternative path of inquiry. It is accepted that
a Super-K like detector performs very well at lower energies (such as at 0.8
GeV at K2K and eventually the JHF). It may be interesting to consider moving
such a detector even further off the NuMI axis, tuning the baseline and angle
to second oscillation maximum. This would be at a lower peak energy, more
suited to quasi-elastic kinematic reconstruction, and with considerably
easier $\pi^0$ identification topologies.

\def\({ \left( }
\def\){ \right) }
\def\b{\begin{equation}}
\def\e{\end{equation}}
\def\={\ =\ }
\def\ap{\ \approx\ }
\def\+{\ +\ }
\def\-{\ -\ }
\def\lc{\ell_c}
\def\Ls{\cal L \rm}
\def\ls{\cal l \rm}
\def\mumu{$\mu^+\mu^-$}
\def\ee{$e^+e^-$}
\def\rr{}
\def\bb{}


\section{Liquid Argon TPC}

The recent successful operation of the ICARUS 300-ton liquid-argon 
time-projection-chamber prototype \cite{Rubbia77,ICARUS,icarus_tm_01_09} 
indicates that it is timely to
review the possibilities for large-scale application of this technology for
accelerator-based neutrino physics, neutrino astrophysics, and proton 
decay \cite{lanndd,franco1,icarus_02_02}. 

\begin{figure}[htp]  
\begin{center}
\includegraphics*[width=5in]{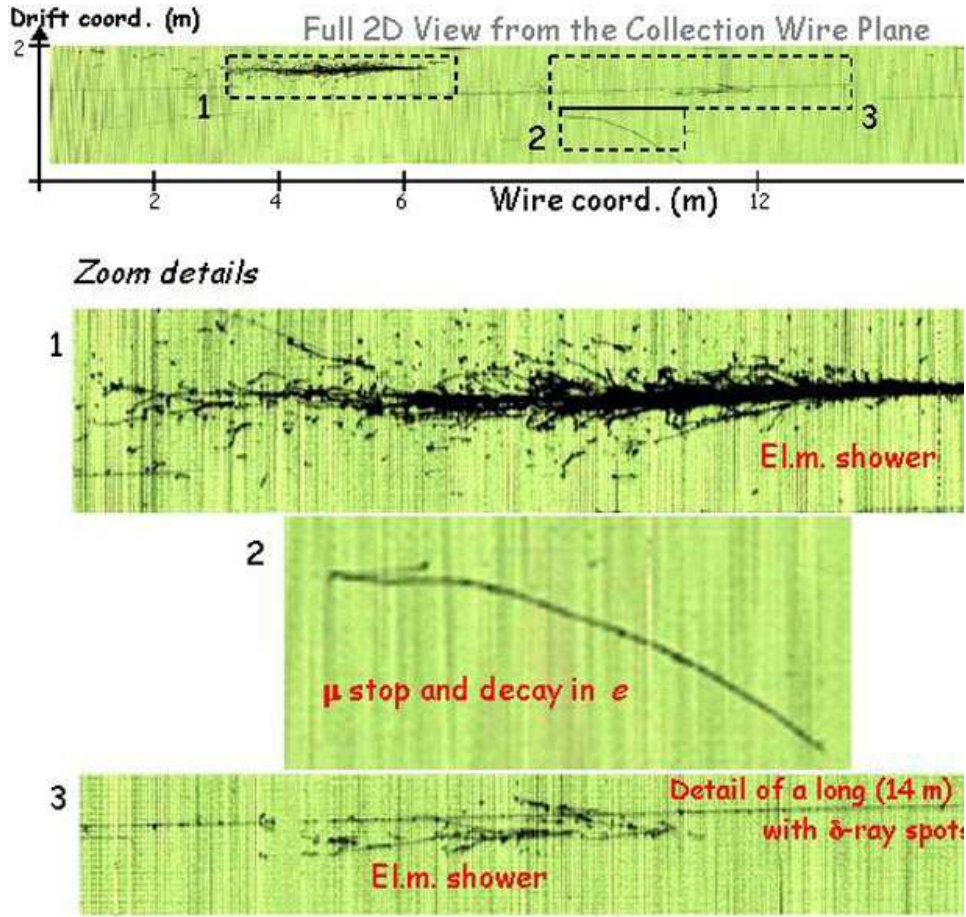}
\parbox{5.5in} 
{\caption[ Short caption for table of contents ]
{\label{icarus2} An event from the recent cosmic-ray test run of ICARUS  
\cite{ICARUS},
showing excellent track resolution over long drift distances in zero magnetic 
field.
}}
\end{center}
\end{figure}

A liquid argon detector for neutrino physics is a total absorption calorimeter with
time-projection readout via the signal of drifting electrons collected in crossed
planes of wires.  The effective pixel size is about $5 \times 5 \times 5$ mm$^3$,
compared to the radiation length of 14 cm and nuclear interaction length of 55 cm.
At a drift field strength of 500 V/cm, the drift velocity is about 1 mm/$\mu$s, so the
drift time over, say, 5 m would be 5 ms.  Even if operated at the Earth's surface with
no shielding, a liquid argon TPC has only about 1 (localized) cosmic-ray track per
m$^2$ of horizontal surface per drift time, so events appear very clean, as shown in
Fig.~\ref{icarus2}.

\begin{figure}[htp]  
\begin{center}
\includegraphics*[width=4in]{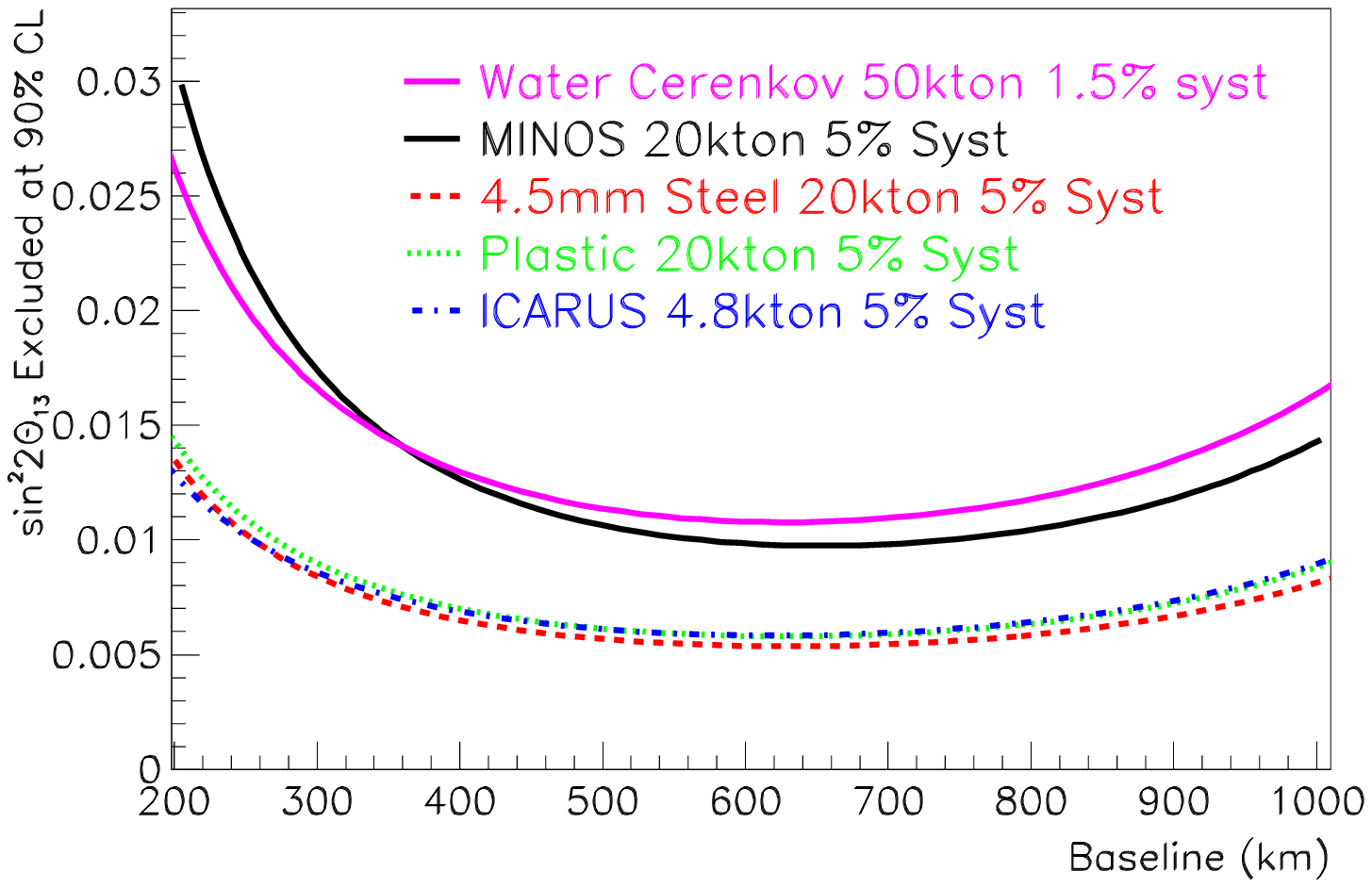}
\parbox{5.5in} 
{\caption[ Short caption for table of contents ]
{\label{harris} Comparison of several types of detectors in measuring
$\sin^2 2\theta_{13}$ in the presence of backgrounds typical of
a pion-decay neutrino beam at intermediate baselines \cite{harris_nnn02}.
The detector labeled ICARUS \cite{ICARUS}
is a liquid argon time projection chamber.  With a 25-kton liquid
argon detector, and an off-axis NuMI neutrino beam, 
the sensitivity to $\sin^2 2\theta_{13}$ would be at least 0.002.
}}
\end{center}
\end{figure}

Because a liquid argon TPC is a total absorption detector with fine-grained sampling,
it offers superior separation of charged-current and neutral-current neutrino interactions,
as we as excellent identification of $e/\mu/\pi/K/p$, included $\pi^0$'s.  Hence, it
is the most effective detector per unit mass for $\nu_\mu \to \nu_e$ appearance measurements
(of $\sin^2 2 \theta_{13}$, the sign of $\Delta m^2_{23}$, CP violation, ...), as shown
in Fig.~\ref{harris}.

\begin{figure}[htp]  
\begin{center}
\parbox{3in}{\includegraphics*[width=3in]{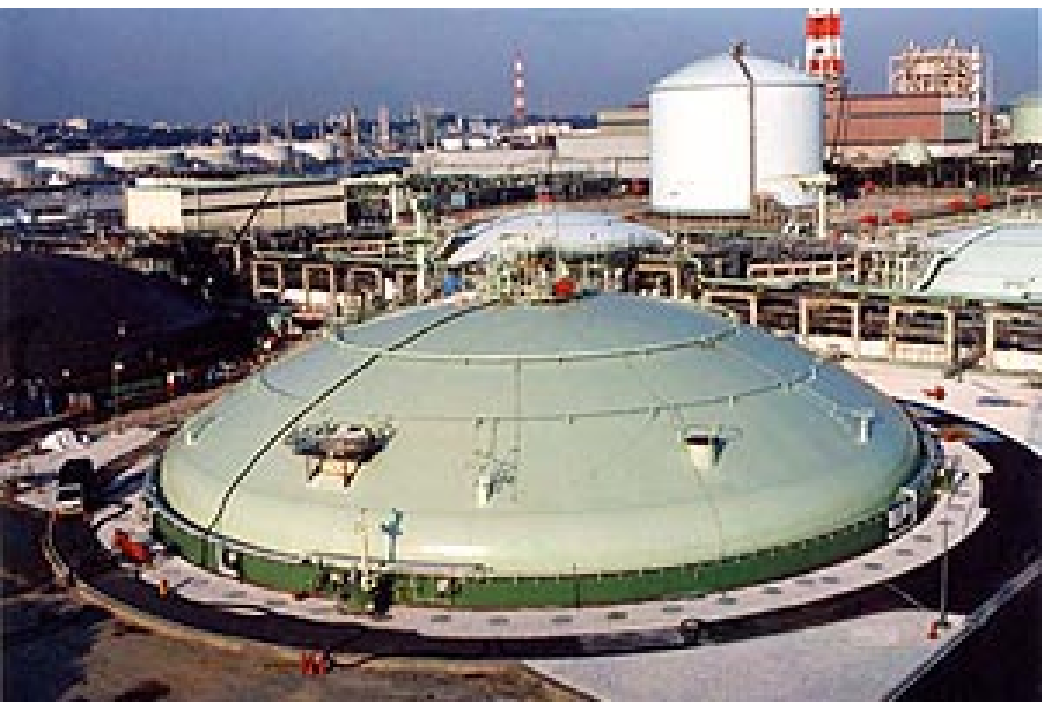}}
\parbox{3in}{\includegraphics*[width=3in]{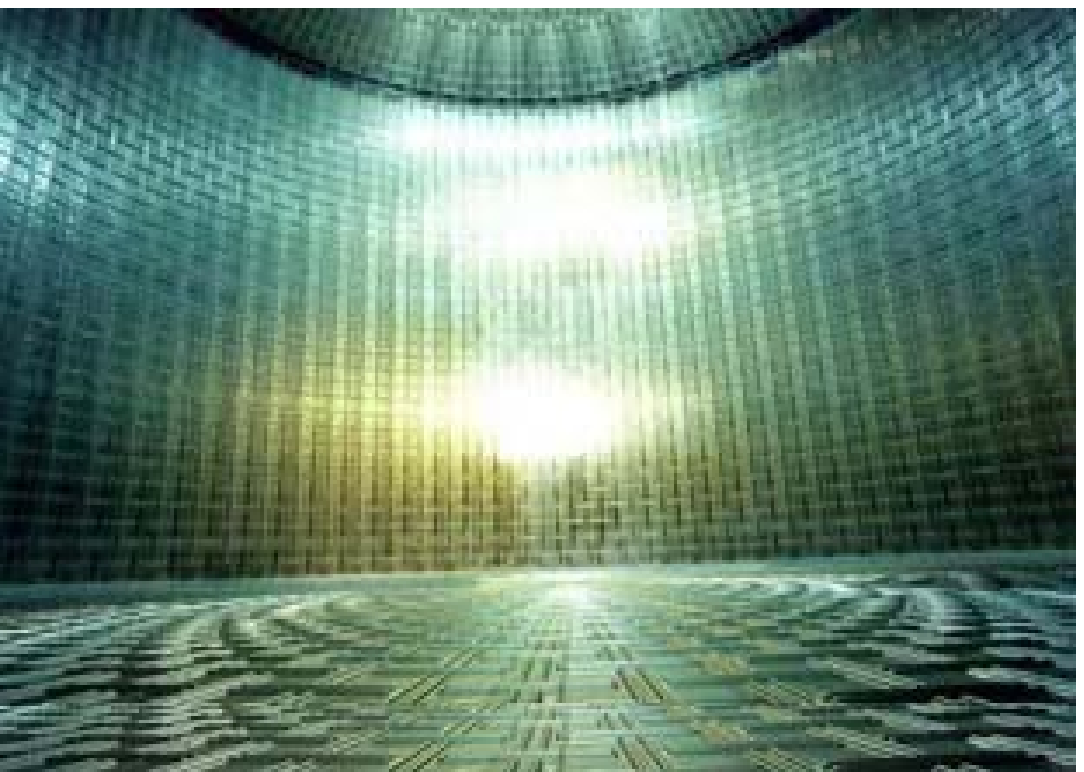}}
\parbox{5.5in} 
{\caption[ Short caption for table of contents ]
{\label{larfig1} Photographs of cryogenic storage tanks of volumes $\approx 100,000$
m$^3$.
}}
\end{center}
\end{figure}

To obtain economies of scale, a large liquid argon detector should be implemented in a
single cryostat, such as those commonly used in the liquefied natural gas industry.
Cryogenic volumes of up to 200,000 m$^3$ (= 280 kton if liquid argon) are now in
use, as sketched in Fig.~\ref{larfig1}.

\begin{figure}[htp]  
\begin{center}
\includegraphics*[width=3in]{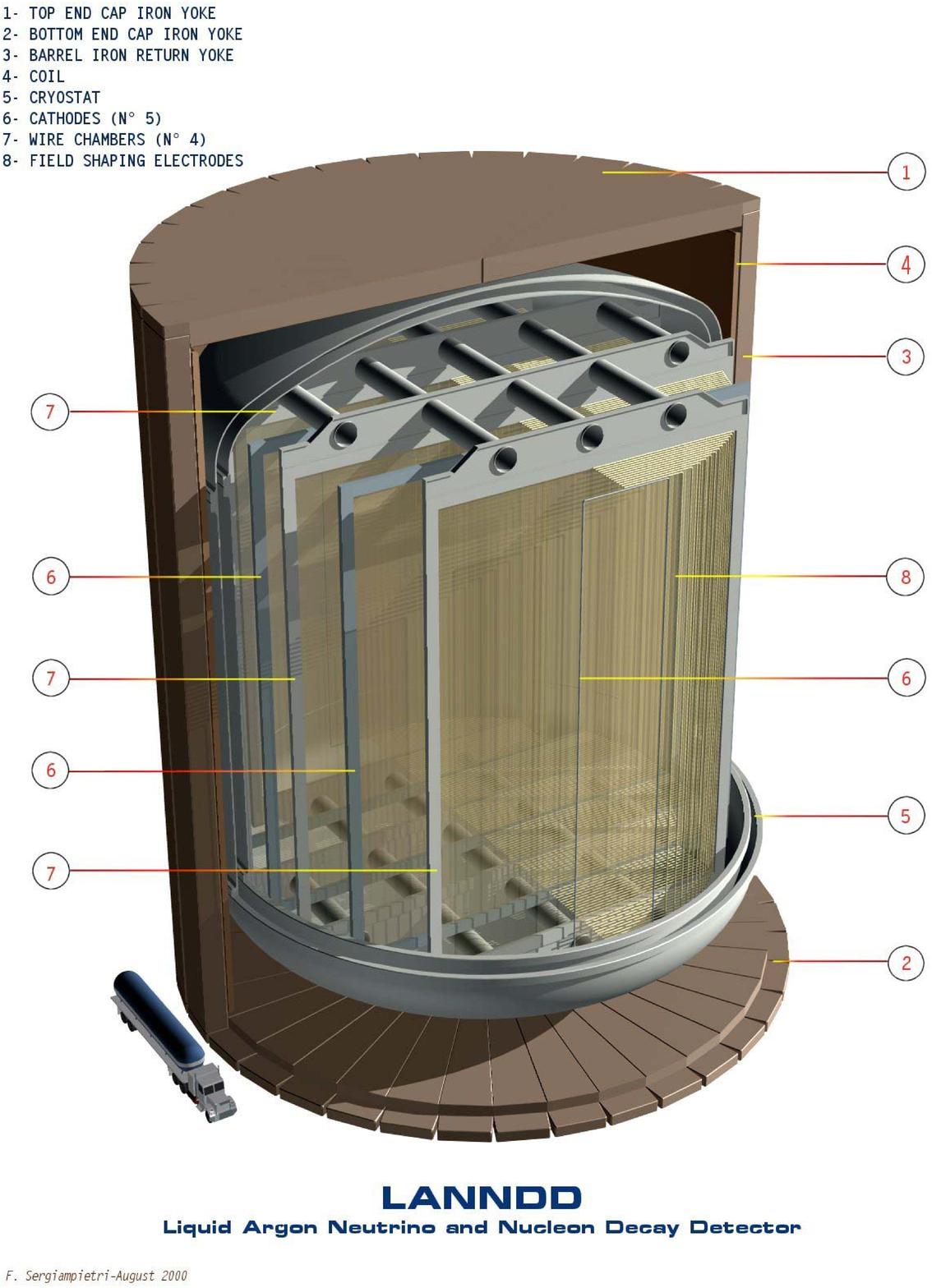}
\parbox{5.5in} 
{\caption[ Short caption for table of contents ]
{\label{lanndd1} Concept of a 70-kton Liquid Argon Neutrino and Nucleon Decay
Detector (LANNDD) \cite{lanndd,franco1}.
}}
\end{center}
\end{figure}

An overall concept of a large magnetized liquid argon detector is shown in 
Fig.~\ref{lanndd1}.  
Anticipating the possibility that neutrino beams are eventually
sent to it from more than one accelerator, the magnetic field is vertical so the
trajectories of secondary particles are generally orthogonal to the magnetic field.

\begin{figure}[htp]  
\begin{center}
\includegraphics*[width=2.5in]{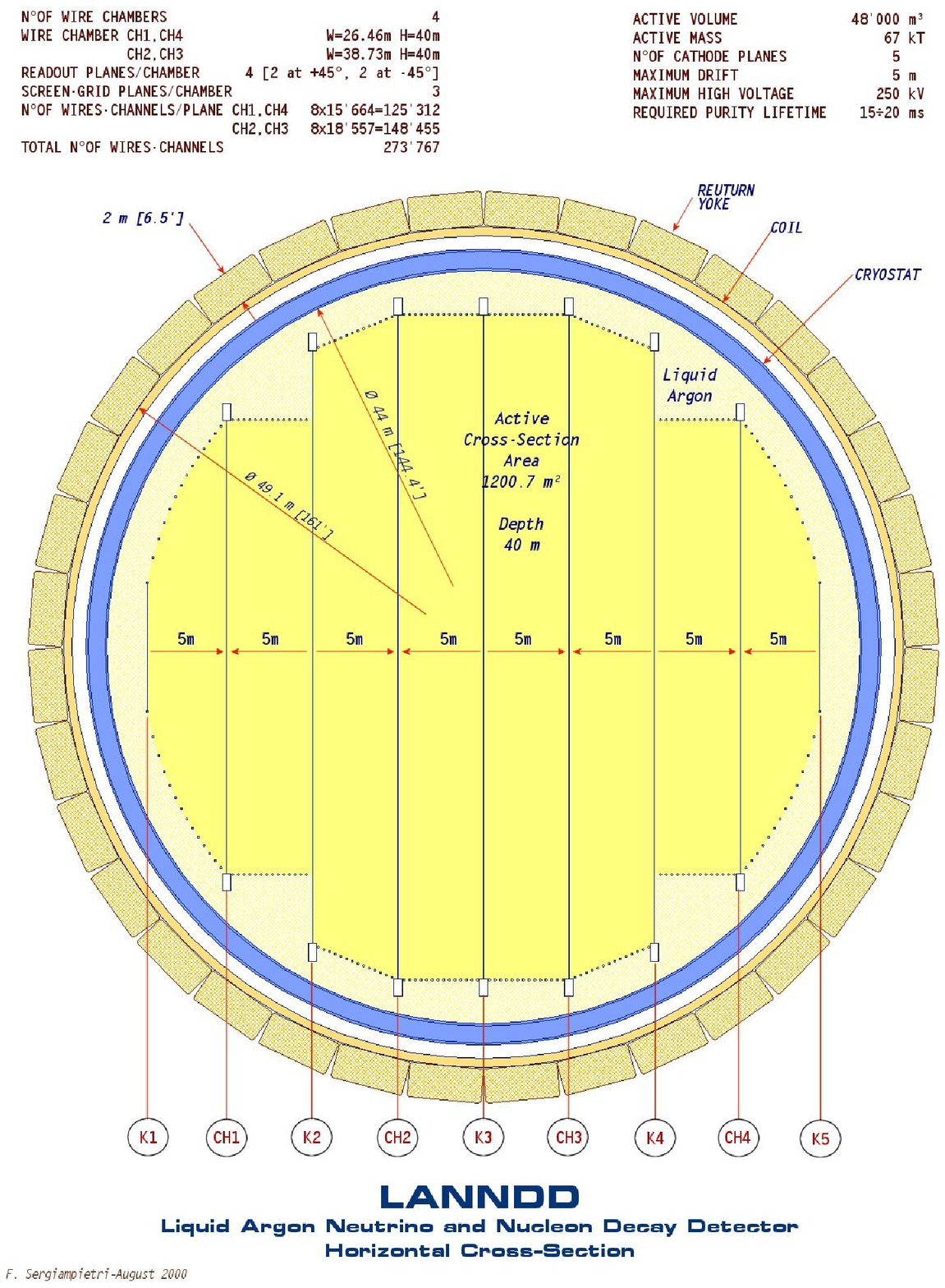}
\parbox{5.5in} 
{\caption[ Short caption for table of contents ]
{\label{lanndd_top} Top view of the electrode arrangement of a 70-kton 
Liquid Argon Neutrino and Nucleon Decay
Detector (LANNDD) \cite{franco1}.
}}
\end{center}
\end{figure}

The usable drift distance in a liquid argon detector is limited by oxygen impurities to
2-5~m.  The diameter of a large liquid argon detector will be greater than (twice) the 
maximum drift distance, so the readout must consist of a set of parallel anode and
cathode planes that subdivide the detector, as shown in Figs.~\ref{lanndd1} and
\ref{lanndd_top}.

A large liquid argon detector has excellent capability to search for nucleon decay
\cite{larpdk},
particularly because it is almost fully efficient for the decay $p \to K^+ \bar\nu_\mu$
that is favored in many SO(10) SUSY models \cite{Pati}.

The data acquisition for a liquid-argon TPC can operate in a pipelined, deadtimeless mode,
with zero-suppression \cite{icarus_daq}.  This could permit fully live
operation at the Earth's surface for nucleon decay studies, in addition to triggered
data collection of neutrino interactions from a pulsed accelerator beam.  The data rate
would, of course, be quite high in this case.  It may nonetheless be less costly to
implement a high rate data-acquisition system than to reduce the un-triggered data rate by
siting the detector deep underground.

\begin{figure}[htp]  
\begin{center}
\includegraphics*[width=4in]{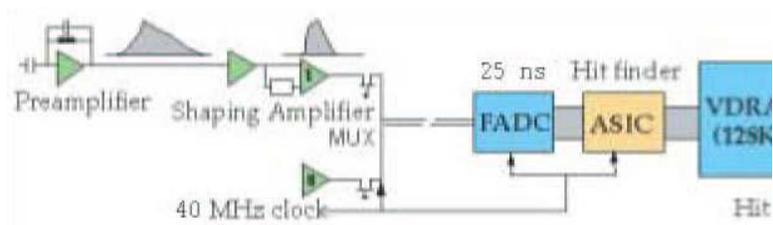}
\parbox{5.5in} 
{\caption[ Short caption for table of contents ]
{\label{frontend} Scheme of the front-end electronics of the ICARUS detector
\cite{icarus_daq}.
}}
\end{center}
\end{figure}

\begin{figure}[htp]  
\begin{center}
\includegraphics*[width=4in]{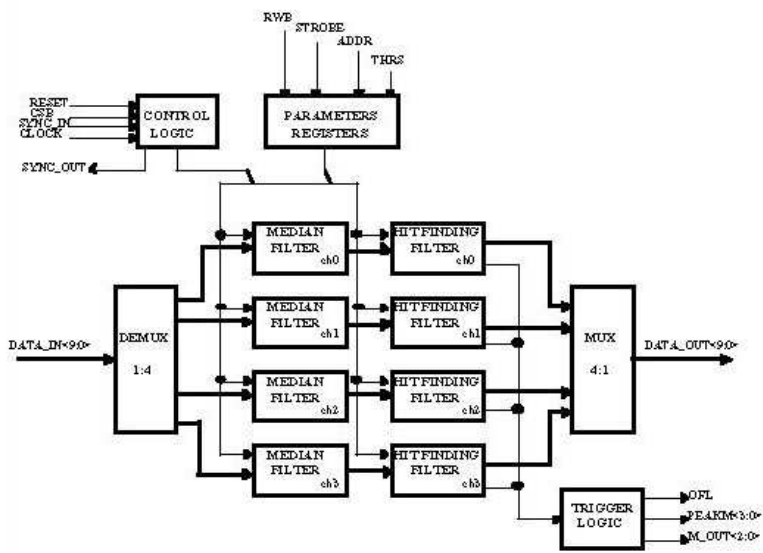}
\parbox{5.5in} 
{\caption[ Short caption for table of contents ]
{\label{backend} Scheme of the back-end electronics of the ICARUS detector
\cite{icarus_daq}.
}}
\end{center}
\end{figure}

\begin{figure}[htp]  
\begin{center}
\includegraphics*[width=3in]{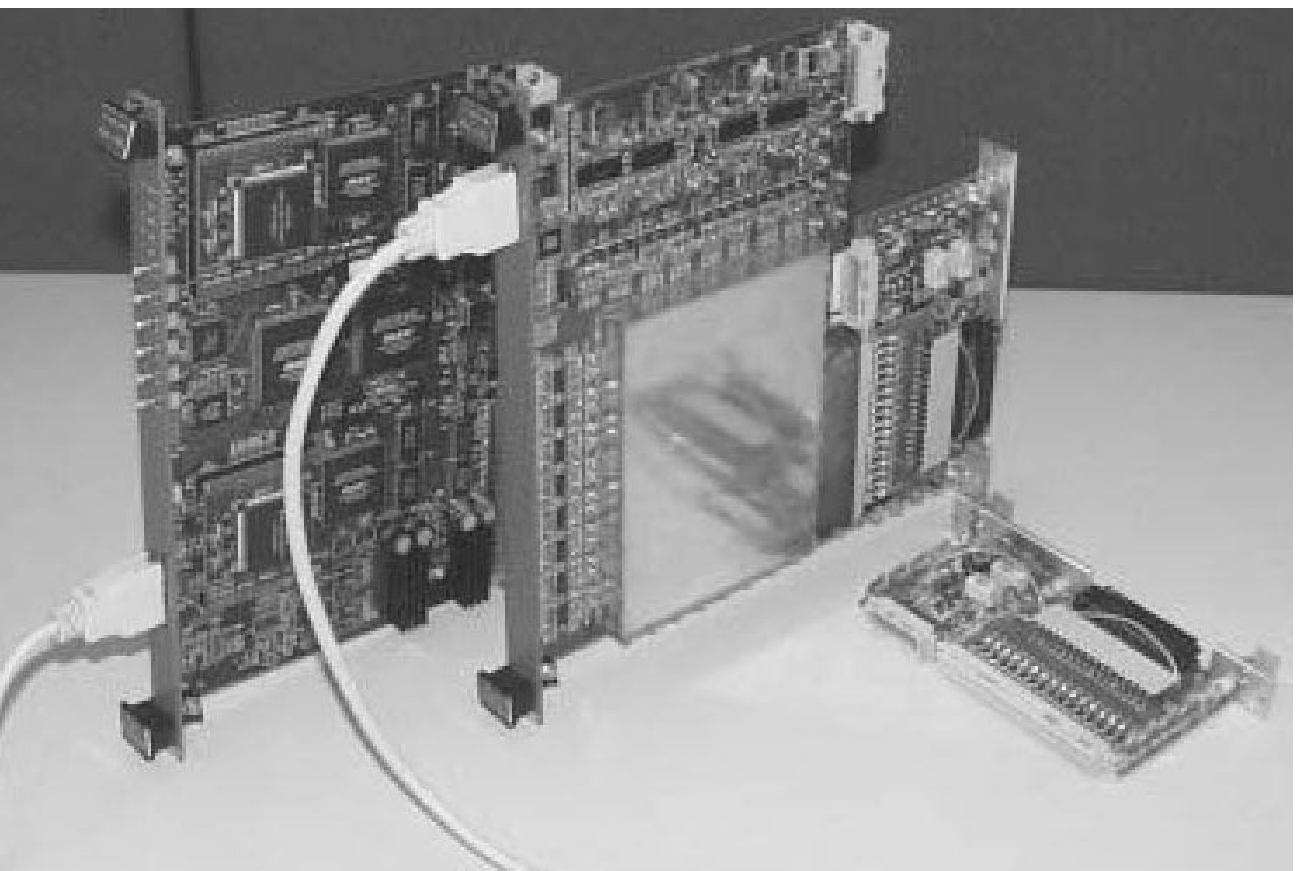}
\parbox{5.5in} 
{\caption[ Short caption for table of contents ]
{\label{modules} Readout electronics of the ICARUS detector
\cite{icarus_daq}.
}}
\end{center}
\end{figure}

In the future, neutrino beams from pion decay may be supplanted by those from muon decay
\cite{study2}, in which cases beams of $\nu_\mu$ and $\bar\nu_e$ or $\bar\nu_\mu$ and $\nu_e$
would be available.  To take advantage of such beams for studies of CP violation, the
sign of the final state lepton much be determined in the detector, which must therefore
be immersed in a magnetic field.  A liquid argon TPC is compatible with a superimposed
magnetic field.  The sign of muons could be well determined with a field of only 0.1 T,
and the sign of electrons up to a few GeV energy could be determined in a field of 0.5 T
by analysis of the curvature of tracks in the electromagnetic shower 
\cite{venice,argonprop}. 

As well as being the highest-performance large detector for neutrinos, a liquid argon TPC
is also one of the least costly.  Liquid argon costs about \$1M per kton, or \$20m for a
20 kton detector.  Such a detector would require about 100k readout channels, 
at a cost of about \$10M.  The greatest cost uncertainty at present is in the storage tank
and associated cryogenic/purification system.  A initial engineering study on this
issue is underway \cite{Mulholland}.  If the cryo system cost proves to be of order the
cost of the argon itself, then a 20 kton detector would cost about \$50M -- which would make
a liquid argon TPC the price/performance leader for large neutrino detectors.

\chapter{ Possible Sites for a NuMI Off-Axis Detector}

\section{Criteria for Site Selection}

There appears to be general agreement that because of the 10  $\mu s$
beam spill, a NuMI off-axis detector can be located
in a surface or near-surface laboratory.
Besides the general physics considerations 
 discussed in the preceeding section there are more specific 
requirements that determine whether a give site is adequate. The most 
important ones are listed below.                 

\begin{enumerate}

\item The site should be at least $450~ m$ along an axis pointing
towards Fermilab and $100~ m$ along the transverse access.
Such a site will accommodate the largest possible detector, earth
berms for shielding, if they are required, and space for ancillary
facilities. Access 
to the entire site should be protectable by lease or options,
even though only a portion of the site is necessary for the first
stage laboratory.

\item The site should permit cost-effective construction of the detector
and associated structures, if any. The detector may need shielding
and good availability of material for shielding is an advantage.

\item The site should have good highway access. A location close
to Soudan is not a requirement. However, a site close enough
to Soudan to enable easy staff travel between the laboratories
would facilitate laboratory management and result in savings
in operating costs.

\item The site should have good access to $\sim 1$ MW of electrical
power and telecommunications, including fiber optic data links.
The site needs access to water in quantities that depend on detector
design. Existing sewage treatment access would be a plus for
any site.

\item An ideal site would have low environmental and political risk,
that is, not adjacent to large populations, to conservation or
park areas or to areas of natural beauty, such as shorelines
or other features. An ideal site would also have soil and rock
with low content of sulfur and heavy metals, in order to minimize
environmental issues due to leaching.

\item Depending on future information about
information about $\Delta m^{2}$, matter effects or CP violation,
the detector may be moved at some future time. Although such a relocation
might need to be far, in order to substantially change the baseline,
another possibility is a short move transverse to the beam. Thus,
a site near other ideal sites located at different
transverse dimensions from the beam center line would have some advantage.

\item An ideal site would have gravity drainage, even for an underground
laboratory location. Such drainage reduces both operating cost
and risk of flooding due to electrical or mechanical failure.

\end{enumerate}

\section{Potential Sites}
We have initial examination of possible sites which might satisfy the 
above requirements. So far, only one has been sufficiently studied. This 
site, at $712~ km$ from Fermilab is a former mining site owned by LTV.
 Preliminary discussions with the appropriate parties lead us to believe that 
there would be a positive response to our potential request to site the 
detector there.  The principal negative factor is that the site is not as 
far away from Fermilab as one would desire ideally. It is probably adequate 
if $\dmsq23$ is above $2.5\times 10^{-3}$. The optimal energy for 
$\dmsq23 =2.5\times10^{-3}$ would 
be somewhat below $1.5 GeV$. Such a low energy might make the rejection of NC 
background somewhat easier. The site is described in more detail in the 
Appendix B.

From examination of the map of Ontario, there are two potential site areas 
further north, in Canada. The first one lies along the 
southern Trans Canadian Highway, and is about 850 km away from Fermilab. A 
somewhat more distant variant in this site area would be along Rt 502 which 
runs north by north east from the main highway and could offer a site about 
$20 - 40~ km$ further. The other site area would be at about $985~ km$, 
paralleling 
the northern branch of the Trans-Canadian Highway. There are also two railroad
 tracks along the highway, one just south of it and the other one about 
$20~ km$ 
north of it. This site would be a little too far if $\dmsq23$ is above 
$2.5 \times 10^{-3} eV^2$. 
At this time we do not know whether it would be possible 
to operate a neutrino detector in either one of these areas.

\begin{figure}
\begin{center}
\includegraphics[scale=0.5]{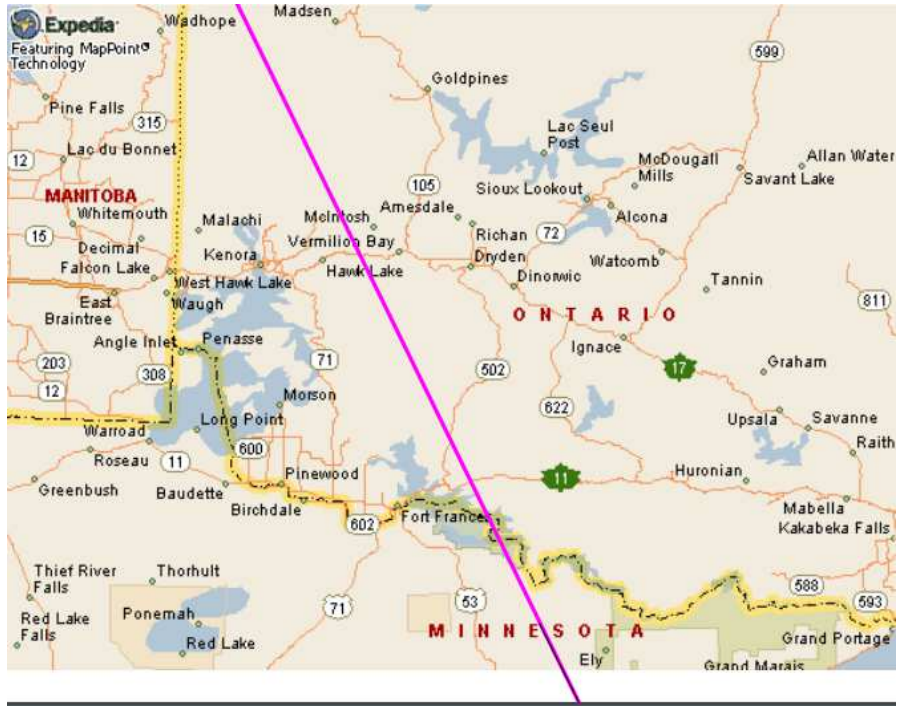}
\caption{Map of southern Ontario and the NuMI beam direction} 
\label{canada}
\end{center}
\end{figure}

Looking further into the future one might  look at possible sites for 
Phase II, perhaps at the $L/E$ corresponding to the second oscillation 
maximum. This would involve going 
further north, to a distance of about $1100~ km$ and energy around $1~ GeV$. 
A possible site is in the neighborhood of Red Lake. This is a pretty 
desolate area and on-site investigation will be required to see if it 
could be suitable. A map of this area of Canada, with the 0 deg beam line 
superimposed, is shown in Fig.~\ref{canada}. A list of the potential sites for an off-axis experiment is shown in Table~\ref{sites}.

\begin{table}
\begin{center}
\footnotesize
\tabcolsep 0.15cm
\footnotesize

\begin{tabular}{lcccc}
\hline
Area & Distance & Depth & Access & Comments\\
& to Fermilab& & & \\
\hline
Iron River  & 612 km & 6.5 km & Wisconsin & Near mouth of \\
 & & & Highway 13 & Iron River; \\
 & & & & rural, forest \\
\hline
Two Harbors & 643 km & 5.2 km & U.S. Highway & Two Harbors \\
(Lake Superior & & & 61 & MN; mixed \\
North Shore) & & & &urban and rural \\
\hline
Tower-Soudan-Embarass & 720 to 735 km & 1 km & MN Highway & Off-axis \\
 & & & 169 & locations rural,\\
 & & & & forest \\
\hline
 Buyck & 775 km & -1.8 km & Orr-Buyck-Ely & Off-axis \\
  & & & Road (Echo & locations rural, \\
  & & & Trail) & forest \\
\hline
Ash River & $\sim 800$ km & -3.4 km & Ash River Trail, & Kabetogama \\
 & & & east of U.S. 53 & State Forest \\
 & & & & south side of \\
 & & & & Voyageur's \\
 & & & & National Park \\
\hline
Farrington & $\sim 850$ km & -6.9 km & Ontario & \\
 & & & Highway 11 & \\
 & & & east of Fort & \\
 & & & Frances & \\
\hline
Otukamamoan Lake & $\sim 900$ km & -10.9 km & Ontario & \\
 & & & Highway 502 & \\
 & & & northeast of Fort & \\
 & & & Frances & \\
\hline
Vermilion Bay & $\sim 950$ km & -15.3 km & Ontario & \\
 & & & Highway 17 & \\
 & & & east of Kenora, & \\
 & & & west of Dryden & \\
\hline
\end{tabular}
\label{sites}
\caption{Possible sites for an off-axis experiment}
\end{center}
\end{table}

\section{LTV sites }

While the investigation of possible sites continues, the sites
for which we have the most information already more than meet
the minimum criteria for an acceptable site. In particular, we
have identified two sites that lie in Minnesota's Iron Range,
a mixed hematite, magnetite and taconite deposit several kilometers
in width that extends $\sim 200$ km ENE from Grand Rapids MN
to Babbitt MN. These two sites are just north of Erie Pit No. 1
in the former LTV Steel Mine about 10 km north of Aurora MN.
Both sites consist of tailings piles, each $\sim 15-20$ m
in height, 500 m in the beam direction and $>200~m$
transverse to the beam direction. The westerly of the two sites
is 10.3 km from the beam axis; the easterly site is 9.1 km
from the beam axis. Both sites are $\sim 712$ km from
Fermilab. These sites are currently owned by Cliffs-Erie LLC,
an iron mining company. There is neither active mining nor plans
for mining within several kilometers of either site. Both sites
are accessible by a private two-lane paved highway from MN Highway
135. The sites are located $\sim 20$ miles by state highway
from Soudan.

Both LTV sites have good access to mine electrical power. An
existing power line connects the two sites. Public utility power
is available along Highway 135 and it may be possible to reconfigure
the existing line to provide public utility power to either site.
Fiber optic telecommunications exist along Highway 135 and connect
to cell phone towers located on the peaks of the Embarrass Mountains
less than 2 km north and west of the LTV sites. Erie Pit No.
1, which is about 6 km in length, is a potential source of water
for a detector and also provides natural drainage for both sites.
The LTV Mine is $\sim 20$ km transverse to the beam line. It
extends past the beam centerline to the east and about 5 km to
the west of these two sites. Numerous tailings piles can be found
on the LTV site if moving the detector becomes desirable at some
later time. Tailings piles also exist further west along Highway
135, both east and west of Biwabik MN. Biwabik is $\sim 20$
km west of the NuMI beam centerline.

Preliminary information from Michael Johnson, the site manager
for Cliffs-Erie indicates low levels of sulfur and heavy metals
at these sites. (The other end of the LTV property, east of the
beam centerline is believed to have copper-nickel-precious metal
deposits in a sulfide ore, with new mining activity possible
during the next several years.) Thus, we believe from current
information the environmental risk associated with this property
is low. Because of previous mining activity, the political risk
is also low. Indeed, the IRRRB, the regional state economic development
agency, has already indicated positive political and possible
financial support for these sites. The positive reputation of
the Soudan Lab is also strong in this area, which also minimizes
political risk. Mr. Johnson on behalf of Cliffs-Erie has indicated
a willingness of the company to discuss various acquisition options
any time we are ready.

In summary, the LTV sites located 9.1 and 10.3 km transverse
to the beam centerline, 712 km from Fermilab provide high-quality,
low-risk areas for an off-axis detector, which could be accessed
in a short period of time.


\begin{thebibliography}{3}
\bibitem{nufact}
there is a huge number of papers on neutrino factories covering the machine 
and experimental aspects as well as the physics potential. A dedicated series
of conferences NuFactXX is covering the progress of the project. Proceedings 
of these conferences and the references cited therein give a good indication 
of the amount of effort devoted.
\end{thebibliography}
\end{document}